\documentclass[]{pasj02}
\usepackage[switch,mathlines]{lineno} 
\usepackage{xcolor}
\jyear{2024}
\Received{}
\Accepted{}

\begin{document} 
\title{Rotation and stability of the circumnuclear gas disk in the Galactic Center potential by the ALMA CMZ Exploration Survey (ACES)} 
\def\afmark{\altaffilmark}
\def\aftext{\altaffiltext} 
\def\scr{\scriptsize}  
\author{
Yoshiaki \textsc{Sofue}\afmark{1}\orcid{0000-0002-4268-6499}\email{sofue@ioa.s.u-tokyo.ac.jp}\aftext{1}{\scr Institute of Astronomy, The University of Tokyo, Mitaka, Tokyo 181-0015, Japan},
%
\orcid{0000-0001-6353-0170}{Steven N. \textsc{Longmore}}\afmark{3,4}\aftext{3}{Astrophysics Research Institute, Liverpool John Moores University, IC2, Liverpool Science Park, 146 Brownlow Hill, Liverpool L3 5RF, UK}\aftext{4}{CCosmic Origins Of Life (COOL) , https://coolresearch.io},
%
\orcid{0000-0001-7330-8856}{Daniel \textsc{Walker}}\afmark{5}\aftext{5}{{\scr UK ALMA Regional Centre Node, Jodrell Bank Centre for Astrophysics, Oxford Road, The University of Manchester, Manchester M13 9PL, United Kingdom}},
%
\orcid{0000-0001-6431-9633}{Adam \textsc{Ginsburg}}\afmark{6},\aftext{6}{Department of Astronomy, University of Florida, P.O. Box 112055, Gainesville, FL 32611}
%
\orcid{0000-0001-9656-7682}{Jonathan D. \textsc{Henshaw}}\afmark{3,7},\aftext{7}{Max Planck Institute for Astronomy, K\"{o}nigstuhl 17, D-69117 Heidelberg, Germany}
%
\orcid{0000-0001-8135-6612}{John \textsc{Bally}}\afmark{8},\aftext{8}{Center for Astrophysics and Space Astronomy; Department of Astrophysical and Planetary Sciences; University of Colorado, Boulder, CO 80389, USA}
%
\orcid{0000-0003-0410-4504}{Ashley T. \textsc{Barnes}}\afmark{9},\aftext{9}{European Southern Observatory (ESO), Karl-Schwarzschild-Strasse 2, D-85748 Garching, Germany}
%
\orcid{0000-0002-6073-9320}{Cara \textsc{Battersby}}\afmark{10},\aftext{10}{Department of Physics, University of Connecticut, 196A Auditorium Road, Unit 3046, Storrs, CT 06269, USA}
%
\orcid{0000-0001-8064-6394}{Laura \textsc{Colzi}}\afmark{11},\aftext{11}{Centro de Astrobiología (CAB), CSIC-INTA, Carretera de Ajalvir km 4, Torrejón de Ardoz, 28850 Madrid, Spain}
%
Paul \textsc{Ho}\afmark{12},\aftext{12}{AS/NTU Astronomy-Mathematics Building,  Roosevelt Rd, Taipei 10617, Taiwan}\orcid{0000-0002-3412-4306}
Izaskun \textsc{Jimenez}-\textsc{Serra}\afmark{11}
\orcid{0000-0003-4493-8714},
%
\orcid{0000-0002-8804-0212}J.~M.~Diederik \textsc{Kruijssen}\afmark{4},
%
Elizabeth \textsc{Mills}\afmark{13},\aftext{13}{Department of Physics and Astronomy, University of Kansas, 1251 Wescoe Hall Drive, Lawrence, KS 66045, USA}\orcid{0000-0001-8782-1992}
\orcid{0000-0002-6362-8159}Maya A. \textsc{Petkova}\afmark{14},\aftext{14}{Space, Earth and Environment Department, Chalmers University of Technology, SE-412 96 Gothenburg, Sweden}
%
\orcid{0000-0001-6113-6241}{Mattia C. \textsc{Sormani}}\afmark{15},\aftext{15}{Como Lake centre for AstroPhysics (CLAP), DiSAT, Universit{\`a} dell'Insubria, via Valleggio 11, 22100 Como, Italy}
%
\orcid{0000-0000-0000-0000}Jen \textsc{Wallace}\afmark{10},
\orcid{0000-0003-3341-6144}{Jairo  \textsc{Armijos-Abenda\~no}}\afmark{17},\aftext{17}{Observatorio Astron\'omico de Quito, Observatorio Nacional, Escuela Polit\'ecnica Nacional, Interior del Parque La Alameda, 170136, Quito, Ecuador}
%
\orcid{0009-0004-0121-1560}{Zi-Xuan Feng}\afmark{15,29},\aftext{15}{Como Lake centre for AstroPhysics (CLAP), DiSAT, Universit{\`a} dell'Insubria, via Valleggio 11, 22100 Como, Italy}\aftext{29}{Universit\"{a}t Heidelberg, Zentrum f\"{u}r Astronomie, Institut f\"{u}r Theoretische Astrophysik, Albert-Ueberle-Str.\ 2, 69120 Heidelberg, Germany}
%
\orcid{0000-0002-9409-8322}{Karl \textsc{Fiteni}}\afmark{15,32},\aftext{32}{Institute of Space Sciences \& Astronomy, University of Malta, Msida MSD 2080, Malta}
%
%
%
%
\orcid{0000-0002-8586-6721}{Pablo \textsc{Garc\'ia}}\afmark{21,22},\aftext{21}{Chinese Academy of Sciences South America Center for Astronomy, National Astronomical Observatories, CAS, Beijing 100101, China}\aftext{22}{Instituto de Astronom\'ia, Universidad Cat\'olica del Norte, Av. Angamos 0610, Antofagasta, Chile}
%
\orcid{0000-0002-1313-429X}{Savannah \textsc{Gramze}}\afmark{6},
%
%
\orcid{0000-0002-7495-4005}{Christian \textsc{Henkel}}\afmark{24},\aftext{24}{MPIfR, Auf dem H\"ugel 69, Bonn, Germany}
\orcid{0000-0001-9155-3978}{Pei-Ying \textsc{Hsieh}}\afmark{19},\aftext{19}{National Astronomical Observatory of Japan, 2-21-1 Osawa, Mitaka, Tokyo 181-8588, Japan}
%
%
%
%
Ralf S.\ \textsc{Klessen}\afmark{29,30},\aftext{29}{Universit\"{a}t Heidelberg, Zentrum f\"{u}r Astronomie, Institut f\"{u}r Theoretische Astrophysik, Albert-Ueberle-Str.\ 2, 69120 Heidelberg, Germany}\aftext{30}{Universit\"{a}t Heidelberg, Interdisziplin\"{a}res Zentrum f\"{u}r Wissenschaftliches Rechnen, Im Neuenheimer Feld 225, 69120 Heidelberg, Germany} \orcid{0000-0002-0560-3172}
%
%
%
%
%
%
Francisco \textsc{Nogueras-Lara}\afmark{31},\aftext{31}{Instituto de Astrof\'isica de Andaluc\'ia (CSIC), Glorieta de la Astronom\'ia s/n, E-18008 Granada, Spain} \orcid{0000-0002-6379-7593}
%
%
%
\orcid{0000-0002-5811-0136}{Dylan M. Par\'e}\afmark{33},\aftext{33}{Joint ALMA Observatory, Alonso de Cordova 3107, Vitacura, Casilla 19001, Santiago de Chile, Chile}
%
%
%
%
\orcid{0000-0002-2887-5859}{Víctor M. \textsc{Rivilla}}\afmark{11}, 
%
%
%
%
%
%
%
%
%
and 
\orcid{0000-0002-3078-9482}{\'Alvaro S\'anchez-Monge}\afmark{42,43}\aftext{42}{Institut de Ci\`encies de l'Espai (ICE), CSIC, Campus UAB, Carrer de Can Magrans s/n, E-08193, Bellaterra, Barcelona, Spain}\aftext{43}{Institut d'Estudis Espacials de Catalunya (IEEC), E-08860, Castelldefels, Barcelona, Spain}
}  


\maketitle

 
\def\be{\begin{equation}}\def\ee{\end{equation}}\def\vlsr{v_{\rm LSR}} \def\Vlsr{\vlsr} 
\def\vr{v_{\rm r}}\def\deg{^\circ}  \def\d{^\circ}
\def\vrot{V_{\rm rot}} \def\Vrot{\vrot}\def\co{$^{12}$CO } \def\coth{$^{13}$CO $(J=1-0)$} \def\Xco{X_{\rm {^{12}}CO}}   \def\Tb{T_{\rm B}} \def\Tp{T_{\rm p}}\def\Htwo{H$_2$} \def\htwo{H$_2$}  \def\Kkms{K km s$^{-1}}  \def\Hcc{{\rm H \cm^{-3}}} \def\hcc{{\rm H \cm^{-3}}} \def\kms{km s$^{-1}$} \def\Ico{I_{\rm CO}}  \def\Kkms{K \kms }\def\mH{m_{\rm H}}  \def\Ico{I_{\rm ^{12}CO}} \def\Icoth{I_{\rm ^{13}CO}} 
 \def\htwo{H$_2$} \def\Tb{T_{\rm B}}   \def\mH{m_{\rm H}} \def\ekms{~{\rm \ km \ s^{-1}}~}  \def\epc{{\rm \ pc} }  \def\Hii{HII} \def\apj{ApJ} \def\aap{A\&A} \def\mnras{MNRAS} \def\pasj{PASJ} \def\aj{AJ} \def\xcounit{H$_2$ cm $^{-2}$ [K km s$^{-1}]^{-1}$}  \def\log{{\rm log}}  \def\tc{t_{\rm C}} \def\fc{f_{\rm C}} \def\SFR{{\rm SFR}} \def\sfr{{\rm SFR}}\def\tc{t_{\rm c}}\def\lc{l_{\rm c}}\def\vc{v_{\rm c}}\def\tp{t_{\rm p}}\def\rc{r_{\rm c}}\def\nc{n_{\rm c}}\def\pcc{p_{\rm c}}  
\def\Msun{M_{\odot \hskip-4.8pt \bullet}}  
\def\Lsun{L_{\odot \hskip-4.8pt \bullet}}        
\def\kms{km s$^{-1}$}  \def\deg{^\circ}   \def\Htwo{H$_2$\ }  \def\fmol{f_{\rm mol}} \def\Fmol{ $f_{\rm mol}$ }  \def\sfu{\Msun~{\rm y^{-1}~kpc^{-2}}} \def\sfuvol{\Msun~{\rm y^{-1}~kpc^{-3}}}\def\log{{\rm log}}
\def\hcc{{\rm H~cm^{-3}}} \def\Hcc{ $\hcc$ }\def\Htot{ H$_{\rm tot}$ } \def\ssfr{\Sigma_{\rm SFR}} \def\vsfr{\rho_{\rm SFR}} \def\sfr{{\rm SFR}}\def\H{{\rm H}}\def\cm{{\rm cm}}\def\kpc{{\rm kpc}} \def\bc{\begin{center}}\def\ec{\end{center}} 
\def\xcounit{\Htwo cm$^{-2}$ [K \kms]} \def\pc{{\rm pc}}\def\My{{\rm My}} \def\kpc{{\rm kpc}}\def\rc{r_{\rm c}} \def\vc{v_{\rm c}}
\def\urho{\Msun \pc^{-3}}\def\urhohtwo{{\rm H_2} \cm^{-3}} \def\nc{n_{\rm c}} 
\def\vexpa{v_{\rm expa}} \def\rbub{r_{\rm b}}\def\x{\times}\def\xfour{\times 10^4}\def\xthree{\times 10^3}\def\xfive{\times 10^5}\def\xfifty{\times 10^{50}}\def\xmtwe{\times 10^{-20}} \def\sigv{\sigma_v} \def\Rbow{R_{\rm bow}} \def\Rzero{R_0} \def\Lcone{L_{\rm cone}}  \def\Rhii{R_{\rm HII}} \def\nuv{n_{\rm UV}} \def\ni{n_{\rm i}}  \def\ar{a_{\rm r}}
\def\Te{T_{\rm e}}\def\Tn{T_{\rm n}}\def\cosh{{\rm cosh}}\def\({\left(} \def\){\right)}\def\[{\left[} \def\]{\right]}\def\Hcc{H cm$^{-3}} \def\Hsqcm{H cm$^{-2}$} \def\L{\mathcal{L}}\def\Rc{R_{\rm c}} \def\rhom{\rho_{\rm m}} \def\rhoc{\rho_{\rm c}} \def\V{V_{\rm rot}} \def\Vp{V_{\rm pat}} \def\vpat{V_{\rm pattern}}\def\red{\textcolor{red}} 
\def\ss{\subsection}\def\sss{\subsubsection}
\def\hcnaces{H$^{13}$CN $(J=1-0)$}
\def\hcn{H$^{13}$CN $(J=1-0)$}
\def\cs{CS ($J=2-1$)}
\def\csaces{CS ($J=2-1$)}
\def\hcnaste{HCN ($J=4-3$)}
\def\dvdl{d\vlsr/dl}
\def\sgrastar{Sgr A$^*$} 
\def\vex{V_{\rm ex}}
\def\Jybeam{Jy beam$^{-1}$}
\def\cc{cm$^{-3}$}
\def\asec{''\!\!}
\def\degd{^\circ\!\!}
\def\htwocc{\htwo cm$^{-3}$ }
\def\G02{G+0.02-0.02+100}
\def\vex{V_{\rm expa}}
\def\vexpa{V_{\rm expa}}
\def\ekpc{{\rm kpc}}
 \def\Htwocc{\Htwo~cm$^{-3}$}
 \def\degp{\deg\!.}
 \def\Jyb{Jy beam$^{-1}$}
 \def\lw{\linewidth}
 \def\be{\begin{equation}} \def\ee{\end{equation}}
 \def\nH2{n_{\rm H_2}}
 \def\epc{{\rm pc}} 
 \def\rhomass{\rho_{\rm dyn}} 
 \def\h40{H40$\alpha$} 
\def\lw{\linewidth}  

\begin{abstract} 
We investigated the gravitational potential and mass distribution in the Galactic Center by examining the morphology and kinematics of the circumnuclear gaseous disk revealed by the molecular line data from the ALMA CMZ Exploration Survey (ACES). We obtain an estimate of the shape of the potential {within the central $\sim 20$ pc} to reproduce the observed properties of the circumnuclear gas disk (CND) by simulating the motion of test particles for various axial ratios and show that the potential is approximately spherical. We construct a rotation curve by applying the terminal velocity method to the position-velocity diagrams, and calculate the mass distribution in the Galactic Center. The distribution of mass density is found to be of cusp type, approximated by $\rhomass \sim 1.56\times 10^5(R/1 \epc)^{-1.9}~\Msun \epc^{-3}$, where $R$ is the distance from the nucleus.
We discuss the tidal effect caused by the gravitational potential that produces the rotation curve and show that the gas disk is stable against self-gravitational contraction within a critical radius of $
R_{\rm T}\sim 14 ~(\rho_{\rm gas}/10^5 {\rm H_2~cm^{-3}})^{-1/2}~{\rm pc}$. This suggests suppression of star formation and a top-heavy IMF in the circmunuclear region.
\end{abstract}
 
 
\section{Introduction}
\label{intro}  

The central molecular zone (CMZ) of our Galaxy is a high-density molecular gas disk with moderate star formation \citep{h16,henshaw+2023,2022MNRAS.516..907S,2025ApJ...984..157B}, which rotates in the deep gravitational potential of the Galactic bulge and the central supermassive black hole.
Extensive studies of the kinematics of the CMZ have revealed a large-scale multi-arm structure \citep{1995PASJ...47..527S,2022MNRAS.516..907S} and/or a twisted ellipse structure \citep{2011ApJ...735L..33M,lon13,k15,h16,2025ApJ...984..159L}. 
Recently, we \citep{sofue+25a,sofue+25b} showed that the inner CMZ is composed of several rotating arms and rings using the molecular line cube data taken by the large project ACES (ALMA CMZ Exploration Survey) \citep{ACESI,ACESII,ACESIII,ACESIV,ACESV}, where the kinematic properties of the molecular gas in the position-velocity diagrams (PVD) are shown to provide information about the gravitational potential in which it is orbiting. 

ALMA observations have also provided detailed individual studies of the circumnuclear disk (CND) \citep{hs21}.
However, the innermost region within 20 pc around \sgrastar, one of the major parts of the CMZ, has not yet been thoroughly studied, and in fact multiple arm and ring structures have been recognized \citep{sofue+25a,sofue+25b}.
The high-resolution and high-sensitivity data from ACES with uniform mapping quality across the entire CMZ makes it possible to investigate these individual structures from a more general perspective.

The innermost region of the CMZ is known to be the CND identified as Arms V and VI in \citet{sofue+25a}, which are high density molecular tori of radii $\sim 2$ to $\sim 10$ pc \citep{2011ApJ...732..120O,2012ApJ...756..195L,2013ApJ...779...47M,2017ApJ...850..192M,2017ApJ...847....3H,hs21,2018PASJ...70...85T}. 
The CND exhibits two-fold dynamical properties that it is a torus that being fueled from the outer arms of the CMZ and fuels the more inner region, including the nucleus
and the minispiral
\citep{2004A&A...426...81P,2012A&A...538A.127K,2012ApJ...756..195L,2017A&A...603A..68M,2016PASJ...68L...7T,2017ApJ...842...94T}.

Extensive hydrodynamical simulations have been performed to study the evolution and gas dynamics in the Galactic Center 
\citep{k15,kruijssen+2019,2017MNRAS.466.1213K,2019MNRAS.486.3307D,2017MNRAS.469.2251R,2022MNRAS.514L...1S,2020MNRAS.499.4455T,2024A&A...692A.216H}.
In the current work, there have been two types of density profiles to create the gravitational potential as listed below:
(i) Plateau or finite peak type density distribution as $\rho\propto 1/(1+(R/a)^n)$ \citep{1975PASJ...27..533M} or $\rho\propto {\rm exp}(-b(R/a)^n)$ \citep{2002A&A...384..112L},
where $R$ is the radius from the nucleus, $a$, $b$ and $n$ are constants.
(ii) Cusp type with infinite central density as $\rho\propto R^{-n}$ with $n=2$ \citep{binney+1991} 
or $n=1$ \citep{1997ApJ...490..493N}.
In 2D or 3D treatments, the scale radii and height are taken as free parameters to represent the shape of the potential. 

In order to construct a more realistic model to understand the CMZ, which is a gaseous disk orbiting in the gravitational potential, the determination of the underlying potential based on observations is crucial.
The molecular line data of ACES, covering the entire CMZ from the nucleus to its edge at high angular and velocity resolutions, are ideal for this purpose.
In this paper, our aim is to constrain the gravitational potential of the GC region by analyzing the position-velocity (PV) diagrams of the \cs, \hcn\ and \h40\ line data from ACES.

There are two approaches to study the potential or, equivalently, the mass distribution.
One way is to measure the surface density distribution of stars using infrared photometry \citep{2016ApJ...821...44F,2013A&A...549A..57N,2017MNRAS.465.1621P,2020A&A...634A..71G}. 
The other method is to measure the rotation curve (RC) of the gaseous disk and/or the velocity dispersion of the stars \citep{2001ARA&A..39..137S,2020Galax...8...37S}.
We adopted the RC method and applied it to the position-velocity diagrams (PVD) in the molecular and recombination lines of the CMZ taken from the ACES data cubes.

This paper is structured as follows.
In Section \ref{obs} we describe the kinematic properties of inner molecular disks, particularly the circumnuclear disk (CND). 
In Section \ref{simu} we perform test-particle simulations of the kinematical evolution of a molecular cloud and give a constraint on the shape of the potential, which will be shown as spherical.
In section \ref{sec-rc}, we derive the terminal velocity (TV) curve of the CMZ using the longitude-velocity diagrams in the CS and \h40\ lines. 
In Section \ref{sec-mass} we calculate the dynamical mass distribution in the Galactic Center, assuming that the gravitational potential is nearly spherical in the entire CMZ.
Section \ref{sec-jeans} is devoted to the analysis of stability and tidal effect on a molecular clump in the molecular disk with the observed rotation curve. 
{In section \ref{sec-consis} we comment on the implication of the results and the consistency with the current studies.}
We adopted a distance $R_0=8.2$ kpc to the GC, close to the recent measurement \citep{gravity+2019}.

\section{Data and maps}
\label{obs}

We describe the data and maps used in this paper.

\ss{Data}  

The molecular line cubes in this work were taken from the internal release version (2024 August and 2025 October) of the 12m + 7m + TP (Total Power) mode data from the ALMA Cycle 8 Large Program "ALMA Central Molecular Zone Exploration Survey" (ACES, 2021.1.00172.L; Longmore et al. \citep{ACESI}. 
ACES observed the CMZ in ALMA Band 3, covering a frequency range of $\sim$86--101~GHz in six spectral windows of varying spectral resolution and bandwidth. 
In this paper we use the lines \csaces\ (CS) and \hcnaces\ (HCN) to trace the dense molecular gas \citep{2015PASP..127..299S}.
The \cs\ and \hcn\ line data at frequencies of 97.9810 and 86.3399 GHz have angular resolutions of $2\asec.21$ and rms noise of 0.004 \Jybeam (0.10 K) with velocity channel increments of 1.45 and 0.85 \kms, respectively. 
We also used the \h40\ line at 99.02295 GHz to trace the ionized gas of the minispiral in the innermost region around \sgrastar\ at the same angular resolution and a velocity resolution of 1.5 \kms. 
The intensity scales are in \Jybeam, and 1 \Jybeam$=26.1$ and 22.2 K in the main-brightness temperature ($\Tb$) at 98 and 86 GHz, respectively.
In the present work, we cut out the innermost regions from the ACES cubes, which cover a region of the CMZ at $-0\degd.6 \lesssim l \lesssim +0\degd.9$ and $-0\degd.3 \lesssim b \lesssim +0\degd.1$ with spatial and velocity grids of ($0\asec.5 \times 0\asec.5\times 0.15$ \kms) \citep{ACESI}.
 
We mention that the \cs\ line is typically a tracer that may be moderately optically thick. 
\citet{1980ApJ...235..437L} find opacities of 0.3 - 3.0 in classical star-forming regions of the Galactic disk, while \citet{2020A&A...633A.106H} find values of 1.0 - 2.8 in Galactic center clouds. On the other hand, \hcn\ line is relatively optically thin \citep{2023MNRAS.525.4761L}. 
This property will be used to argue for the presence of the central hole in the CND in subsection \ref{ss-cavity}.
 
\begin{figure}   
\begin{center}   
(A) \cs\ moment 0 \\
\includegraphics[width=\lw]{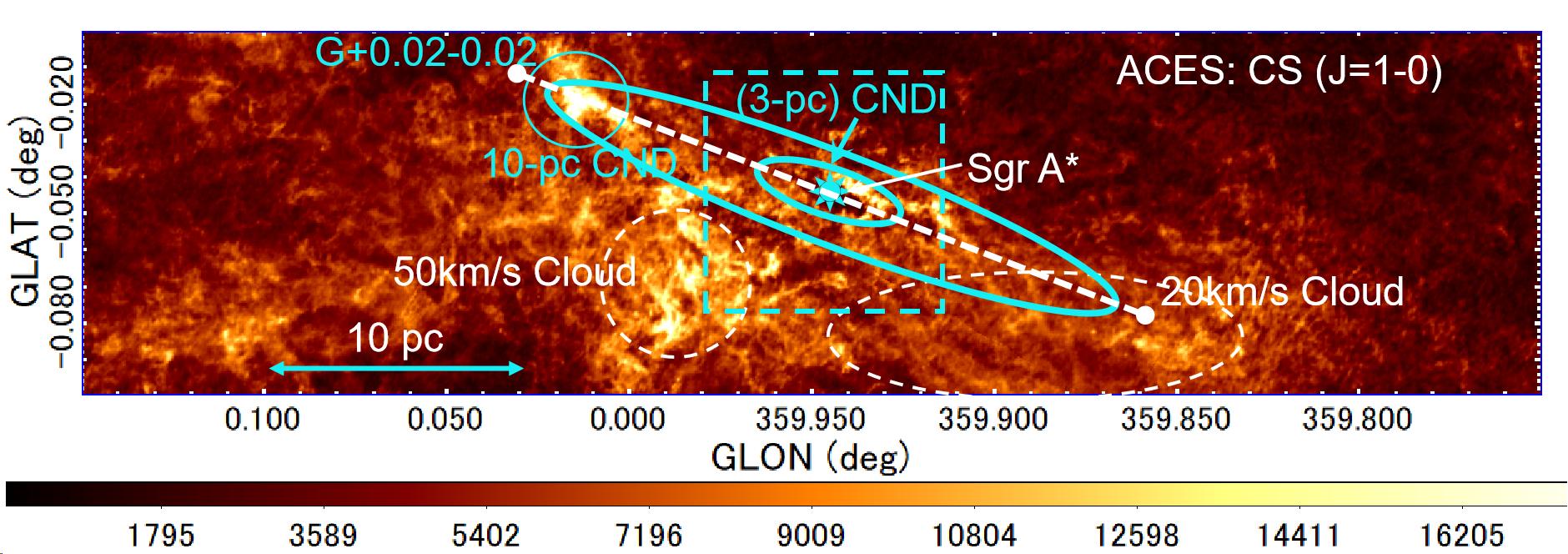}\\
(B) \hcn\ moment 0 \\
\includegraphics[width=\lw]{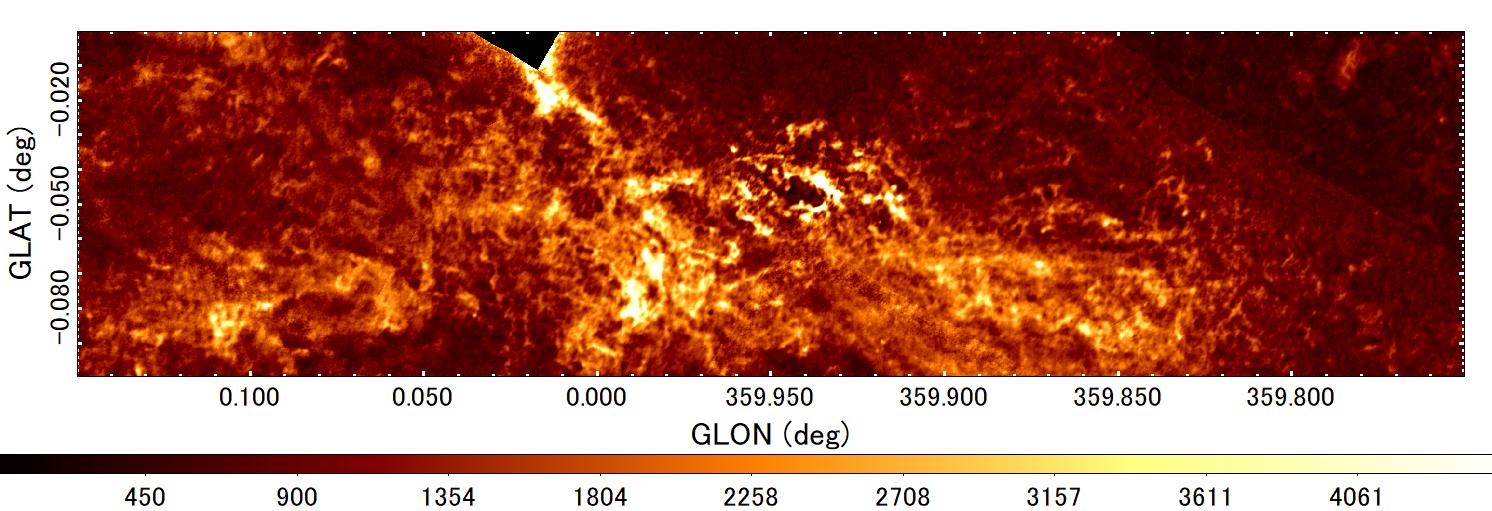}\\
(C) Cross sect. in \cs \\
\includegraphics[width=\lw]{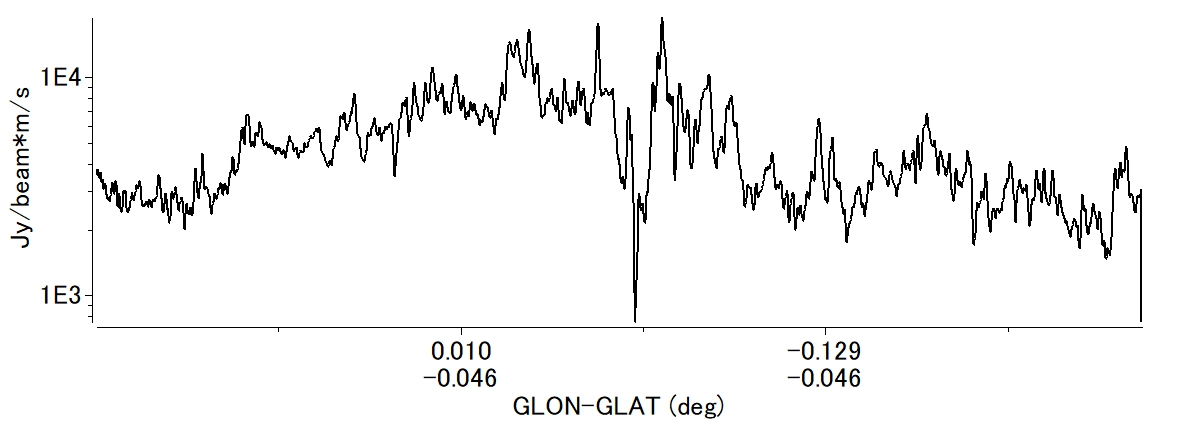}\\
(D) \cs\ moment 1 \\
\includegraphics[width=\lw]{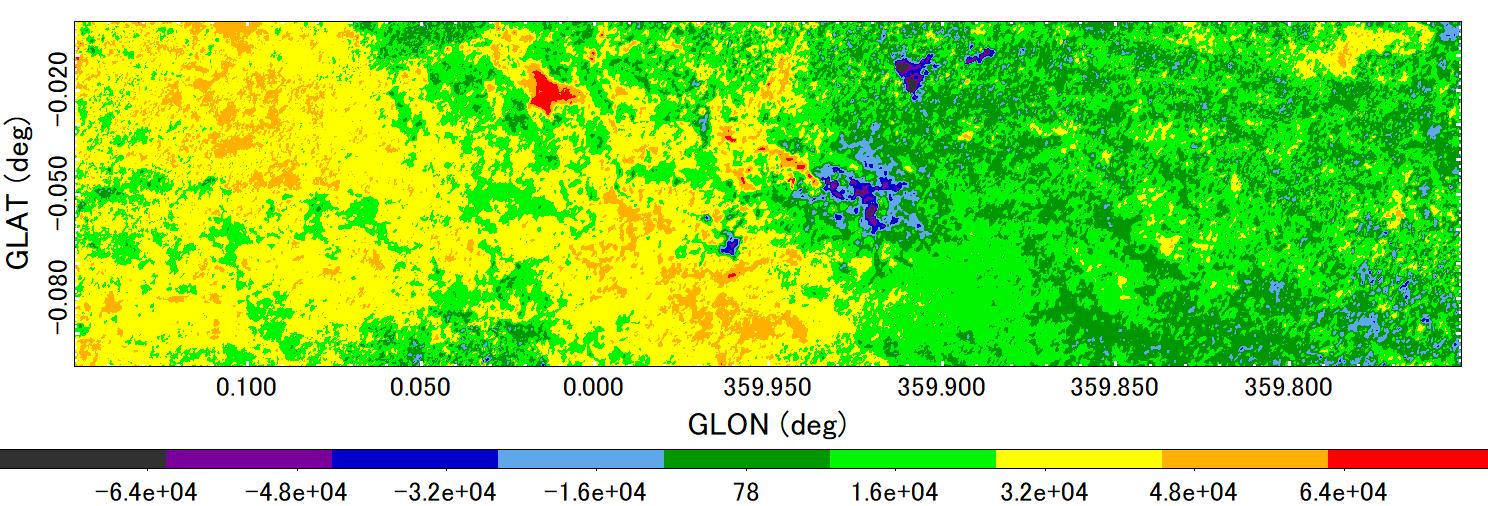}\\
(E) \hcn\ moment 1 \\
\includegraphics[width=\lw]{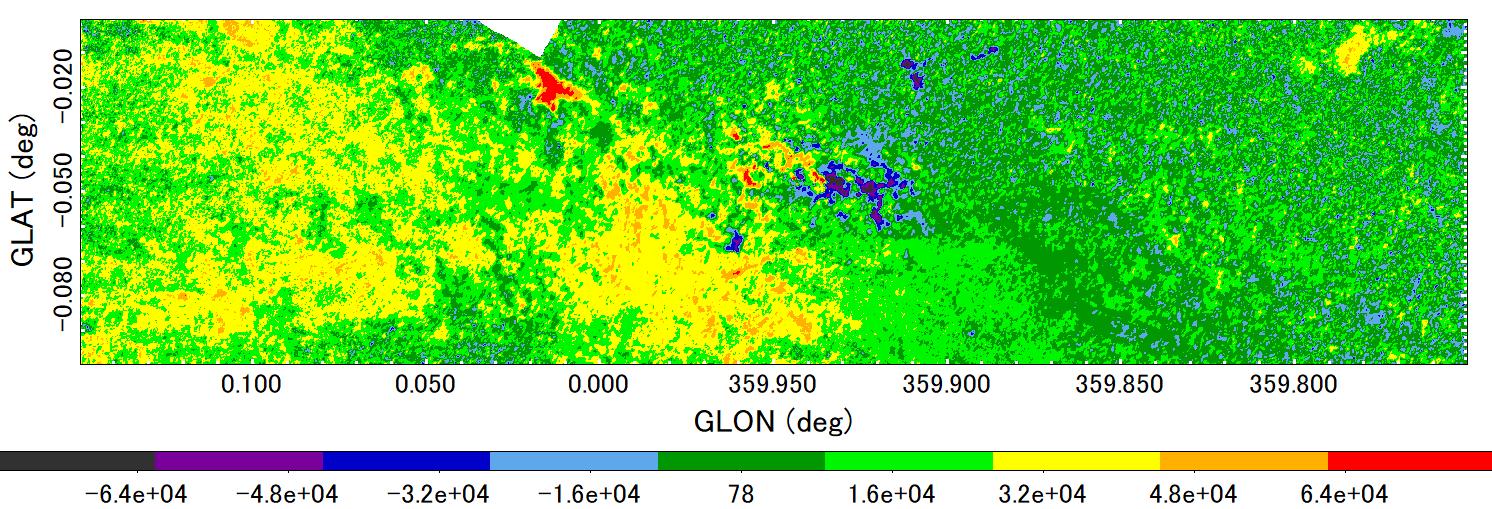}\\
\end{center} 
\caption{
(A) Moment 0 maps of the central $0\deg.4\times 0\deg.1$ region around \sgrastar\ in the \cs\ line by ACES. 
GLON and GLAT stand for $l$ and $b$, respectively.
The structures discussed in the paper are illustrated.
(B) Same, but in \hcn. 
(C) Longitudinal cross section of \cs\ moment 0 map across \sgrastar.
Note the central hole inside the CND, which is enlarged in figure \ref{fig-hole}.
Sharp negative peak is due to absorption of the continuum from \sgrastar.
(D) and (E) Moment 1 maps of \cs\ and \hcn\ lines, respectively, in unit of m s$^{-1}$. 
Note the regular Galactic rotation with positive velocities at positive longitudes and negative velocities at negative longitudes.
{Alt text: Moment 0 maps, intensity cross section, and moment 1 maps in the \cs\ and \hcn\ lines in the central $\pm 0\deg.1$ around \sgrastar. }
}
\label{fig1}  
\label{fig-cnd}  
\end{figure} 

\ss{Maps}

Figure \ref{fig1} shows the moment 0 (integrated intensity) and moment 1 (velocity field) maps of the CND on the lines \cs\ and \hcn.
The CND draws an ellipse in the sky with the major axis at a position angle $PA\sim 70\deg$ and a minor-to-major axial ratio of $b/a\sim 0.2$.
The ellipse is associated with several bifurcated arms trailing outward in the clockwise sense.
The north-east part of the ring is missing due to absorption against the radio continuum of Sgr A, indicating that the upper side of the ring is on the near side of \sgrastar \citep{sofue+25a}.

\begin{figure}     
\begin{center}
\includegraphics[width=.75\lw]{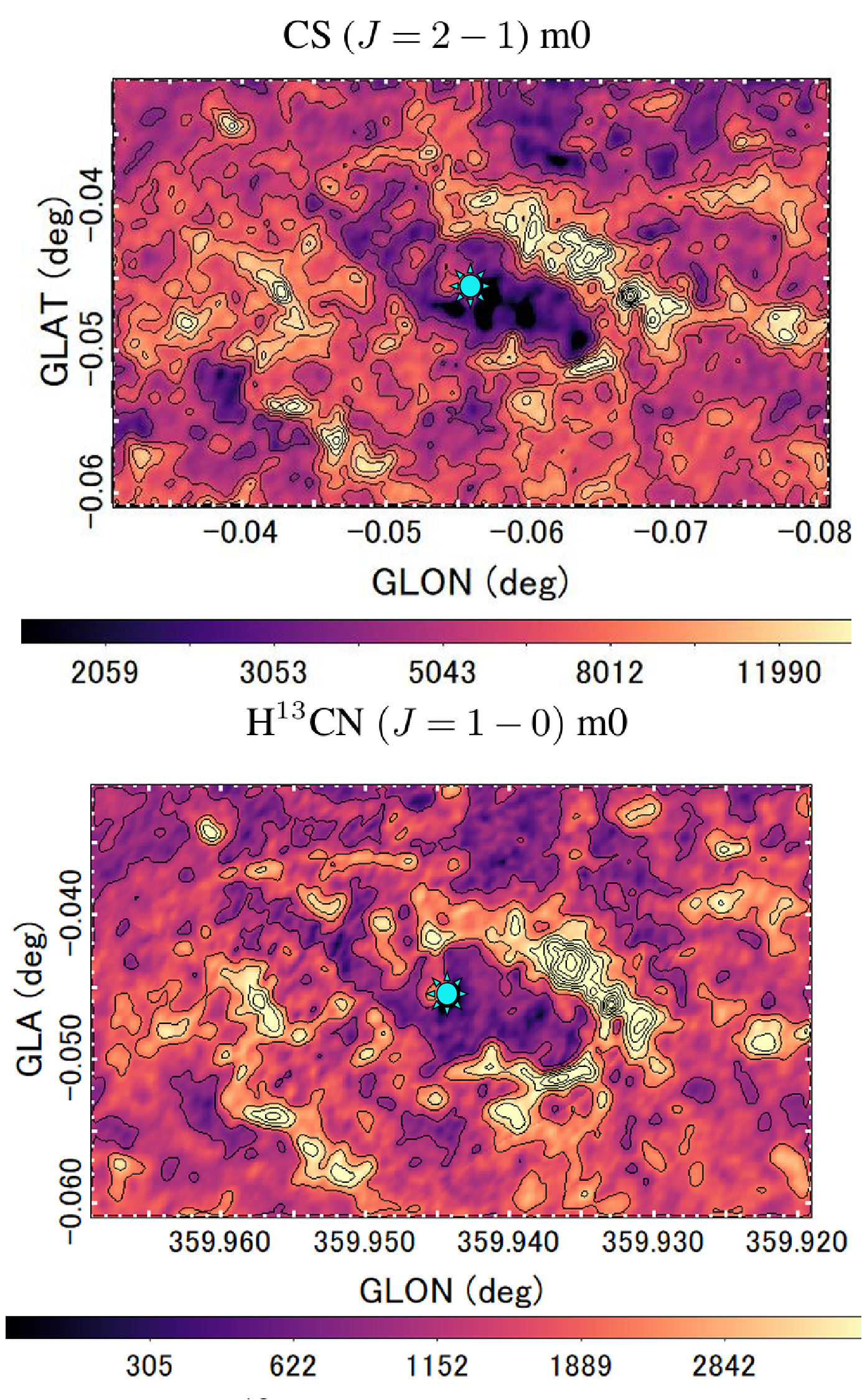} \\  
{H$^{13}$CN/CS ratio (contours=CS m0)}\\
\includegraphics[width=0.75\lw]{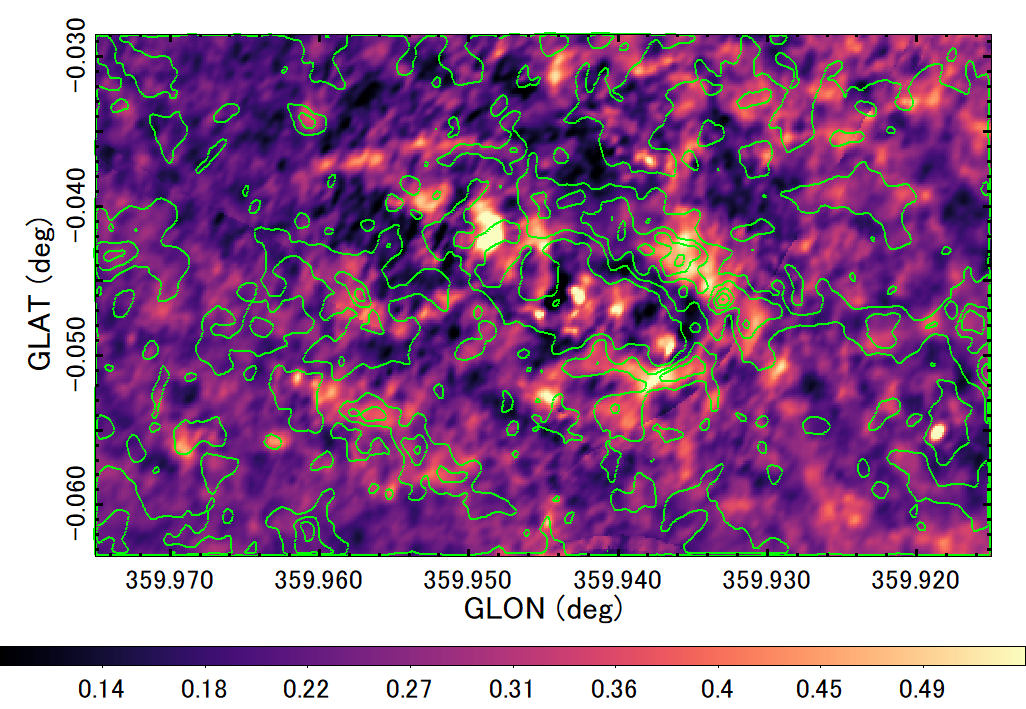} \\ 
{\cs\ m0 cross section.}\\
\includegraphics[width=.7\lw]{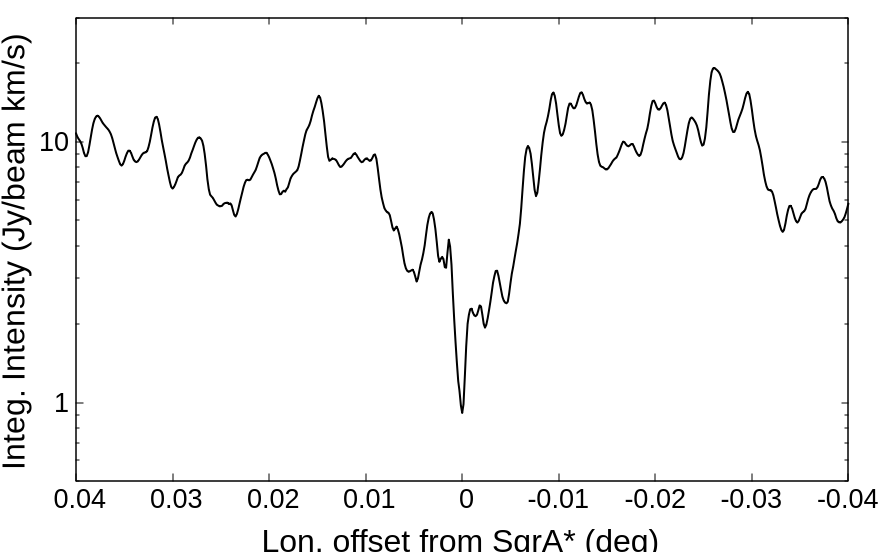}   
\end{center} 
\caption{Close up of the moment 0 maps for the central  $0\deg.05\times 0\deg.03$ region centered on \sgrastar. 
Note the "molecular hole" in the center. 
[Top] \cs\ with contours every 2.5 \Jybeam \kms. 
The blue mark indicates the position of \sgrastar\ at $(l,b)=-0\deg.055835,-0\deg.046110)$ \citep{2022ApJ...940...15X}. 
[2nd] Same, but in \hcn\ with contours every 1 \Jybeam \kms.
[3rd] Ratio of the HCN-to-CS moment 0 maps with contours of CS every 4 mJy beam$^{-1}$ km s$^{-1}$. 
[Bottom] Longitudinal cross section in \cs\ line showing a hole of molecular gas inside $R\lesssim 1.5$ pc.
The sharp negative peak at offset 0 is the absorption of the continuum emission from \sgrastar.
{Alt text: Moment 0 maps of the CND in \cs\ and \hcn\ lines, and a cross section.} 
}
\label{fig-hole} 
\end{figure}  

The interior of the ellipse at $R\lesssim 1.5 \epc$ makes a hole of a depth as low as the $\sim 3$ mJy beam$^{-1}$ \kms, which is more clearly revealed by a cross section of the total intensity map along
the Galactic plane through \sgrastar shown in figure \ref{fig-hole}.
The sharp negative peak in the center is due to absorption against the continuum of \sgrastar.

The hole coincides positionally with the minispiral of the ionized hydrogen gas. 
The electron density has been measured to be  
$n_{\rm e}\sim6\times10^3{\rm cm^{-3}}$
as inferred from observations of the H92$\alpha$ recombination-line \citep{2009ApJ...699..186Z}
and $\sim (7-13)\times10^3{\rm cm^{-3}}$ of the H42$\alpha$ line \citep{2017ApJ...842...94T}.
If a value of $10^{4}$ H cm$^{-3}$ is typical along the main minispiral arms, the total mass of ionized hydrogen amounts only to $\sim 10^2\Msun$ for three arms of 1.5 pc in length and 0.2 pc in width.
This is two to three orders of magnitude smaller than the mass needed to fill the hole by molecular gas with a density comparable to that in the CND. 

\ss{Position-velocity diagrams}

\begin{figure*}    
\begin{center}     
\cs\ \hskip 0.24\lw \hcn\ \\
\includegraphics[width=0.35\lw]{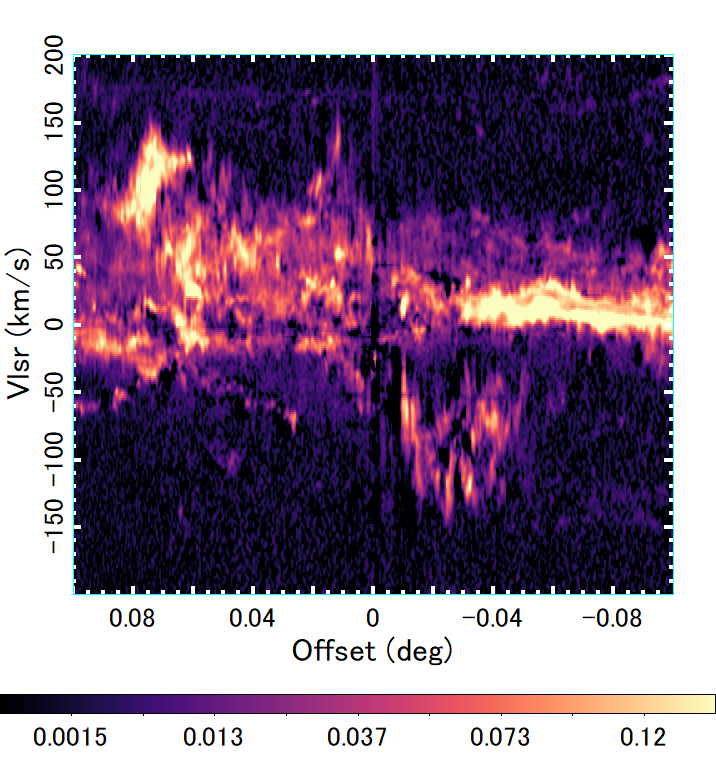} 
\includegraphics[width=0.35\lw]{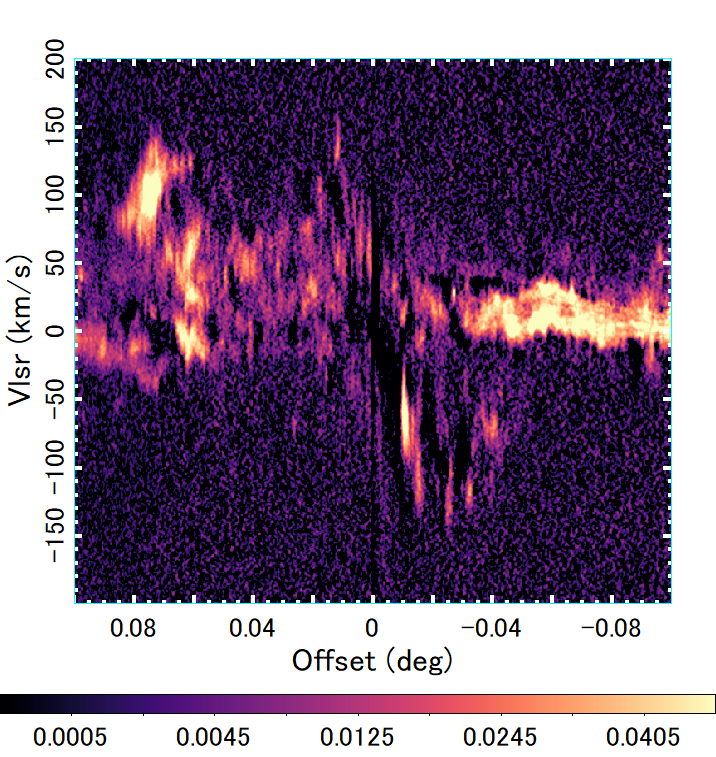}   \\
~~~~~~~~~~\h40\ \hskip 0.22\lw \h40+\cs \\
\includegraphics[width=0.35\lw]{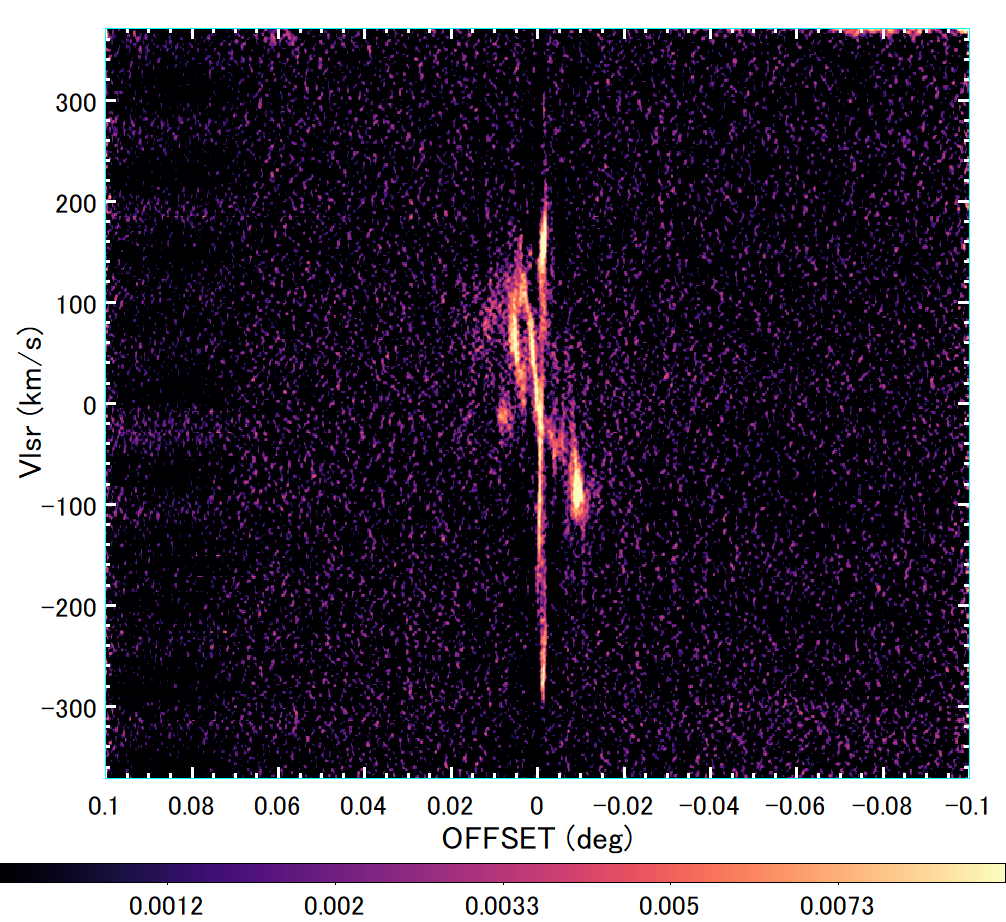} 
\includegraphics[width=.35\lw,height=0.33\lw]{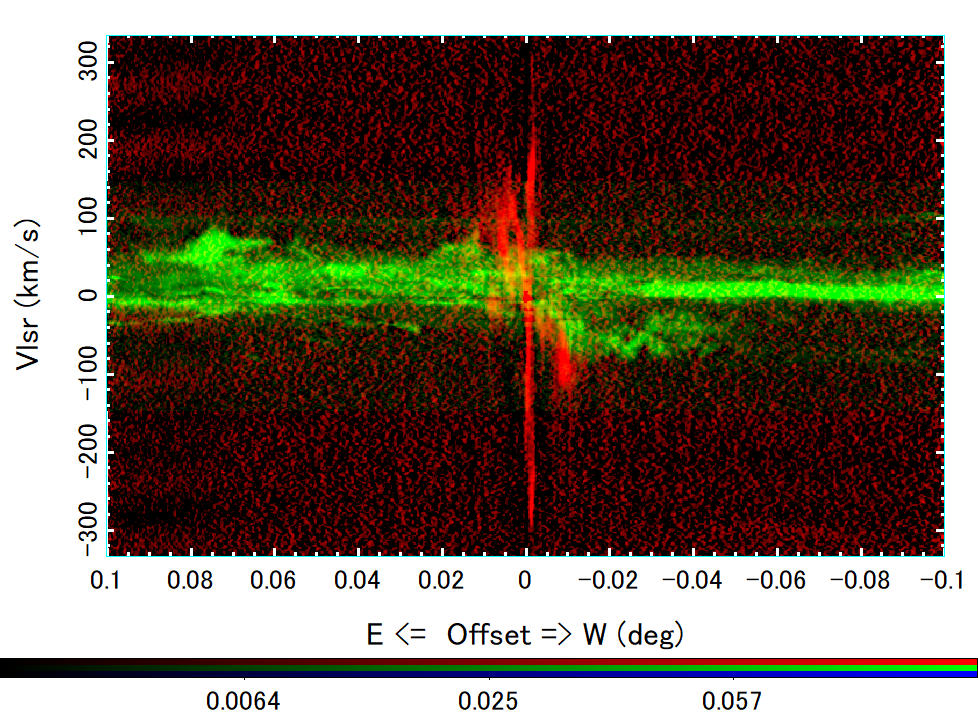}   
\end{center} 
\caption{[Top] Position-velocity diagrams of \cs\, \hcn\ and H40$\alpha$ lines along the major axis of CND at position angle $70\deg$ (white dashed line in the top panel of figure \ref{fig1}) (width 10 pixels for CS, HCN, and 40 pix for H40$\alpha$). 
The horizontal axis is the offset from \sgrastar, positive to the east (left). 
[Bottom] PVD in CS by green overlaid with \h40 by red.
{Alt text: PVD along the major axis of the CND.}}
\label{pvdmjor}  \label{pvh40} 
\end{figure*}

In figure \ref{pvdmjor} we show the position-velocity diagrams (PVD) along the major axis of the CND at position angle $70\deg$ across \sgrastar\ in CS, HCN and H40$\alpha$ lines.
In the appendix (figure \ref{fig-chan}) we show channel maps of the  longitude and latitude-velocity diagrams (LVD and BVD) in the \cs\ line.

The CND is recognized as the broad tilted ridge in the LVD and PVD, obliquely crossing the horizontal ridges due to the fore and background CMZ and Galactic disk. 
The LVD ridge has a mean velocity gradient of $dv/dl\sim 9000$ \kms deg$^{-1}$, as measured for Arm VI (CND) in Paper I, corresponding to a value for a radius $\sim 2$ pc ring rotating at $\sim 120$ \kms. 
The broad width of the LVD ridge indicates a significant non-circular motion of the CND.
BVDs also show a similar property to LVDs, indicating a tilted ridge. 

\ss{CND and central cavity}
\label{ss-cavity}

In our previous paper \citep{sofue+25b} we have shown that there are several arms in the CMZ, named Arms I to VII.
The CND is Arm VI, which is associated with the high-speed cloud G + 0.02-0.02 + 100 \kms and forms the same family with a fainter but more coherent arm of radius 8 pc. 
So, the circum-nuclear region is characterized by multiple rotating rings.
The simulation in the previous section also applies to the 8--10 pc arm (ring) by taking the normalization length to $\sim 10$ pc.
Accordingly, a spherical potential is again preferred in order to keep the ring-like structure of Arm V.

\begin{figure}   
\begin{center}     
Radial profile \\
Obs. \hskip 3cm Sphere\\
\includegraphics[width=0.45\lw]{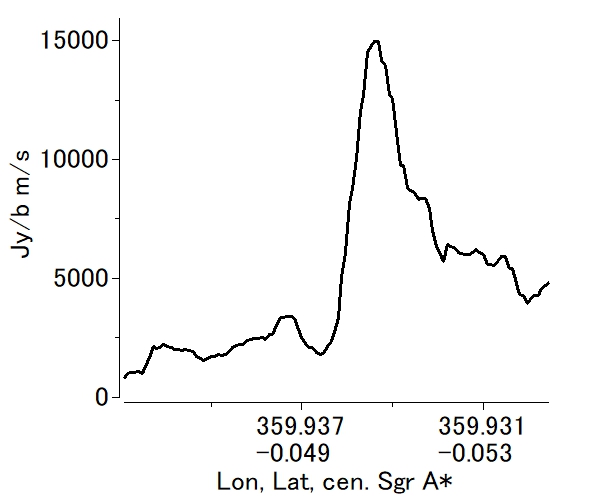}  
\includegraphics[width=0.4\lw]{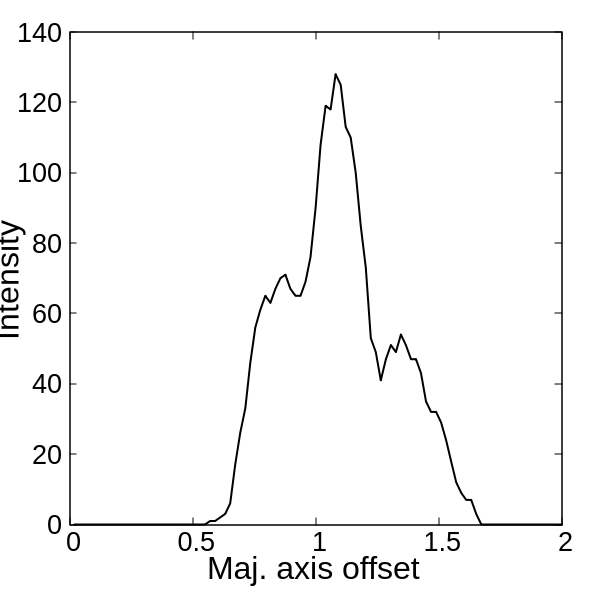}    \\ 
disk \hskip 3cm Bar\\
\includegraphics[width=0.45\lw]{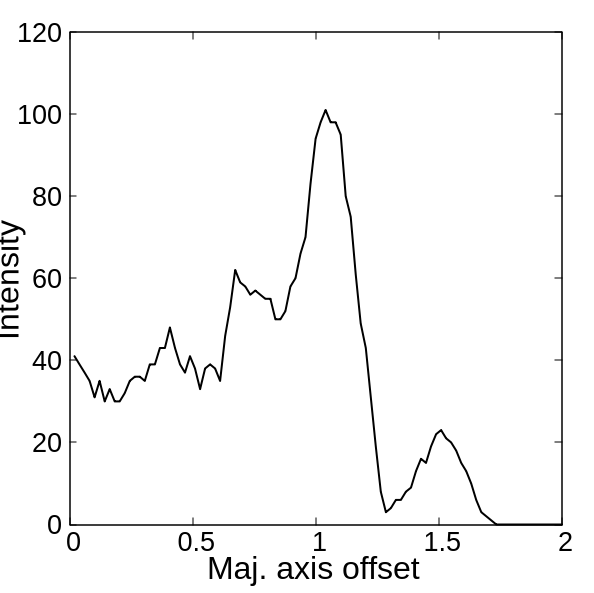}      
\includegraphics[width=0.45\lw]{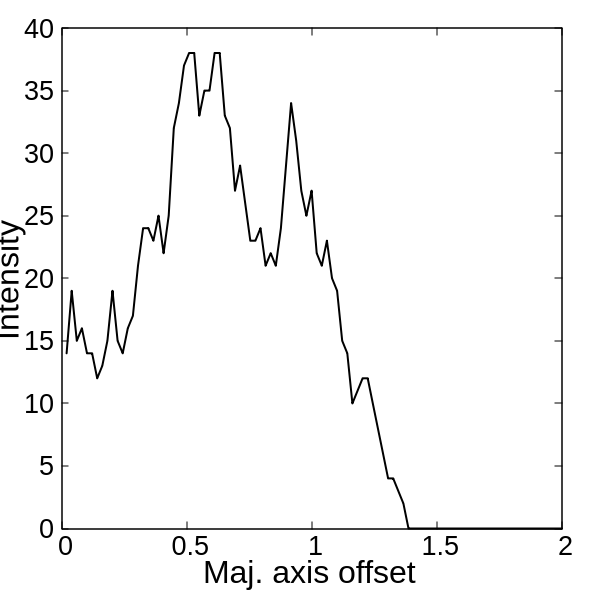}   
\end{center} 
\caption{[Top left] Cross section of the moment 0 map from \sgrastar\ to SW at $PA=240\deg$ showing a clear cut of intensity inside the CND. 
[Top right, Bottom] Simulations for spherical, disk and bar potentials, respectively. 
The central deep hole is reproduced only by the spherical potential.
{Alt text: Cross section of moment 0 map across \sgrastar\ compared with the simulations.}
} 
\label{fig-cut-simu}
\end{figure}

The circum-nuclear disk is defined as the massive molecular torus of several pc radius rotating around \sgrastar
\citep{2011ApJ...732..120O,
2017ApJ...850..192M,
2018PASJ...70...85T,
hs21}.
The torus feature is well recognized by the central hole on the moment 0 map, as shown by the moment 0 map and the cross section of the intensity in figure \ref{fig-hole}%
The feature is naturally explained by the model of a rotating disk in the spherical potential (figure \ref{fig-cut-simu}).
Note that the hole is visible in the model for the spherical potential but not for the disk and bar potentials.
 
It is important to emphasize that, as shown in figure \ref{fig-hole} observationally, both a possibly optically thick \cs\ and an optically thin \hcn\ lines show the central hole.
The hole must be real and exist independently of the opacity of the molecular line considered; even \cs\ becomes invisible there.


\ss{The minispiral}

The minispiral is composed of high speed flows of ionized gas orbiting the central supermassive black hole (SMBH) drawing three elliptical orbits with \sgrastar\ being the common focal point
\citep{2017ApJ...842...94T,2009ApJ...699..186Z}.
Their trajectories in the PV plane provide useful kinematical information about the gravitational potential.

\begin{figure*}     
\begin{center}
\includegraphics[width=.33\lw]{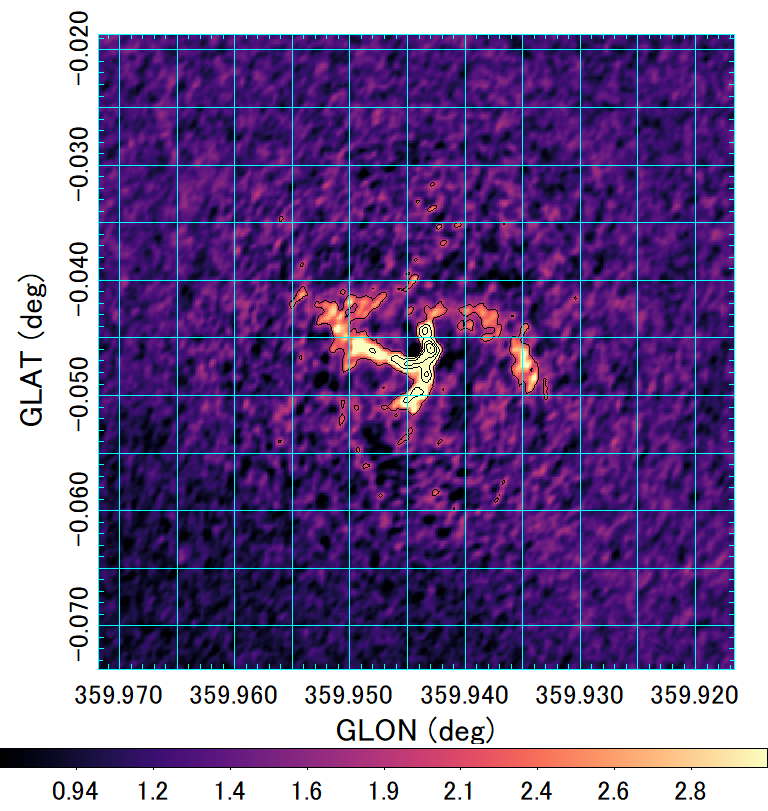}    
\includegraphics[width=.33\lw]{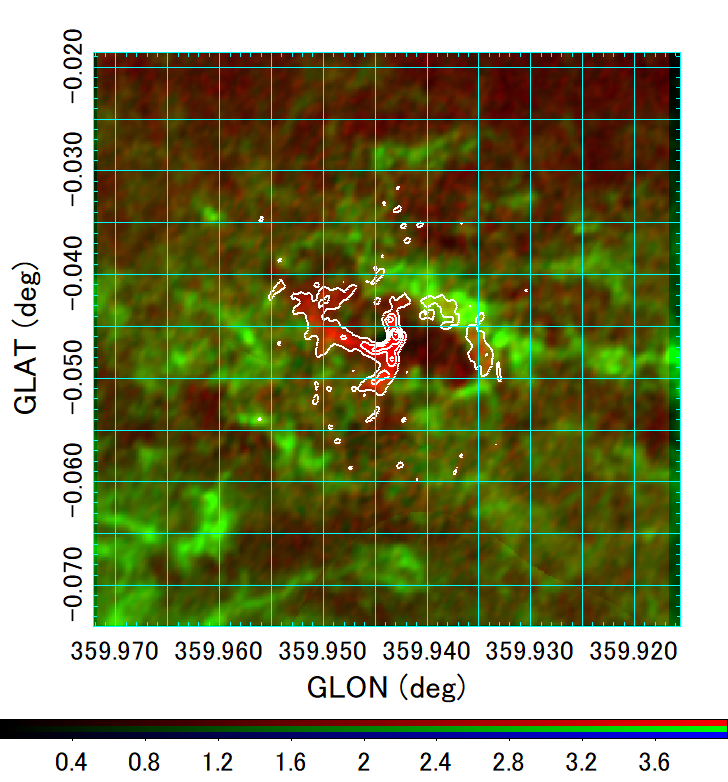} 
\includegraphics[width=.33\lw,height=.35\lw]{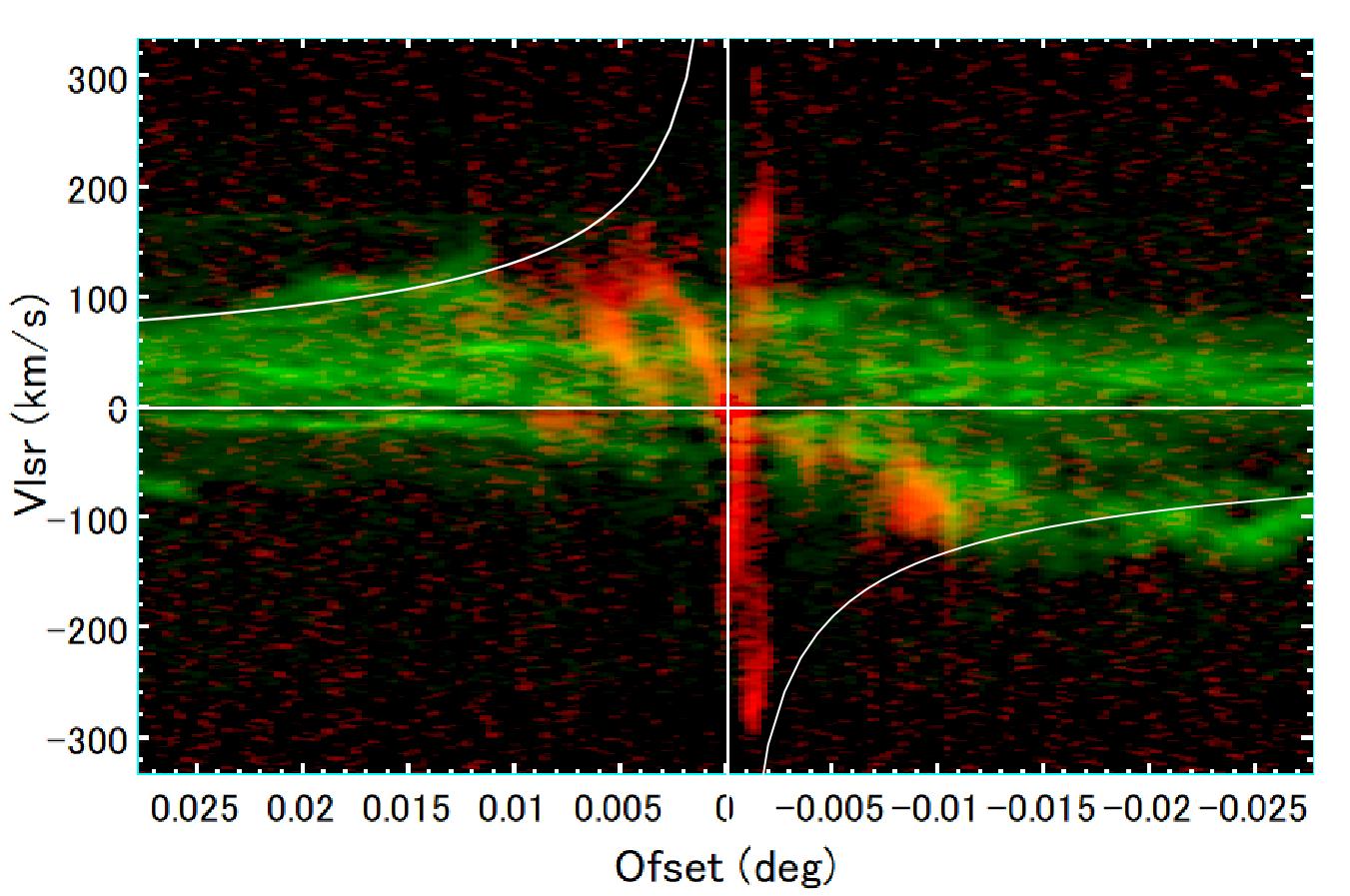}   \\
\includegraphics[width=.33\lw]{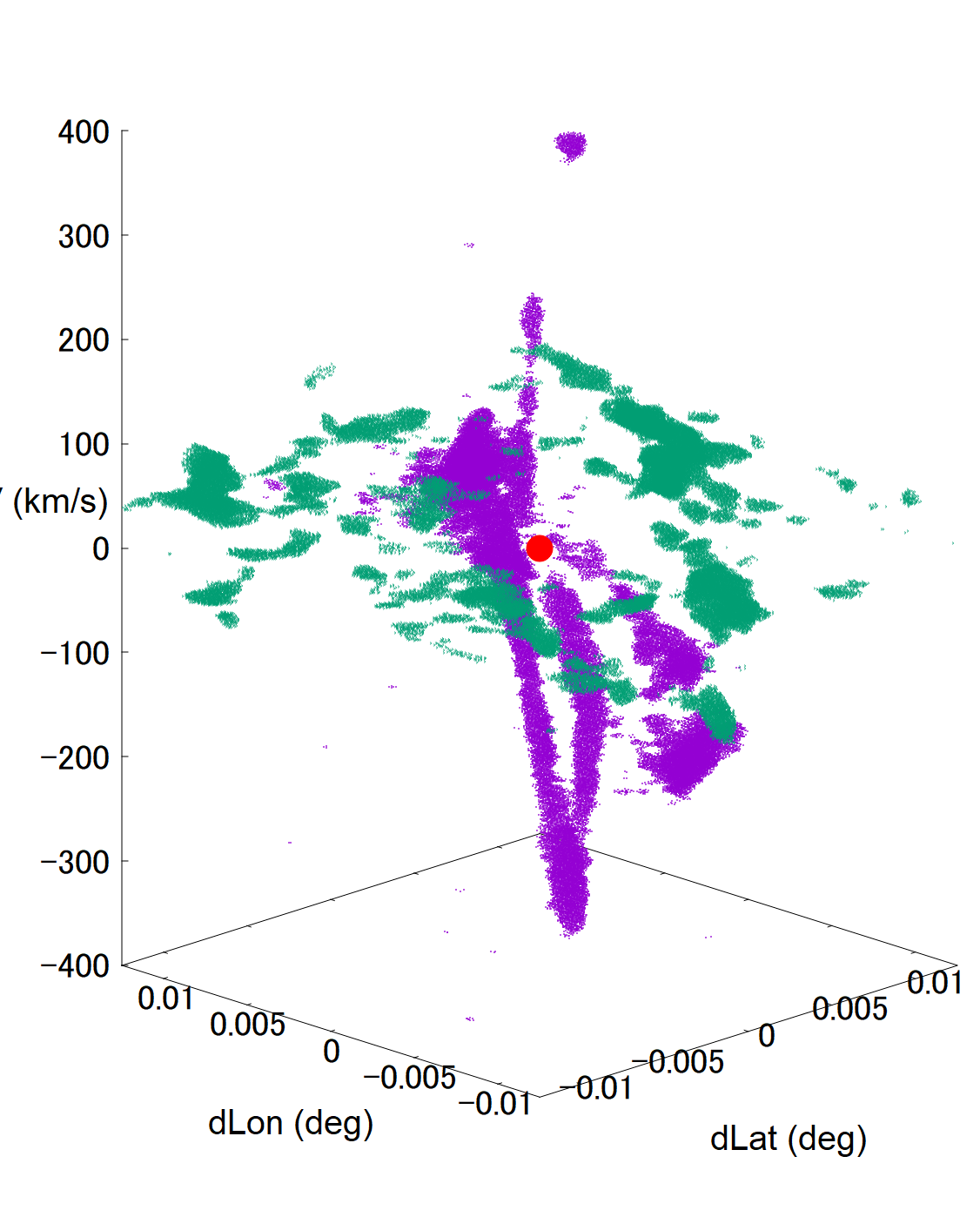}      
\includegraphics[width=.33\lw]{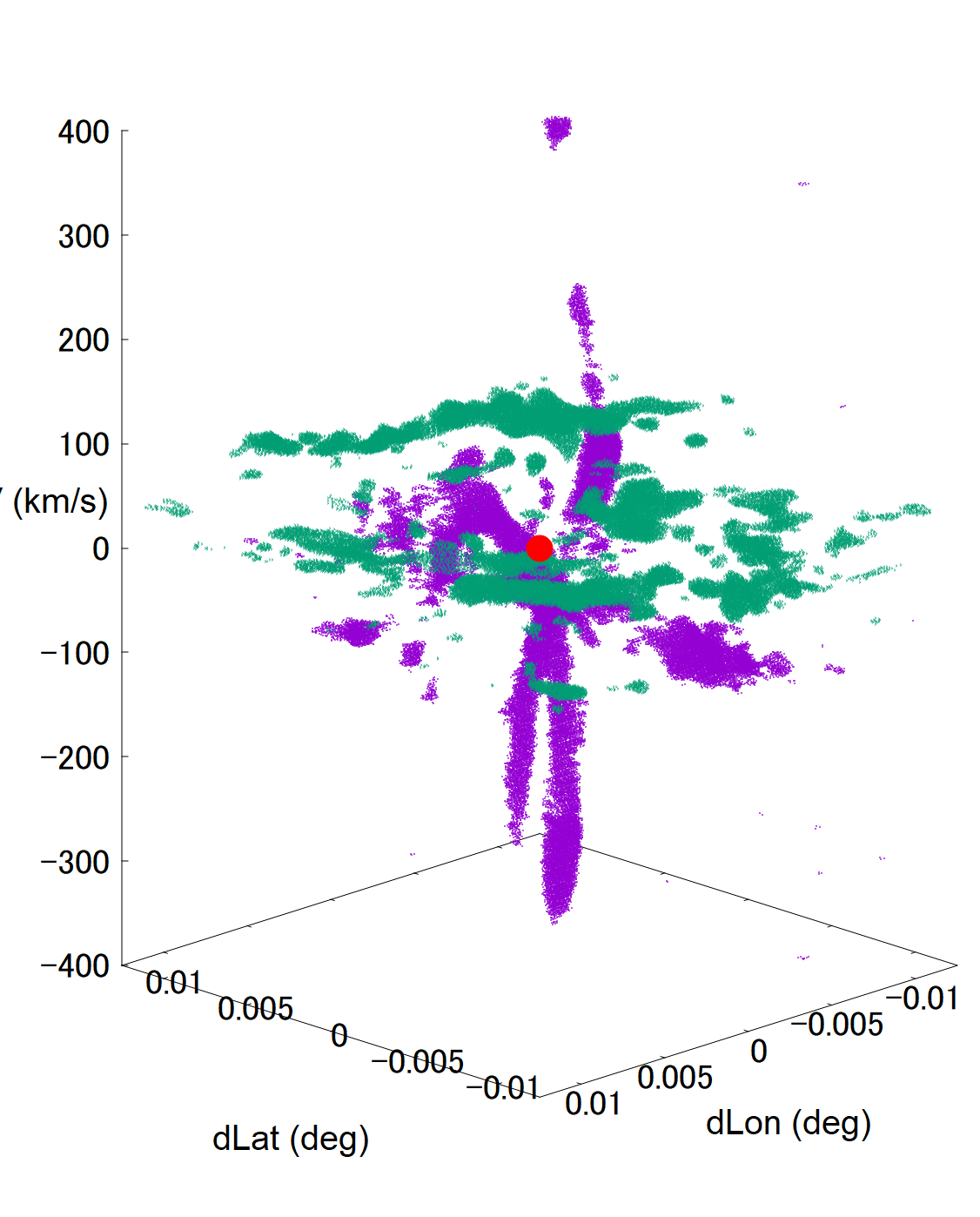}      
\end{center}
\caption{[1st panel] Moment 0 map of the minispiral in the \h40\ line.
[2nd] Same, but with contours overlaid on the \cs\ moment 0 map of the CND.
[3rd] Overlay of LVD of the minispiral in \h40\ on LVD of CND in \cs\. 
The lines indicate Keplerian RC for the central black hole with a mass of $4\times 10^6\Msun$.
[4th panel] Oblique projections of the 3D LBV \h40\ cube in violet from the longitude side superposed with that in the \cs\ line in green.
[5th] Same, but from latitude side.
 {Alt text: Comparison of the minispiral with CND by moment 0 maps and LVDs in \h40\ and \cs\ lines, and 3D projections of the LBV cubes.}}
\label{fig-min}  
\end{figure*}

Figure \ref{fig-min} (1st and 2nd panels) shows the moment 0 map in \h40\ of the minispiral  and superposition with that in CS.
The north-west arms in H40$\alpha$ and CS apparently overlap, but are separate arms as shown by 3D LBV projection below.
The 3rd panel shows a superposition of the LVDs.
The terminal velocity in the CS line is partially connected to the high velocity envelope in \h40, which increases toward the nucleus.
The envelope velocity is approximately presented by the Keplerian law for a point mass of $4\times 10~\Msun$ at the center, as indicated by the white lines in the 3rd panel. 

The 4th and 5th panels of figure \ref{fig-min} show oblique 3D projections of the pixel points with intensity greater than 0.175 Jy beam $^{-1}$ from the LBV cubes in the \h40\ (violet) and \cs\ (green) lines from the longitude and latitude sides.
These diagrams show that the minispiral in \h40\ is a separate system from the CND in phase space.
The molecular gas structure of the CND is not connected to the ionized gas of the minispiral.
 
\begin{figure}    
\begin{center}  
\includegraphics[width=.8\lw]{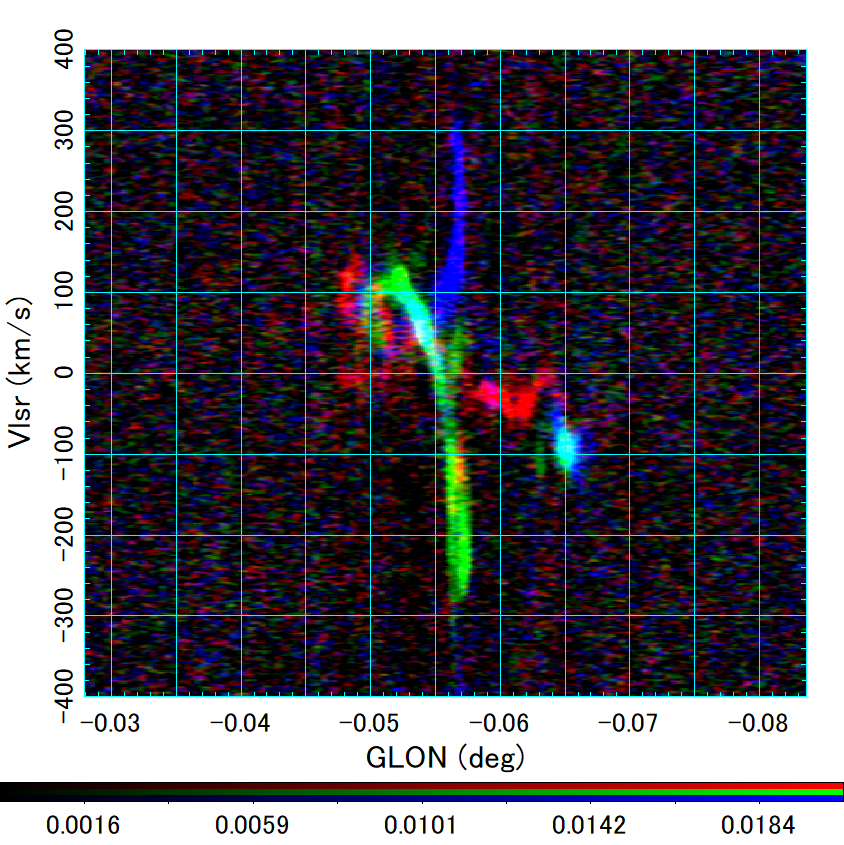}   
\includegraphics[width=.8\lw]{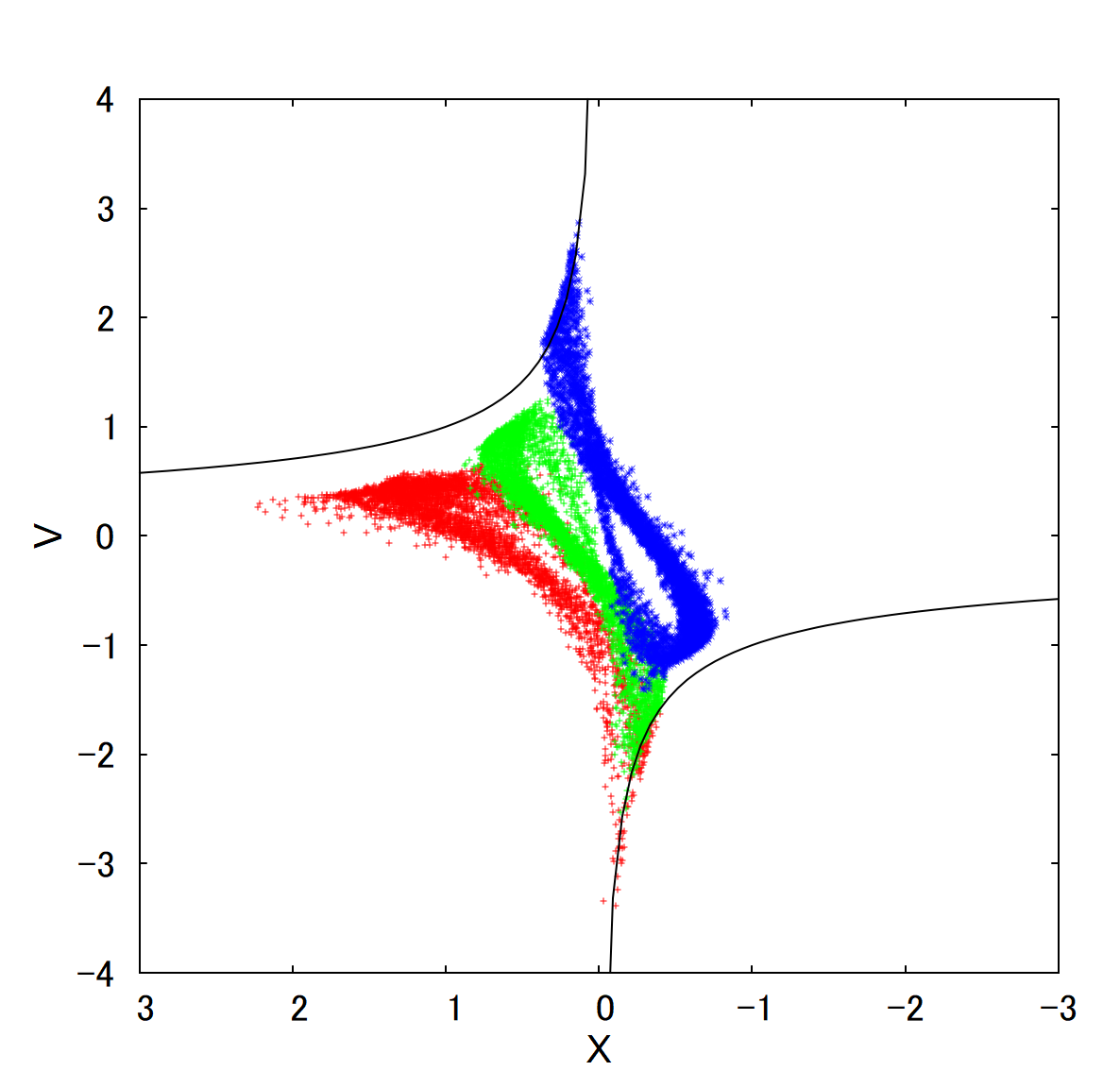} 
\includegraphics[width=.8\lw]{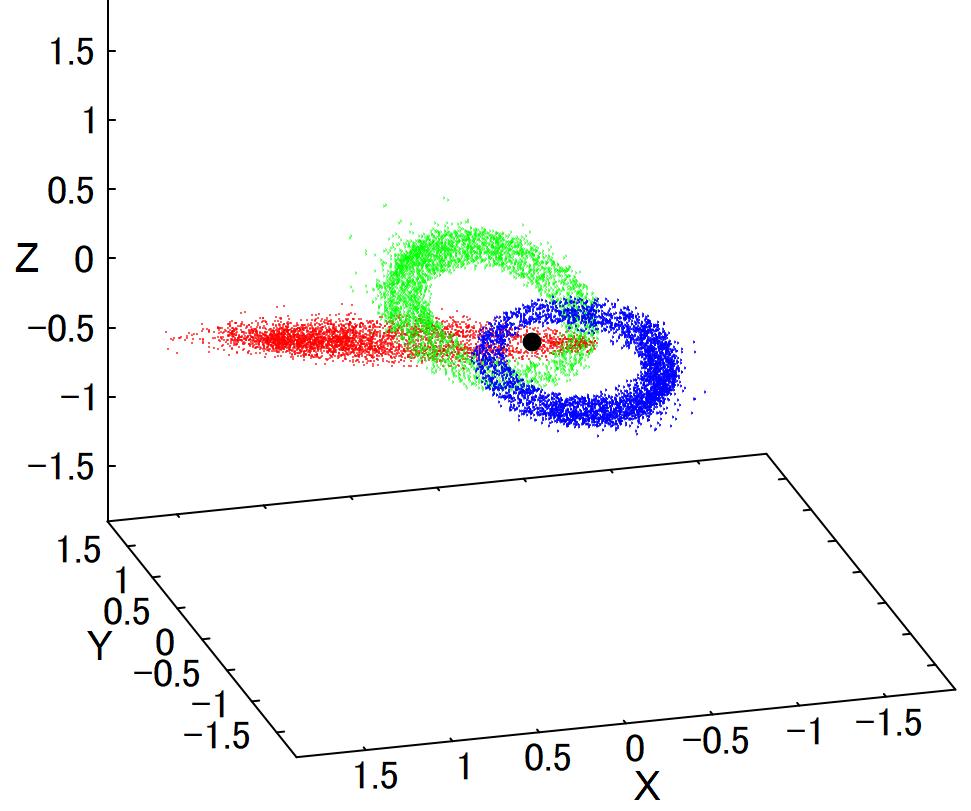}     
\end{center} 
\caption{[Top] Overlay of LVDs in \h40\ of the minispiral at three different latitudes. 
[Middle] Simulated PVD of 3 clouds with arbitrary orbital parameters around a point mass projected on the $(x,v_y)$ plane.
[Bottom] Same, but 3D projection.
{Alt text: PVDs in the CS and H40$\alpha$ lines in different colors around the minispiral. [Middle] PVD by test particle simulation and [bottom] 3D orbits.}}
\label{kepler}  
\end{figure}

Figure \ref{kepler} (top panel) shows the LVDs of the minispiral in \h40\ at three different latitudes.
The 2nd and bottom panels show a simulation of the trajectories of test particles of three clouds orbiting in the point-mass potential by LV and 3D projections.
This simulation qualitatively proves that the observed high-velocity LVDs of the minispiral is explained by the Keplerian motion of the gases rings or arms.

\section{Diagnosis of the shape of gravitational potential by test-particle simulation}
\label{simu}

We examine the sphericity, oblateness, and triaxial ratios of the gravitational potential in the circumnuclear region by simulating the evolution of a molecular cloud. 
Compared with the moment 0 maps and PVDs of the molecular lines, we argue that a spherical potential is preferable in the CND and the disk oblateness is mild in the entire CMZ, but a strong bar may not apply.

\ss{Axial ratio of the potential}

To examine the shape of the gravitational potential, it is convenient to assume the following form \citep{binney+1991},
\be
\Phi=1/2 ~ v_0^2 ~{\rm ln} ~\left[\sum (x_i/q_i)^2 \right],\label{eqbar}
\ee 
where $x_i$ are the Cartesian coordinates of the particle and $q=q_x:q_y:q_z$ gives the axial ratio of the potential. 
When the potential is spherical with $q=1:1:1$, the circular rotation velocity is constant (flat rotation). 
A potential with $q=1:1:q_z$ and $q_z<1$ represents an oblate spheroid or a disk, and $1:q_y:q_z$ with $q_y \ne q_z$ represents a triaxial bar.

In the following, the coordinates are normalized by the radius, $r=1$, at which the circular velocity is also nomaralized to unity, $v=1$. 
The cloud is represented by an ensemble of many test particles centered on the initial orbit of radius 1.
The test particles are distributed in a small radius 0.1 times the orbital radius and velocity dispersion 0.1 times the orbital (rotation) velocity at the unit radius. 
Each test particle moves in a potential of the Galactic bulge and the cloud's gravity represented by a Plummer potential of a scale radius 0.1, so that the Roche radius of the tidal force by the bulge's potential is about equal to 0.1.

\def\sss{\subsubsection}
\sss{Spherical potential - Coplanar rosette orbit}

Figure \ref{fig-cloudevo} panel A shows the result for a spherical bulge potential, showing the evolution of a cloud composed of many test particles initially placed at a radius $r_0=(x_0,y_0,z_0)=(1, 0, 0.3)$ and circular rotation velocity $v_0=(0,1,0)$, plotted every 0.1 rotation period. 
The cloud orbit is circular in the $(x,y)$ plane and draws a straight line in the $(x,z)$ plane as projected on the sky, as shown in the 1st (top-left) and 2nd charts, respectively.
The LVD and BVD show straight rigid body-like behaviors, as shown in the 3rd (bottom-right) and 4th charts.

The cloud's shape suffers from the tidal effect of the Galactic bulge and is stretched around the cloud center.
It is disrupted in one orbital period and is stretched along the circular orbit, finally making a circular ring of radius 1. 
The projections onto the $(x,z)$, $(x,v_y)$ and $(z,v_y)$ planes may be compared with the observed moment 0 map, LVD and BVD in figures \ref{fig-cnd}, \ref{pvdmjor} and \ref{fig-chan}.
Obviously, the present simple ring model cannot reproduce the detailed properties observed in figure \ref{fig-cnd}, although the elongated shape in the sky and the tilted LV and BV ridges observed in figure \ref{fig-cnd} are not inconsistent with the simulation in figure \ref{fig-cloudevo}. 

\begin{figure*}
\begin{center}      
 A. Sphere \hskip 7 cm B. Sphere, Rosette\\
 \includegraphics[width=.4\lw]{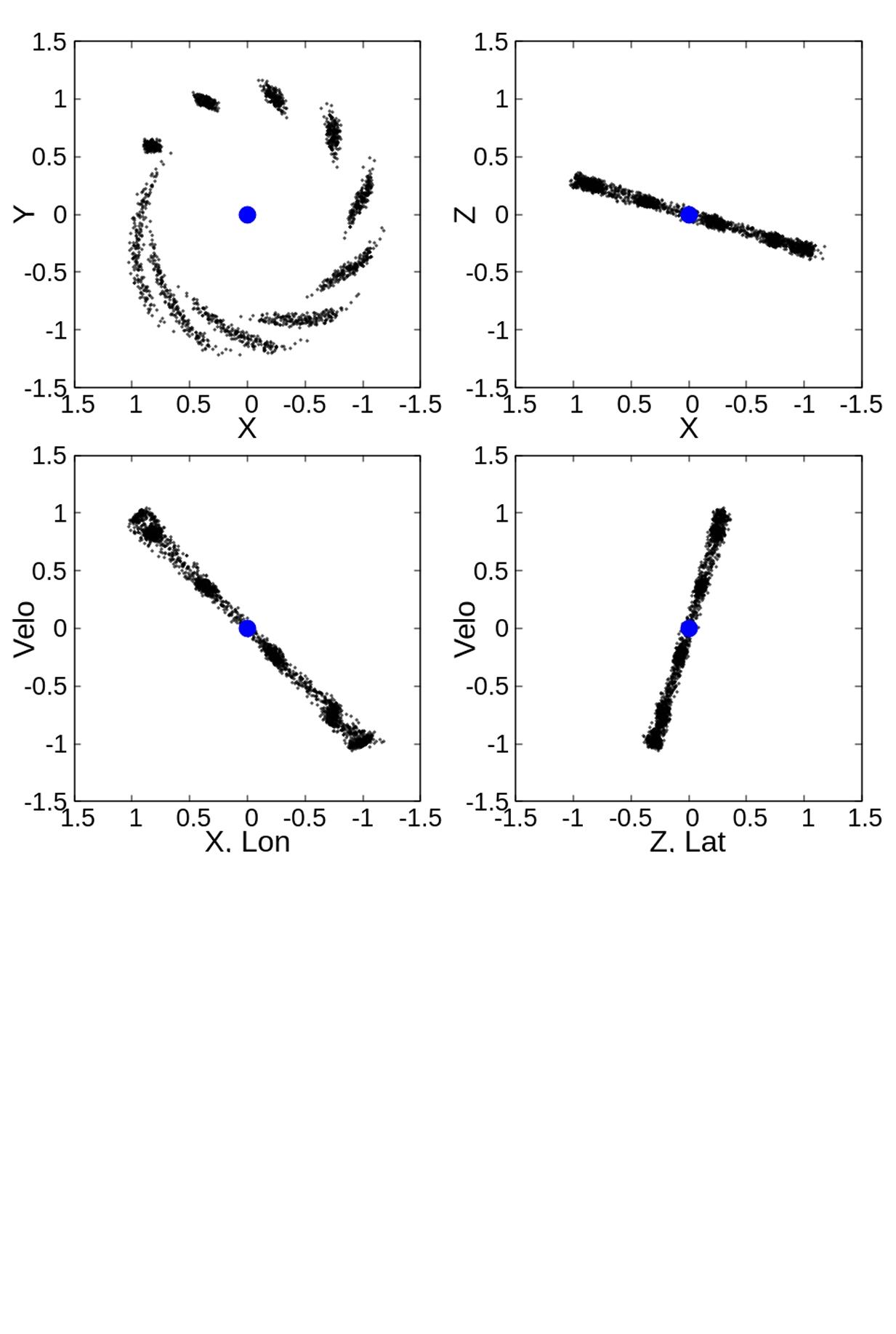} \hskip 8mm
\includegraphics[width=.4\lw]{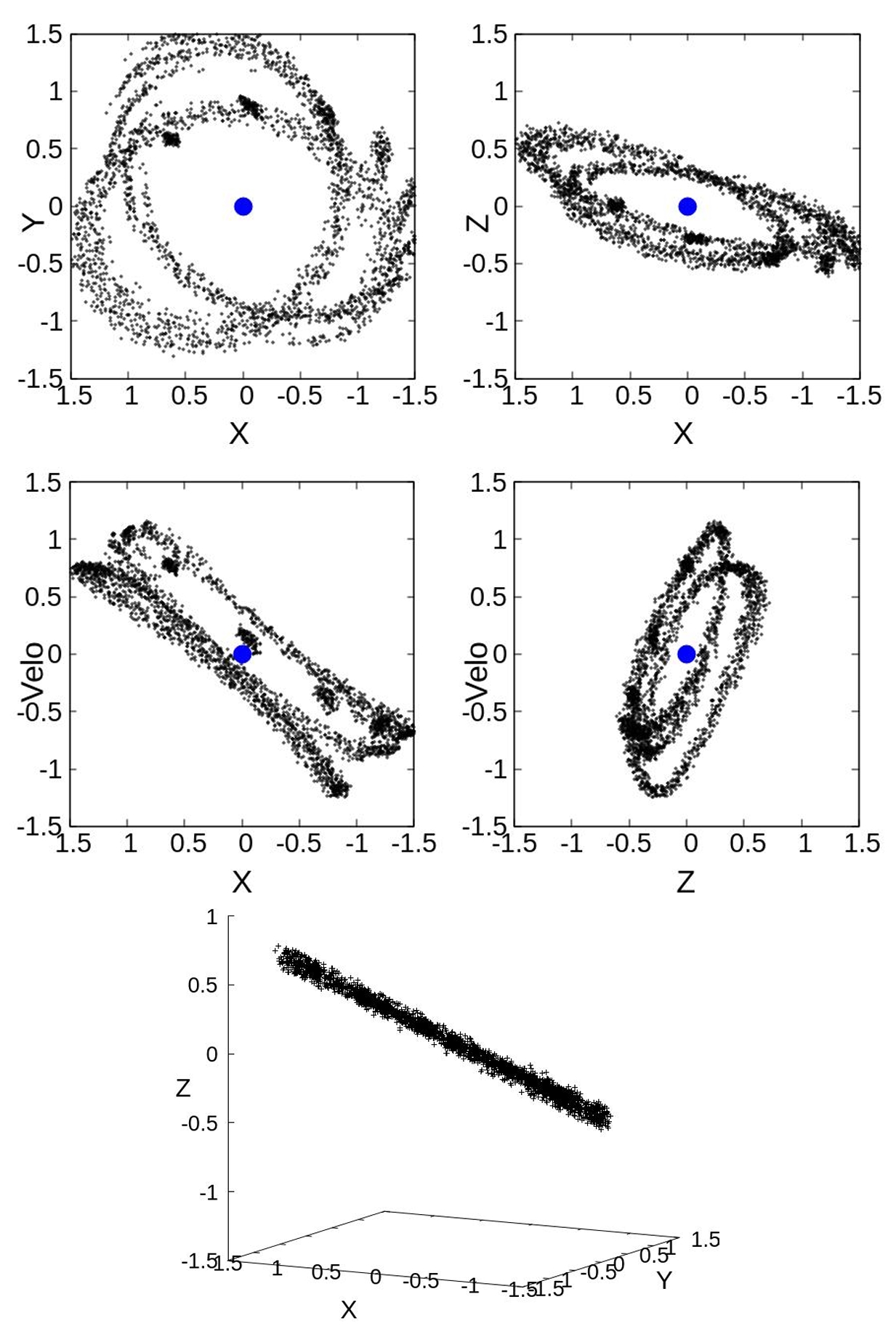} \\
 C. disk \hskip 7.5cm D. Bar\\
 \includegraphics[width=.4\lw]{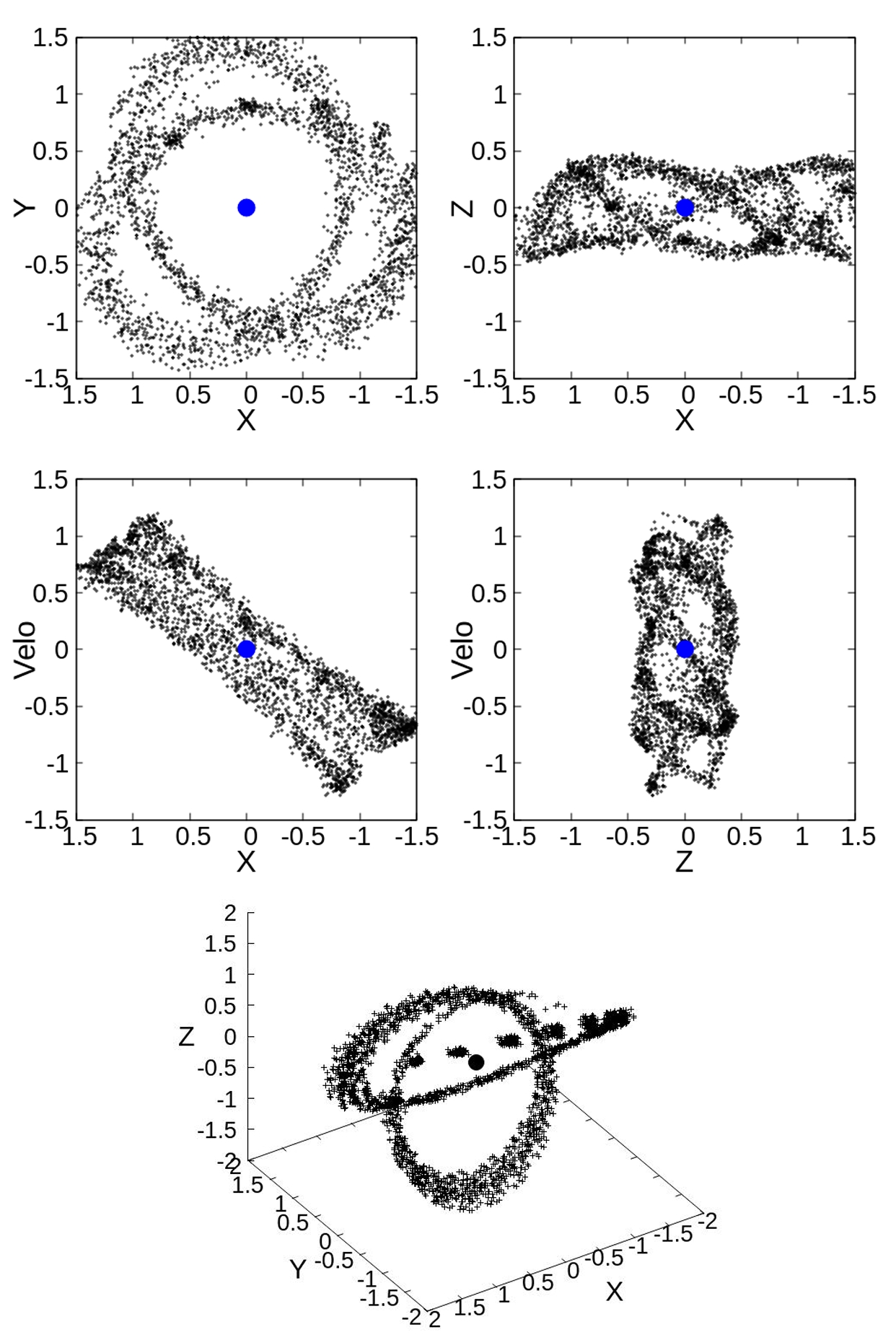} \hskip 8mm
\includegraphics[width=.4\lw]{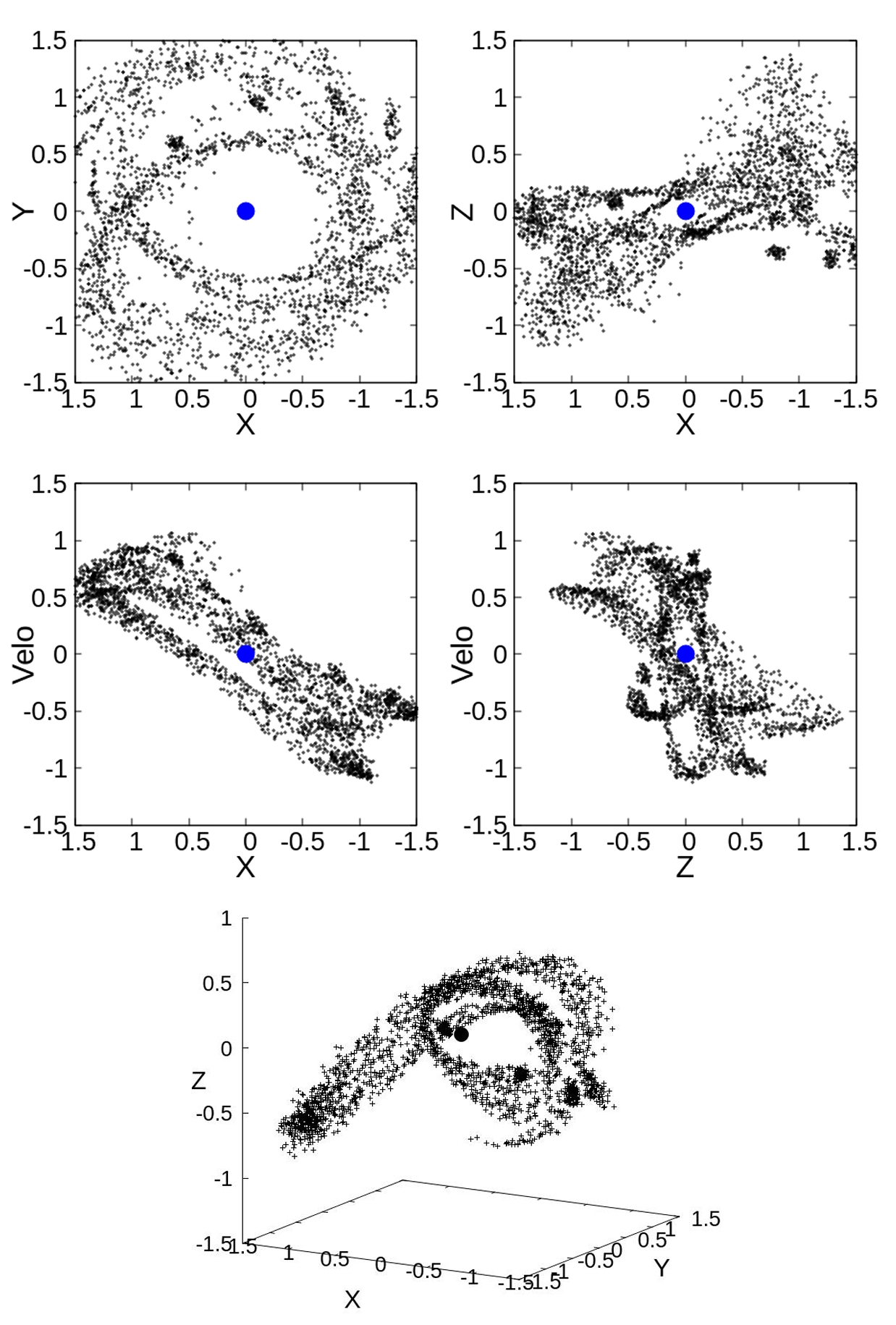} 
\end{center} 
 \caption{
 [A] Circular coplanar orbit in a spherical potential with $q=1:1:1$ 
starting at $(x,y,z; v_x,v_y,v_z)=(    1,   0,   0.3;  -0.3, 1, -0.4).$
[B] Coplanar rosette orbit in a spherical potential for  
$q=    1:1:1$, starting at ($1, 0, 0.3; -0.3, 1,  -0.4$)   
[C] Evolution of molecular cloud in disk potentials with  
$q=1:1:0.6$ starting at  $(x,y,z,v_x,v_y,v_z)=1,   0,   0.3,  -0.3,   1,  -0.3)$.
[D]  Non-coplanar orbits in a bar potential for 
$q=0.8:1:0.8$.
 {Alt text: Test particle simulation of evolution of a cloud in various potential.}  }
\label{fig-cloudevo}\label{fig-cloudevo}\label{fig-cloudevo}\label{fig-cloudevo}
\end{figure*}  
  
The next simplest model is a rosette orbit for a noncircular initial velocity in the spherical potential.
Figure \ref{fig-cloudevo} panel B shows the result for a cloud with the initial velocity oblique to the concentric circle starting from the same place as in panel A and traced for three orbital periods.

The $(x,y)$ projection exhibits a rosette orbit pattern, where the cloud is disrupted within one orbital rotation and is further elongated along the orbital trajectory.
The $(x,z)$ projection, which corresponds to a projection in the sky, yields an elliptical structure with a few bifurcated arms and an empty interior region, mimicking the observed moment 0 map in figure \ref{fig-cnd}.
We point out that the rosette orbit in the spherical potential is coplanar even if the $(x,z)$ projection shows such an elliptical structure.
In the bottom chart of figure \ref{fig-cloudevo} panel B, we confirm this by a 3D projection through the orbital plane of the same simulation result.

The reason why the orbits remain in the initial coplanar plane is that the epicyclic frequencies in the horizontal (radial) and vertical directions are equal due to the spherical potential. 
The reason may also be explained by the conservation of angular momentum in the spherical potential because the gravitational force is kept always radial.
The projections in the $(x,v_y)$ and $(y,v_y)$ planes, corresponding to LVD and BVD, respectively, also reproduce the broad tilted PV diagrams in figure \ref{fig-chan}, although the absorbed regions are not well hidden by this simulation.

\sss{disk potential}

We then examine non-spherical potentials, which do not seem to explain the observations.
Figure \ref{fig-cloudevo}, panel C, shows the result of the evolution of a molecular cloud with the same initial condition as in figure \ref{fig-cloudevo} panel B, but in disk potentials with $1:1:0.6$, and the evolution is traced for 6 orbital periods.
The bottom panel is a 3D projection, demonstrating that the orbit cannot be coplanar because of the non-centrally directed force in the disk potential, which does not guarantee the angular-momentum conservation. 
Therefore, the $(x,z)$ projections, corresponding to the intensity distribution on the sky, do not produce the tilted ring structure with a central hole.
Comparing the result with the observations in figures \ref{fig-cnd} and \ref{fig-hole} we conclude that a disk potential may not be a suitable approximation for the central few pc region of the Galaxy. 

The reason why the orbits do not remain in the initial coplanar plane is that the epicyclic frequencies in the horizontal (radial) and vertical directions are different, so that the nodal points precede fast around the center. 
This is also explained by a rapid precession of the rotation axis around the $Z$ axis and variation of the angular momentum.

\sss{Bar potential}

Figure \ref{fig-cloudevo}, panel D, shows the result for the same initial condition as in figure \ref{fig-cloudevo} panel B, but in bar potentials with $q=1:0.8:0.8$. 
Due to the non-centrally directed force in the bar that does not guarantee the conservation of angular momentum, the orbit cannot remain in a coplanar sheet. 
The bar potential causes stronger precession of the rotation axis in the $x$, $y$ and $z$ axes, producing different epicyclic frequencies in the three directions and a more uniform distribution of the trajectories. 
Compared with the observations, we conclude that the bar potential is also not likely in the central a few pc region of the Galaxy.

\ss{Constraint on the potential shape}

\begin{figure*} 
    \begin{center} 
    Moment 0 \hskip 0.2\lw LVD \hskip 0.2\lw BVD \\
\includegraphics[width=0.26\lw]{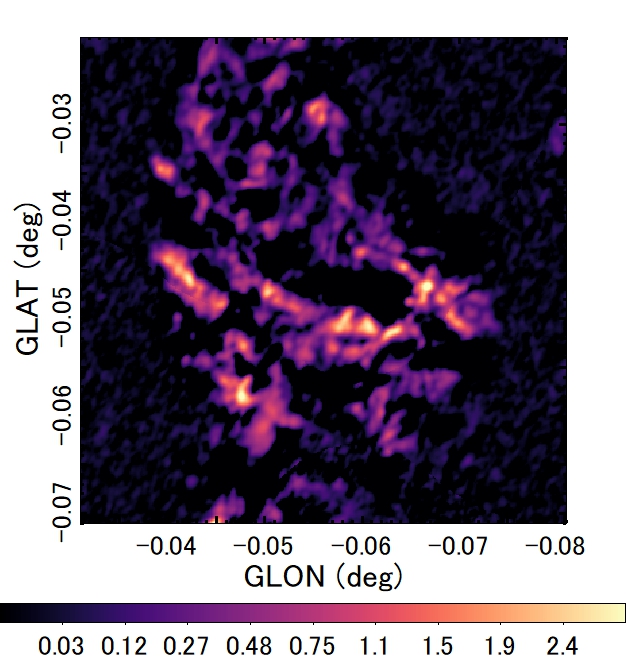}  
\includegraphics[width=0.25\lw]{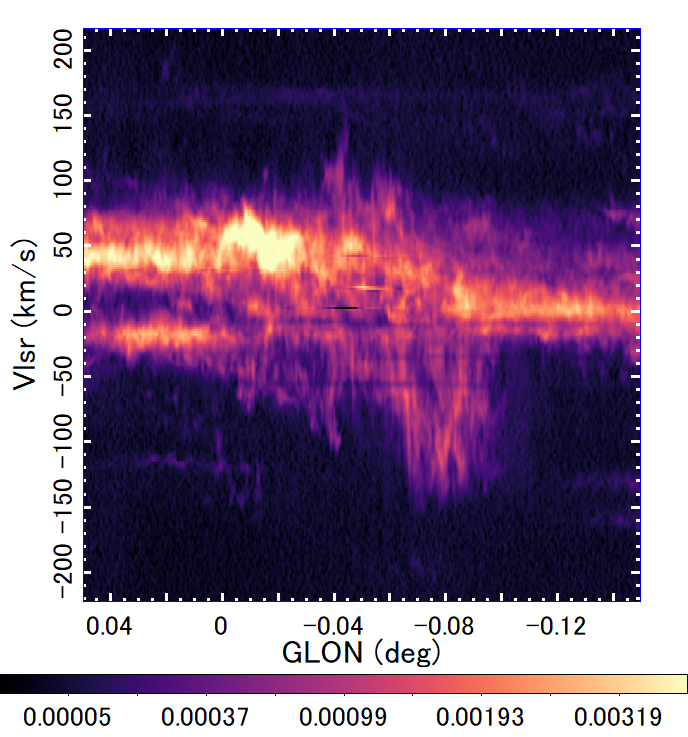} 
\includegraphics[width=0.25\lw]{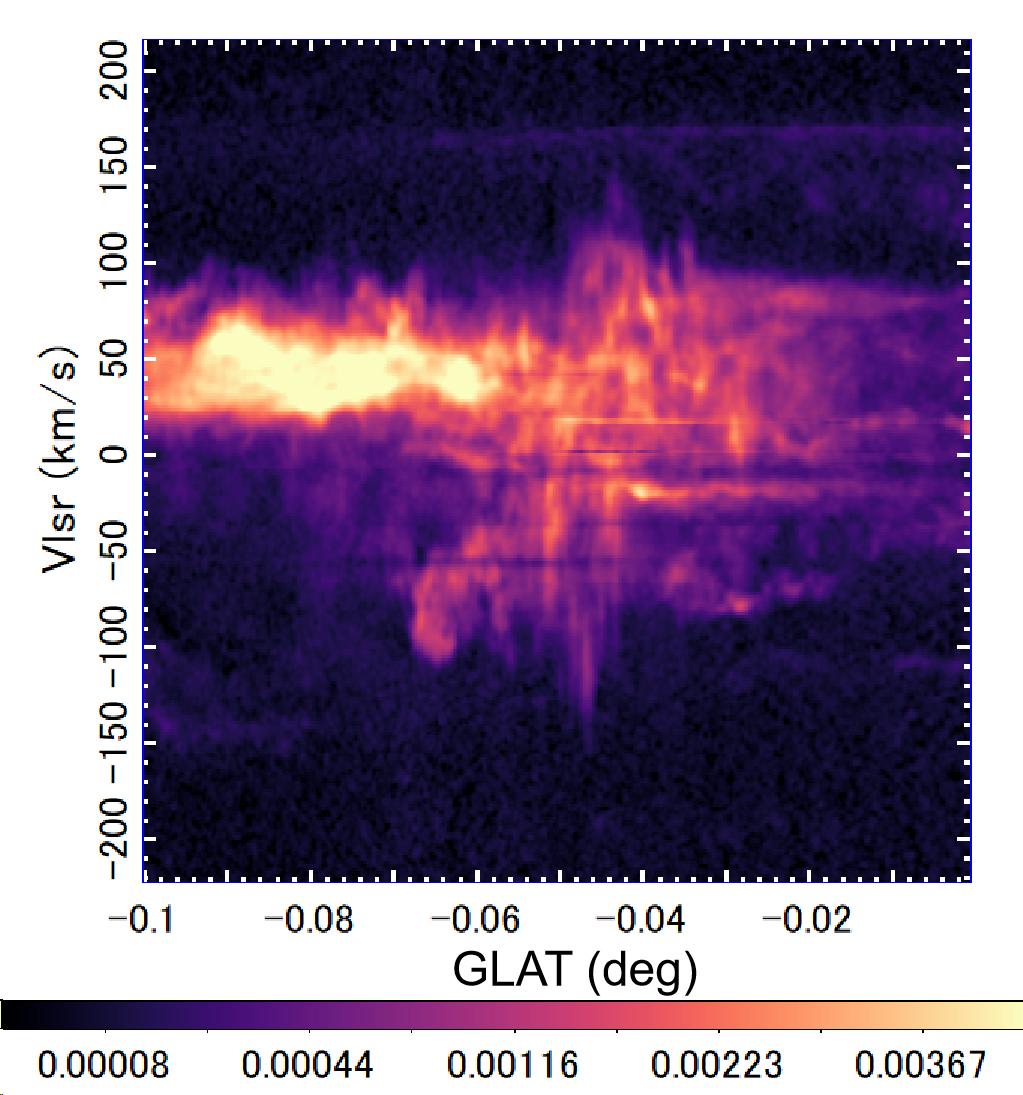}\\ 
\includegraphics[width=0.25\lw]{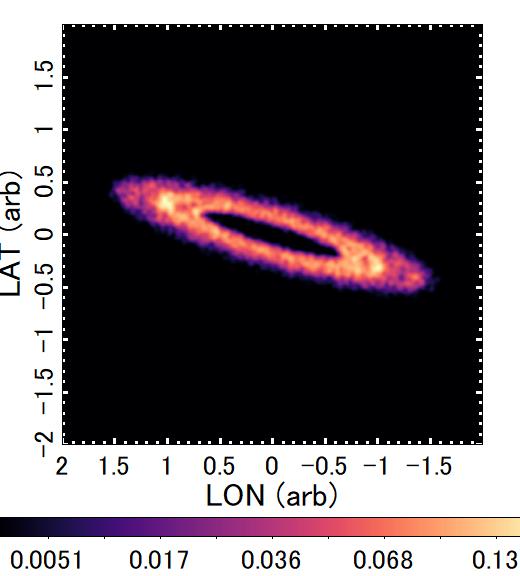}  
\includegraphics[width=0.25\lw]{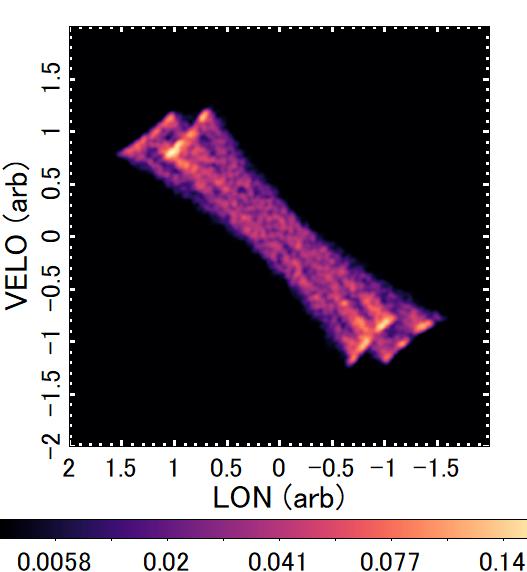}  
\includegraphics[width=0.25\lw]{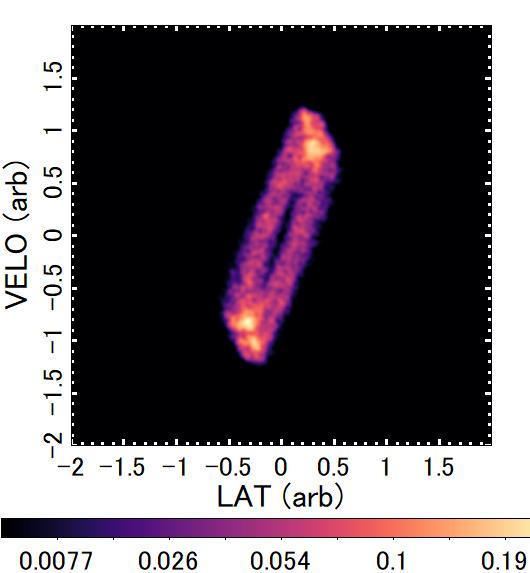}  \\ 
\includegraphics[width=0.25\lw]{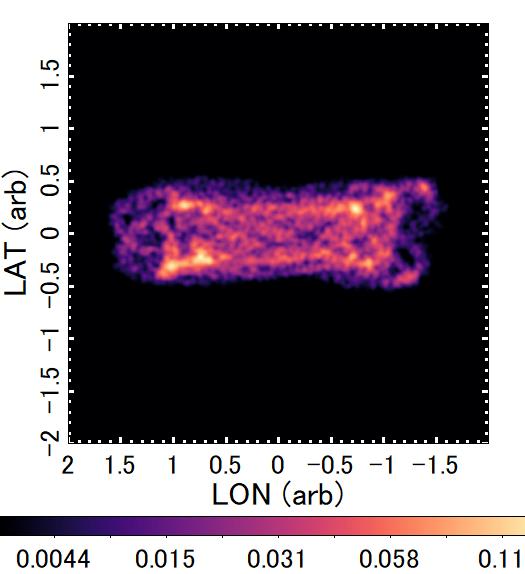}  
\includegraphics[width=0.25\lw]{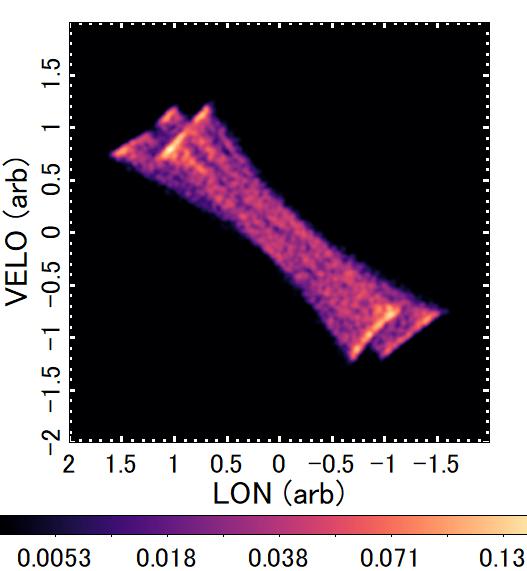}  
\includegraphics[width=0.25\lw]{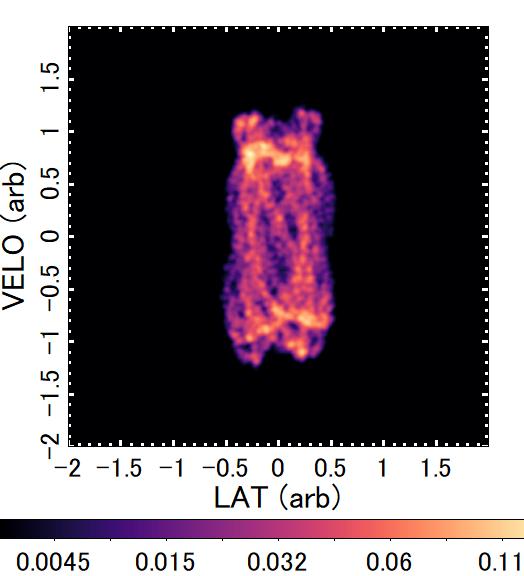}  \\ 
\includegraphics[width=0.25\lw]{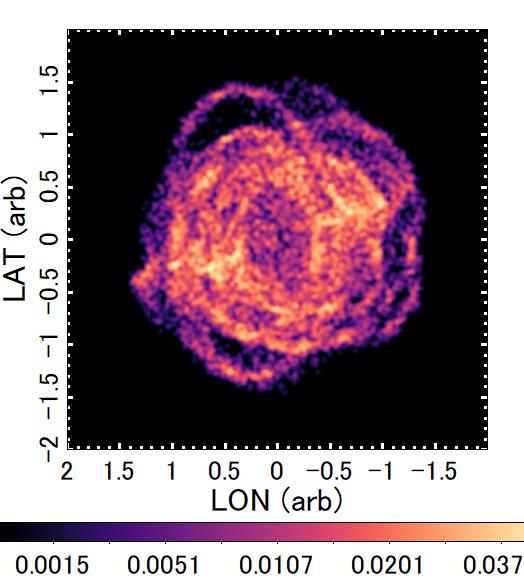}  
\includegraphics[width=0.25\lw]{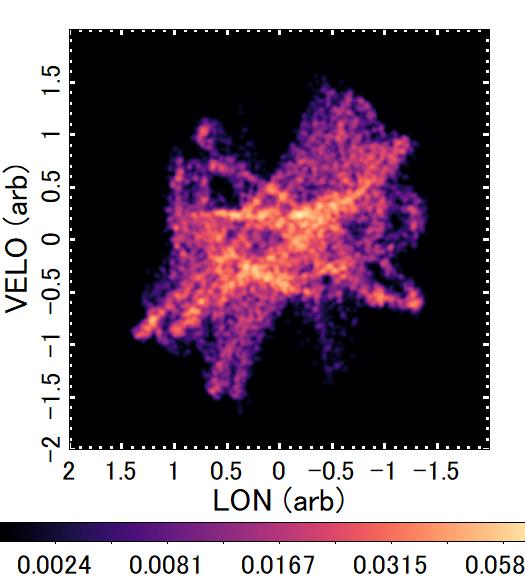}  
\includegraphics[width=0.25\lw]{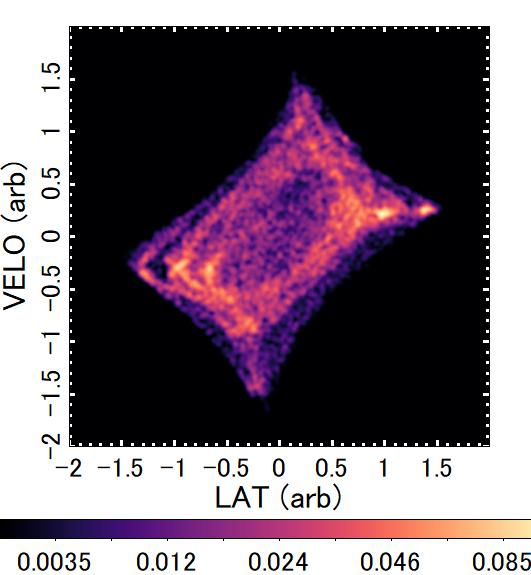}  \\  
\end{center}
 \caption{Summary of the simulation compared with the observation.
 A spherical potential is preferred to keep the ring shaped CND on the sky.
 disk and bar strongly disturb the orbits, and the centrally holed ring totally disappears in $\sim 0.5$ My.
[Top] Observed moment 0 map in the CS line by integrating the emission along the LV ridge, LVD, and BVD.
[2nd] Simulation in a spherical potential for $\sim 10$ rotations, reasonably reproducing the observation.
[3rd] disk potential ($q=1:1:0.8$) is not satisfactory to explain the observation.
[Bottom] Bar ($1:0.8:0.8$), ibid.  
  {Alt text: Summary of simulation compared with observation.}
 }
\label{fig-summary}
\end{figure*}

In figure \ref{fig-summary} we summarize the simulated results for an orbiting period of $\sim 10$ rotations, which corresponds to $\sim 3$ Myr at $R\sim 3$ pc (CND).
After many runs of the test particle simulations with different values of $q_x$ and $q_z$, we obtained the following condition for the central potential to reproduce the observed properties of the CND.
Note that we use the CMZ or the CND as a non-self-gravitating probe to diagnose the external gravitational pontential of the Galactic bulge or the nuclear stellar cluster (NSC) and a massive black hole. 

i) Spherical potential is preferred to keep the ring structure of the CND flat even though the orbits are highly eccentric \citep{sofue+25b}. This applies also to the minispiral, each arm of which traces an ellipse, not mixed to become a 3D cloud. 

ii) In a disk potential, the angular momentum of the flow is not conserved unless the disk is perfectly parallel to the potential plane.
This results in a fat disk with a thickness equal to the height of the orbit.
The central hole disappears when the disk is projected in the sky because of the mixing of nodal points.
The entire CMZ may be in this category.
The disk potential should be sufficiently round with $q_z\gtrsim 0.95$.

iii) A bar potential disturbs the ring shape more strongly as a result of the angular momentum exchange. In order to keep the central hole on the sky for more than a few orbital rotations ($\sim 0.1$ My), the two axial ratios must be $q_z\gtrsim 0.95$ and $q_x\gtrsim 0.95$.

\begin{figure}
\begin{center} 
\includegraphics[width=0.6\lw]{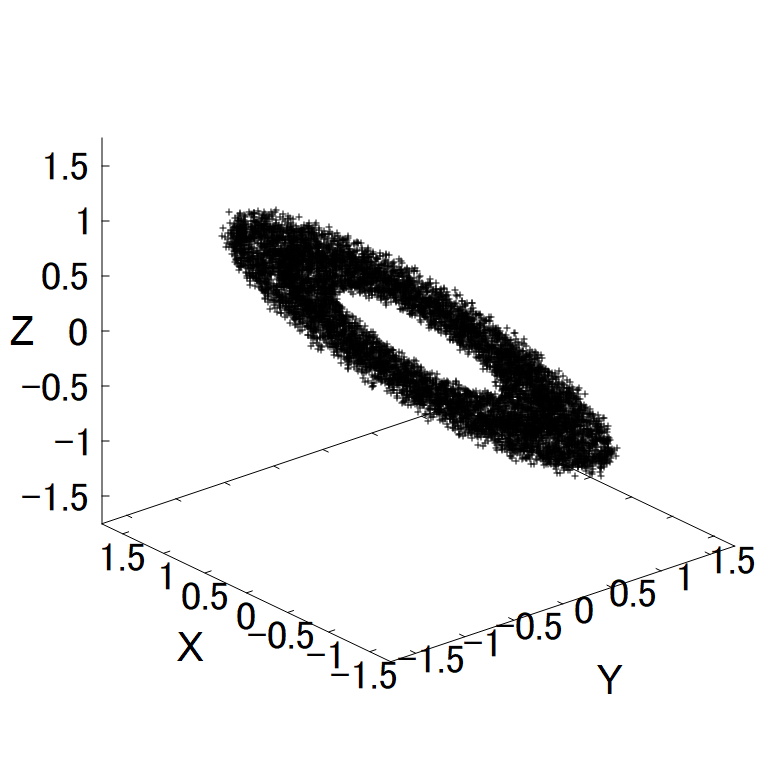} \\
\vskip -4mm\includegraphics[width=0.6\lw]{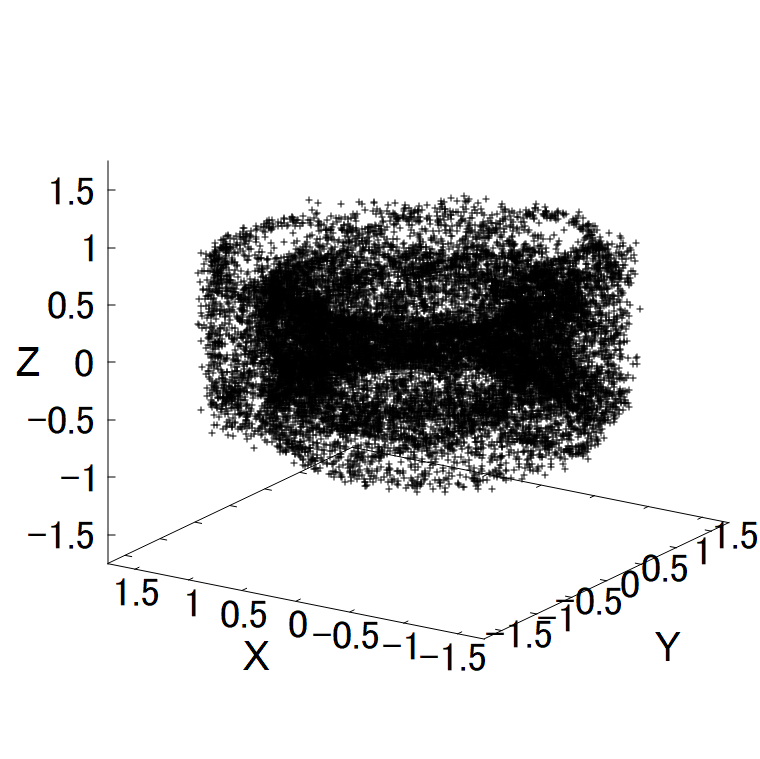}  \\
\vskip  -3mm
\includegraphics[width=0.6\lw]{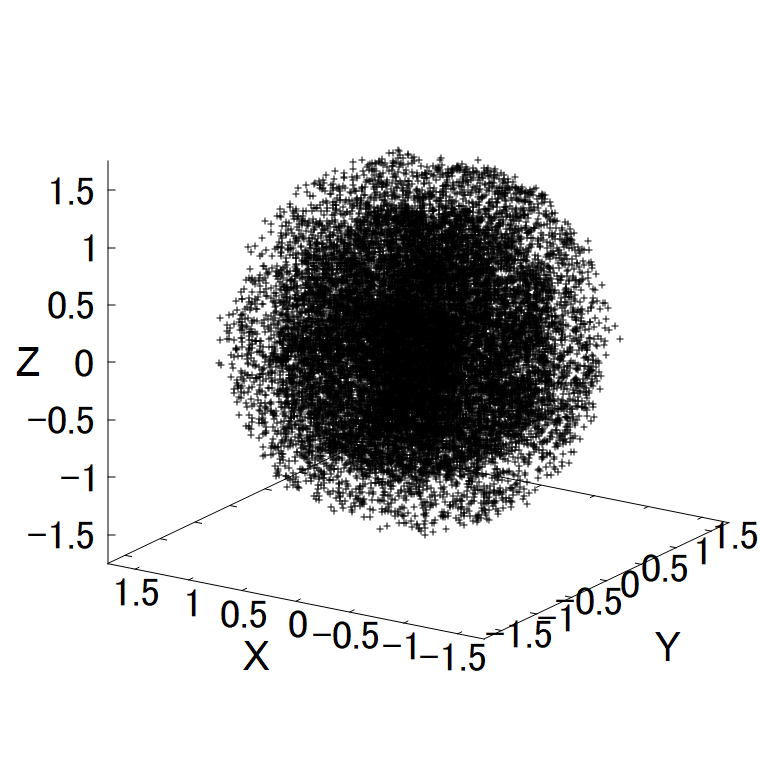} \\ 
\vskip 2mm\includegraphics[width=1\lw]{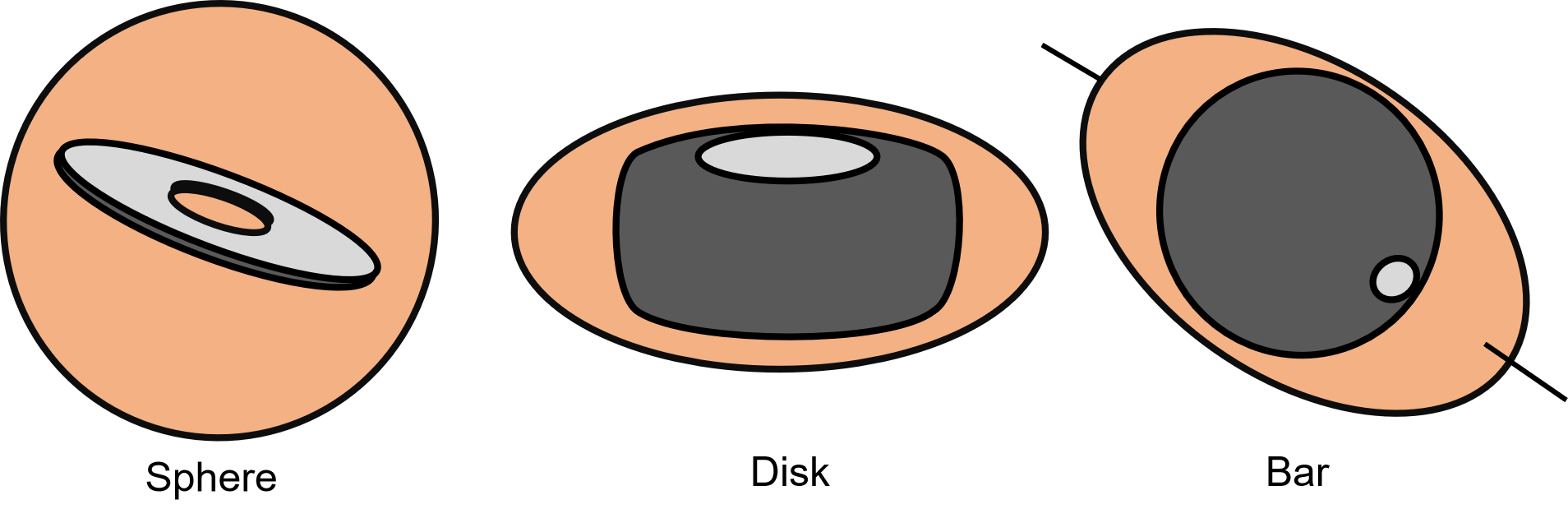} 
\end{center}
 \caption{
 [Top to 3rd] 3D plots of the test particles with initial position and velocity at $\sim$(1.0,   0.0,   0.3:  0.3,   1.0,  -0.5) in spherical ($q=$1:1:1), disk (1:1:0.6) and bar (1:0.6:0.6) potentials, respectively.
 [Bottom] Schematic summary of the final 3D shape of a disk put in a spherical, disk or bar potential.
'Sphere' applies to the CND orbiting in the central cluster and massive black hole, where a tilted disk or a ring can remain flat due to the conservation of angular momentum.
'Disc' may apply to the entire CMZ, but (strong) 'bar' may not apply in the GC.
  {Alt text: Summary of simulation about the final shape of a disk in different external potentials.}
 }
\label{fig-discshape}
\end{figure}

Figure \ref{fig-discshape} summarizes the final shape of the 'disc' in the sphere, disk, and bar potentials.
From (i) as above, we conclude that the gravitational potential in the central $\sim 10$ pc of the Galaxy is almost spherical, having (an) axial ratio(s) greater than 0.95 to reproduce the central cavity structure of moment 0 projected on the sky and the observed LVD and BVD.
This is consistent with the observed infrared luminosity distribution in the central $\sim 10$ pc \citep{2016ApJ...821...44F,2020A&A...634A..71G}.
From (ii) and (iii) we may also conclude that the potential containing the entire CMZ is an axisymmetric disk or at least it is not a strong bar.

This conclusion will be used as the basis for using the rotation curve to calculate the mass distribution in the central region in the next section.
We mention that the present argument applies to 3D potentials, but if the potential is assumed to be 2D or axisymmetric as employed in some simulations, the disk is kept flat by definition (no $z$ extent). 

A concern to be kept in mind is that the simulation does not take into account the gaseous properties such as the pressure and inelastic interaction between the elements.
The former acts to expand the disk, whereas the latter works oppositely.
Therefore, we should consider the result to be qualitative, while the general property of the potential shape does not vary significantly.
 
\ss{Line profiles}

\begin{figure}
\begin{center}      
Line profile\\
Obs. \hskip 3cm Sphere\\
(A)\includegraphics[width=0.44\lw]{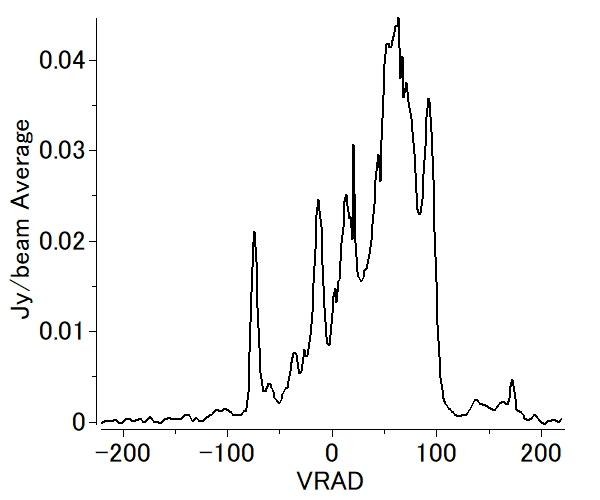}  
(B)\includegraphics[width=0.44\lw]{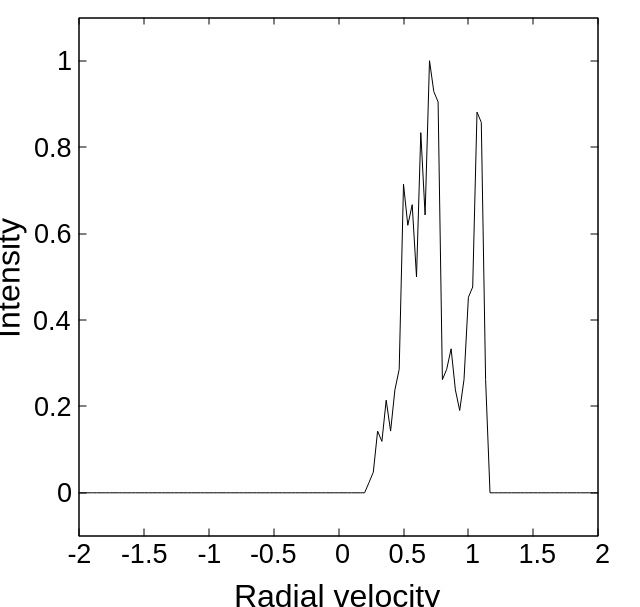}  \\
\vskip 3mm
disk \hskip 3cm Bar\\
(C)\includegraphics[width=0.44\lw]{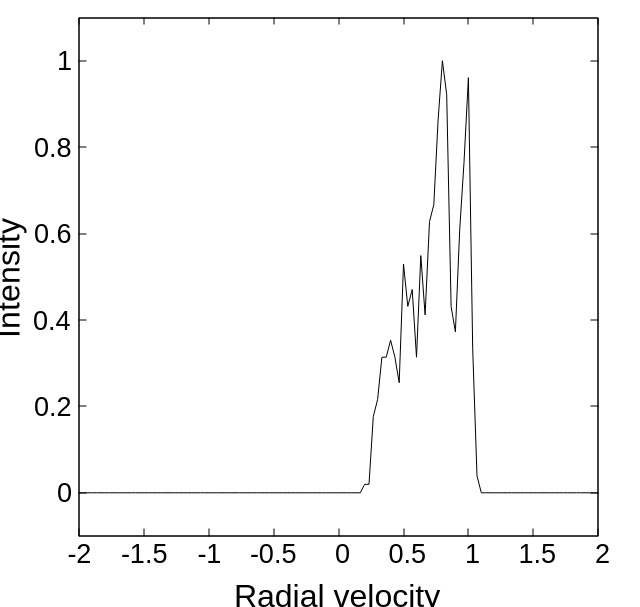} 
(D)\includegraphics[width=0.44\lw]{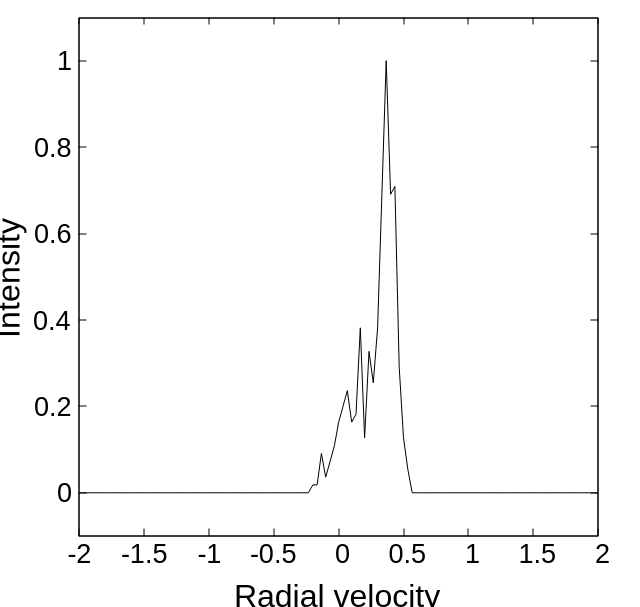} 
 \end{center} 
 \caption{[A] \cs\ line spectrum of the eastern arm of CND. 
 [B] Simulated profile for spherical potential, 
 [C] disk, and 
 [D] bar.  
 {Alt text: Simulated line profiles compared with the observation.}
 }
\label{fig-lineprof}
\end{figure}

In addition, the detailed shape of each line is also useful for constraining the model.
The large widths of the spectral profiles are shown to manifest the kinematical properties of the gas clouds that rapidly rotate around the GC at $\vrot \sim 100-150$ \kms \citep{sofue+25b}.
Figure \ref{fig-lineprof} compares the observed line profile of the eastern clump of CND with the simulated results.

The CS profile exhibits a typical lopsided and broad spectral shape typical for a receding side of a rotating disk, showing a sharp cut at the positive terminal velocity, representing the rotation velocity, and extended outskirts toward the negative velocity.
The simulated profile for the coplanar rosette orbit (panel C) in the spherical potential seems better than the others which all exhibit narrower line widths.
This confirms the conclusion of the previous subsection.

\section{Rotation curve of the GC}
\def\vterm{V_{\rm term}}
\label{sec-rc}

We describe the terminal velocity method using the LVDs, and apply it to the \cs\ and \h40\ line data to obtain the rotation curve of the CMZ and CND.
Based on the argument in the previous section that the central potential is nearly spherical, the rotation curve will be used to derive the mass distribution.

\ss{Terminal velocity method}

The rotation curve is the most fundamental tool for measuring the mass distribution in a disk galaxy \citep{2001ARA&A..39..137S,2017PASJ...69R...1S,2020Galax...8...37S}.
In this section, we apply the RC analysis to diagnose the mass distribution and potential nesting of the CMZ, which is a typical rotating disk.
Among a variety of methods to measure the rotation velocity, the terminal (tangent) velocity method is the most popular and accurate for a gas disk.

The terminal velocity, $\vterm$, was measured as follows: 

[Envelope trace method] A simple method applied to \h40 LVD is the envelope tracing method, which trace the contour drawn at a critical intensity and the errors are given half and twice the intensities of the critical intensity. The critical intensity here was taken three rms. 

[Gaussian deconvolution of line profiles] A more precise method, which we used for the \cs\ lines, is the Gaussian deconvolution of each line profile.
The center velocity of the highest velocity component is taken as the terminal velocity.
Line width is adopted as the standard error of the terminal velocity $\delta \vterm$.

\begin{figure*}    
\begin{center}     
\includegraphics[width=.35\lw]{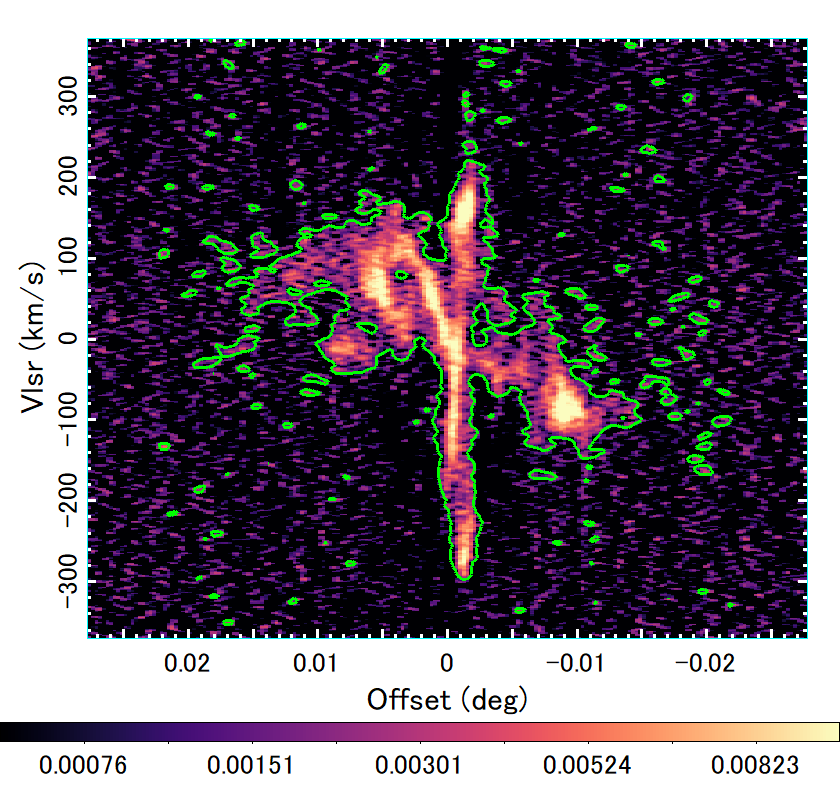}  
\includegraphics[width=.4\lw,height=.33\lw]{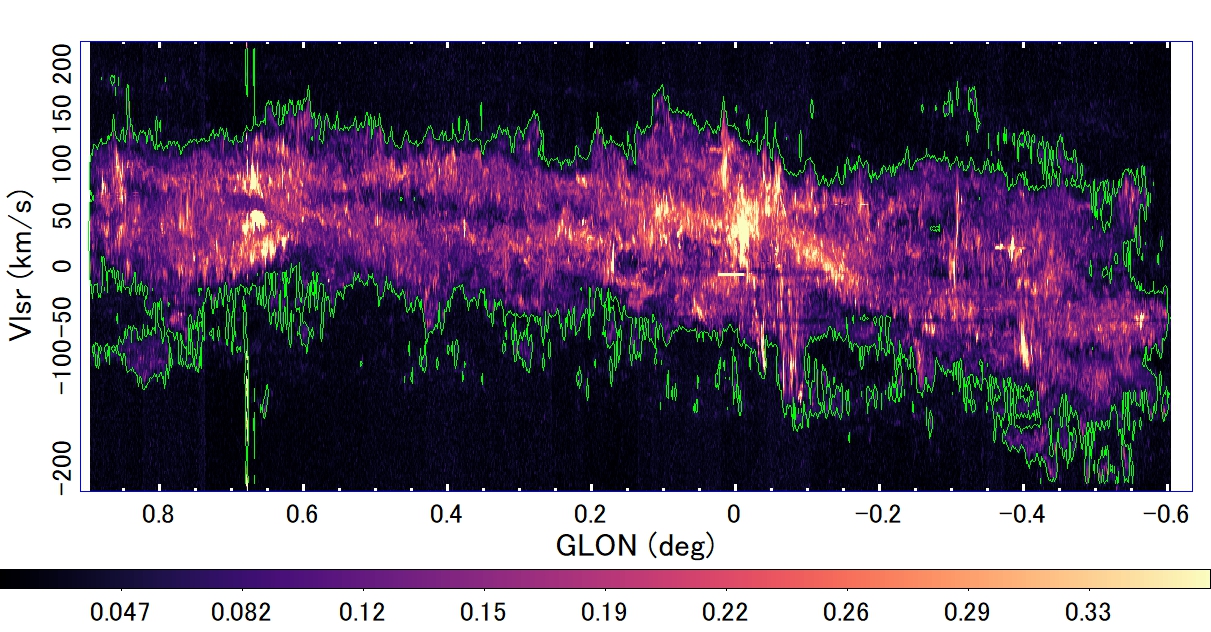}   \\
\includegraphics[width=.4\lw]{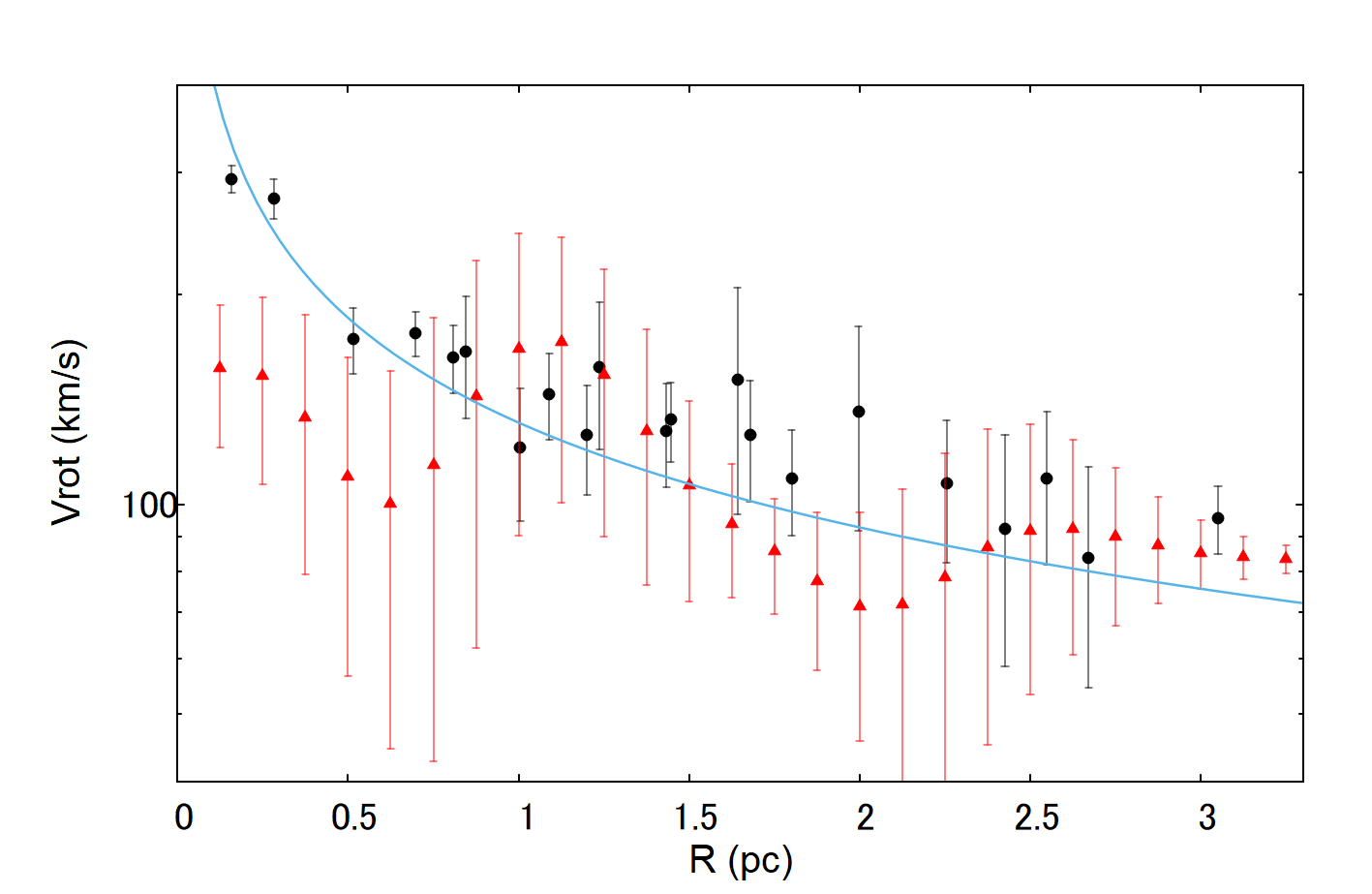} 
\includegraphics[width=.4\lw]{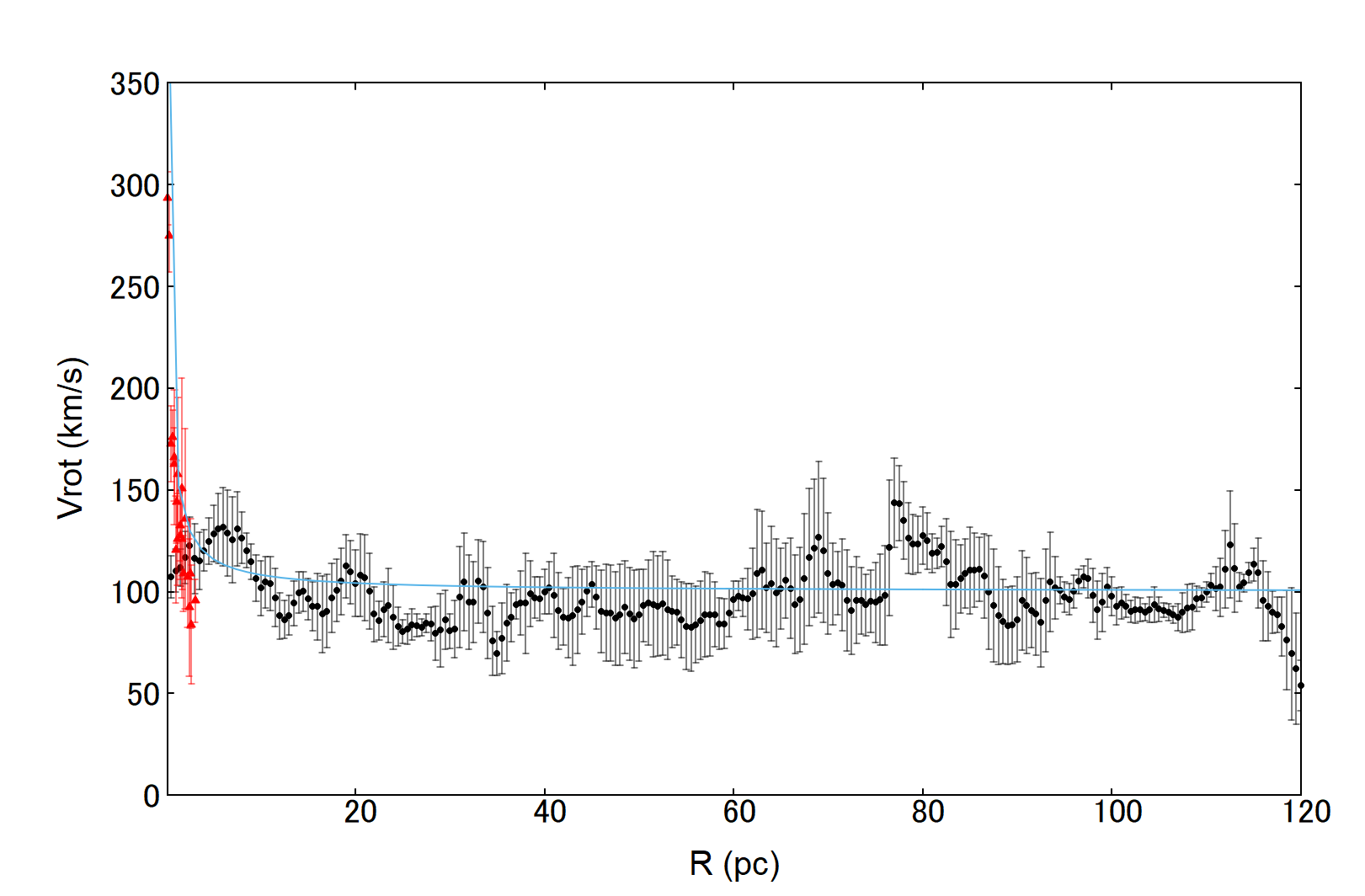}   
\end{center} 
\caption{[Top] Position-velocity diagrams of H40$\alpha$ and \cs\ lines at position angle $70\deg$ with a contour at several rms noise to approximately trace the termminal velocities.  
[Bottom left] Rotation curve (terminal velocity) of the minispiral using PVDs in the top panel. 
Triangles and dots are obtained by the Gaussian deconvolution and envelope tracing method, respectively, while we adopt the latter.
The straight line represent a Keplerian motion around the \sgrastar. 
[Bottom right] RC in the CMZ derived by \cs\ LVD by Gaussian deconvolution.
{Alt text: LVDs with terminal intensity contours of the minispiral in \h40\ and CND in \cs\, and corresponding rotation curves.}} 
\label{rc}  
\end{figure*}

The absolute values of thus determined terminal velocities are plotted as a function of the distance from \sgrastar\ as shown in figure \ref{rc}.  
Finally, the raw plots of the terminal velocities are averaged in each Gaussian bin (here 0.5 pc) of the distance at every 0.5 pc to obtain a rotation curve, $\vrot(R)$.
The standard deviation of the Gaussian running average in each bin is taken as the error of RC, $\delta \vrot$.
The rotation velocity is nearly constant at $\vrot\simeq 120 \ekms$ in the central several pc.

\begin{figure}   
\begin{center}    
\includegraphics[width=.85\lw]{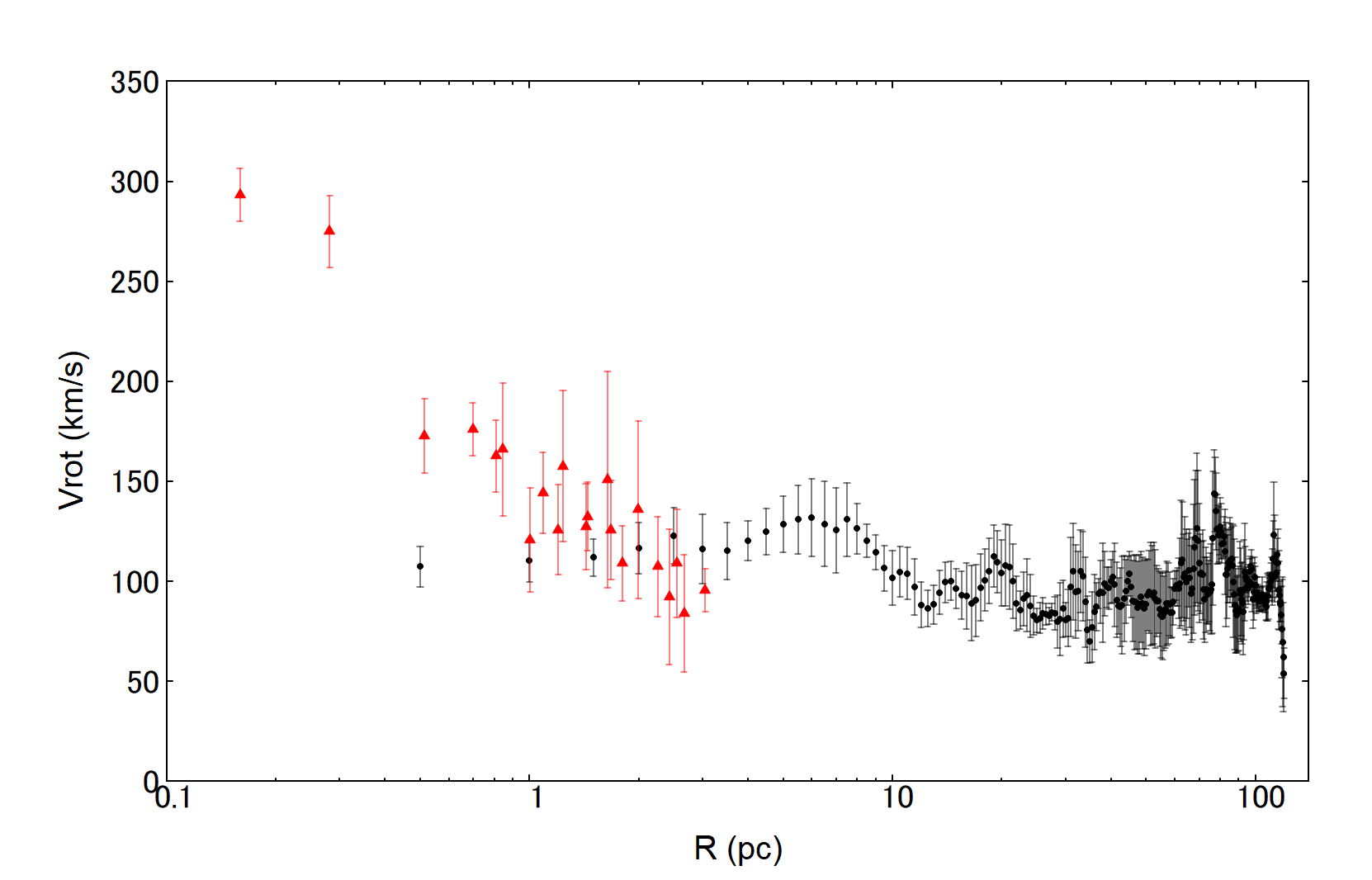} 
\includegraphics[width=.85\lw]{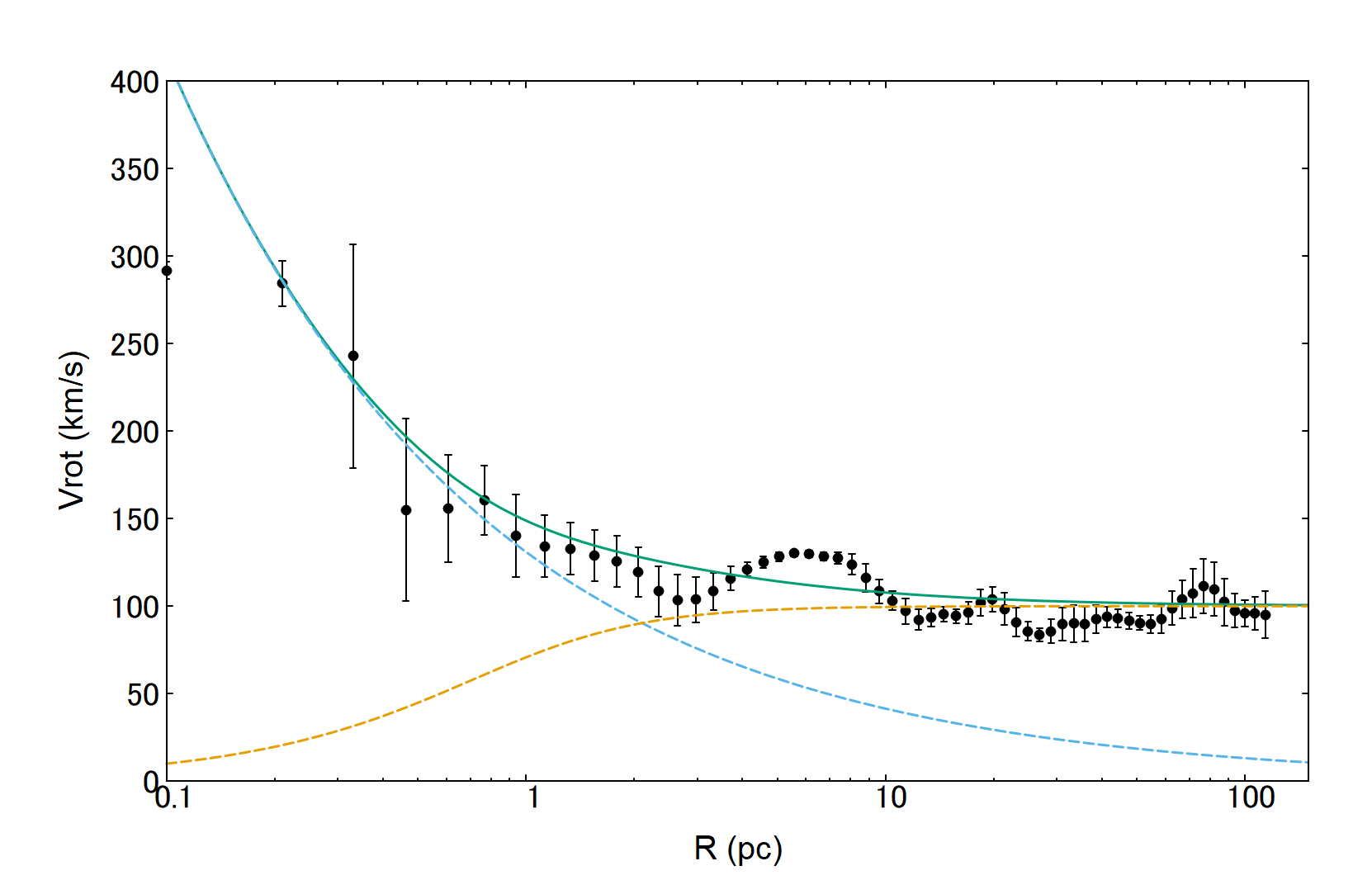} 
\includegraphics[width=.85\lw]{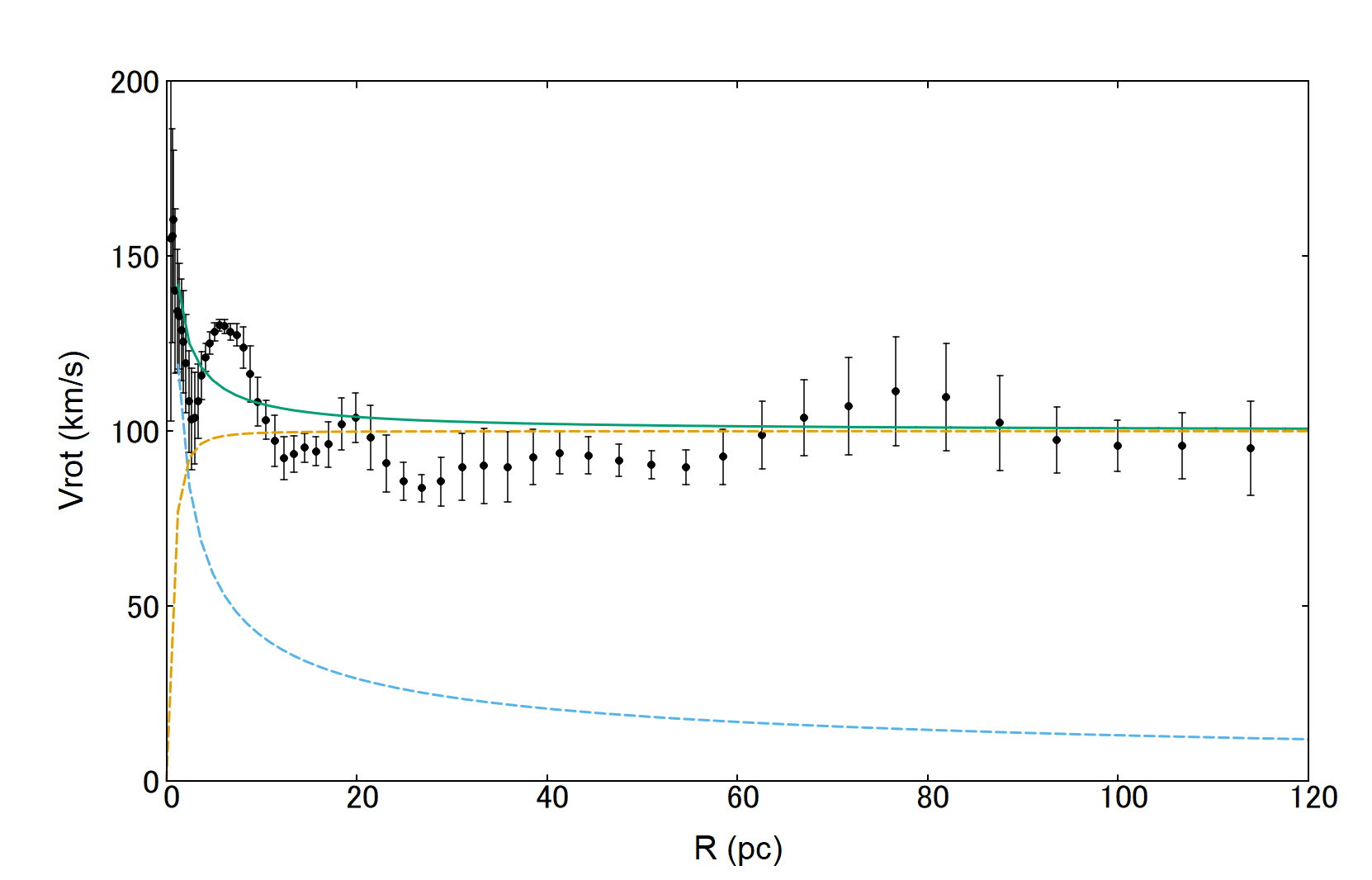}    
\end{center} 
\caption{[Top]
H40$\alpha$ (triangles) + \cs\ (dots) line terminal velocity curve (rotation curve) of the CMZ in semilogarithmic scaling.
[Middle] RC after running averaging.
[Bottom] Same, but in linear scaling.
The curve represents calculated $\vrot$ using equations \ref{eqbh} to \ref{eqvrot}. 
{Alt text: Combined rotation curves of the minispiral and CMZ in semilogarithmic and linear scalings.}
}
\label{rclog}  
\end{figure}

\ss{Relation of the terminal and rotation velocities}
\def\vterm{V_{\rm term}}

Eccentric motion is ubiquitous in the CMZ as inferred from the line-of-sight velocities of $\pm$several tens \kms observed at $l\sim 0\deg$ caused by the extended gravitational potential \citep{k15,sofue+25b}.
It is therefore often argued that the terminal velocity does not necessarily represent the rotation velocity due to the bar potential, where the orbit is not perfectly circular.
We here argue that the terminal velocity represents the rotation velocity in case of a rapidly rotating circumnuclear disk.

The tangential velocity observed as the terminal velocity in the line spectrum of a gas element in an eccentric motion orbiting the Galactic potential is given by
\be
\vterm=\Omega R + \kappa ~r ~\sin \kappa t,
\label{eq-vterm}
\ee
where $\Omega=\vrot/R$ is the angular frequency of the Galactic rotation, and $\kappa$ and $r$ are the epicyclic frequency and radius, respectively, and $t$ is the time. 
In the Galactic disk, a number of gas elements are rotating as ensembles with different epicyclic frequencies, radii, and phases.
Due to the ergodic theorem, the average over the rotational behavior of many individual gas elements should result in an estimate of the rotation curve as
\be
\langle \vterm \rangle = \langle  \Omega R + \kappa~ r~ \sin \kappa t\rangle_t \simeq \Omega R=\vrot(R),
\ee
where the epicyclic term disappears in a couple of rotations or in $t\sim 0.1-0.2$ My.
Therefore, we may safely assume that the measured $\vterm$ represents $\vrot(R)$ in the present circumstances.

\ss{Rotation curves}
\label{ssrc}

Figure \ref{rc} shows a PVD of the \h40\ line near \sgrastar, showing the high-velocity motion of the minispiral.
Using the envelope-tracing method, we determine the terminal velocities and plot them in the lower panel of the figure.
The full line shows the Keplerian law due to the central supermassive black hole with a mass of $M_{\rm BH}=4.2\times 10^6\Msun$ \citep{2008ApJ...689.1044G}.
The innermost terminal velocities within $R\lesssim 1$ pc well fit this Keplerian motion.

Figure \ref{rc} shows a \cs\ LVD covering the entire ACES field from $l=-0\deg.6$ to $+0\deg.9$ (120 pc), where the contours are drawn at $I=0.05$ \Jybeam, approximately following the terminal velocity.
The bottom panel shows the rotation curve in \cs\ obtained by applying the Gaussian deconvolution method.

We then combined the \h40\ and \cs\ rotation curves to obtain a unified RC of the CMZ, as shown in figure \ref{rclog}.
The top panel shows the terminal velocity diagram, and the second and third panels show the results after Gaussian running average in the radius direction.
The rotation curve is well fitted by a superposition of two velocity components of the SMBH and the CMZ disk with a constant rotation velocity represented by
\be
V_{\rm BH}=GM_{\rm BH}/R=131.1 \ekms /\sqrt{X}
\label{eqbh}
\ee
and
\
\be
V_{\rm CMZ}\simeq 100\ekms X/\sqrt{1+X^2},  
\label{eqdisc}
\ee
which is equivalent to a potential given by
\be
\Phi(X)=1/2 v_0^2 \log (1+X^2).
\ee
The rotation velocity is given by 
\be
\vrot=\sqrt{V_{\rm BH}^2+V_{\rm CMZ}^2},
\label{eqvrot}
\ee
where $X=R/1$ pc and $M_{\rm BH}=4\times 10^6\Msun$ \citep{2008ApJ...689.1044G}.

{
\ss{Flat rotation at $\Vrot \sim 100$ \kms }
The rotation curve obtained in this study (figure \ref{rc}) shows that the rotation is nearly flat in the CMZ at $\Vrot= 99 \pm 13 \ekms$. 
This velocity may be compared with the transverse velocity dispersion of the nuclear stellar disk of the order of $\sim 100 \ekms$ inferred from proper motions of the nuclear red clump stars \citep{2022A&A...668L...8N}.  
}
 
\section{Mass distribution}
\label{sec-mass}

Using the rotation curve and assuming that the gravitational potential is nearly spherical, we derive the mass distribution in the GC.
 
 \ss{Mass, density and surface density by spherical assumption}

We infer from the test particle simulations that the morphological and kinematical properties of the CND are reasonably explained by a gas flow orbiting in a spherical potential and drawing a coplanar rosette pattern.
Non-spherical gravitational potentials due to a disk or a bar cannot reproduce the observed molecular-line morphology in the sky.
On the basis of this result, we discuss the mass distribution in the circumnuclear region by using the rotation curve.
Given the spherical potential, the rotation curve of the central bulge is obtained using the terminal velocity along the Galactic plane.

On larger scales, infrared surface photometry indicates that the scale radii of the luminosity profile in the longitude and latitude directions are on the order of $\sim 0.5$ at $R\sim 50-100$ pc \citep{2014A&A...566A..47S}.
So, even in a larger scales such as in the bulge, the photometric shape is mild showing a fat disk.
Considering the long-range gravitational force, this means that the gravitational potential is nearly spherical in the central $\sim 100$ pc.

In this section, we try to determine the enclosed mass within a radius $R$ using the rotation curve obtained in the previous section, assuming that the potential is spherical, and derive the distributions of mass, density, and surface density in the GC.

The mass within $R$ is assumed to be given by
\be
M(R)= R\vrot^2/G.
\label{eqmass}
\ee 
The volume density is calculated by
\be
\rhomass(R)=1/(4\pi R^2) dM(R)/dR,
\label{eqrho}
\ee
and the surface density is obtained by
\be
\Sigma(R)=2\int_0^\infty \rhomass(\sqrt{z^2+R^2}) dz,
\label{eqsig}
\ee 
or approximately
\be
\Sigma(R)\sim M(R)/(\pi R^2),
\label{eqsigpir2}
\ee
where $R$ is the GC distance in the Galactic plane and $z$ is the height from the plane at radius $R$.

\begin{figure}
\begin{center}
\includegraphics[width=.85\lw]{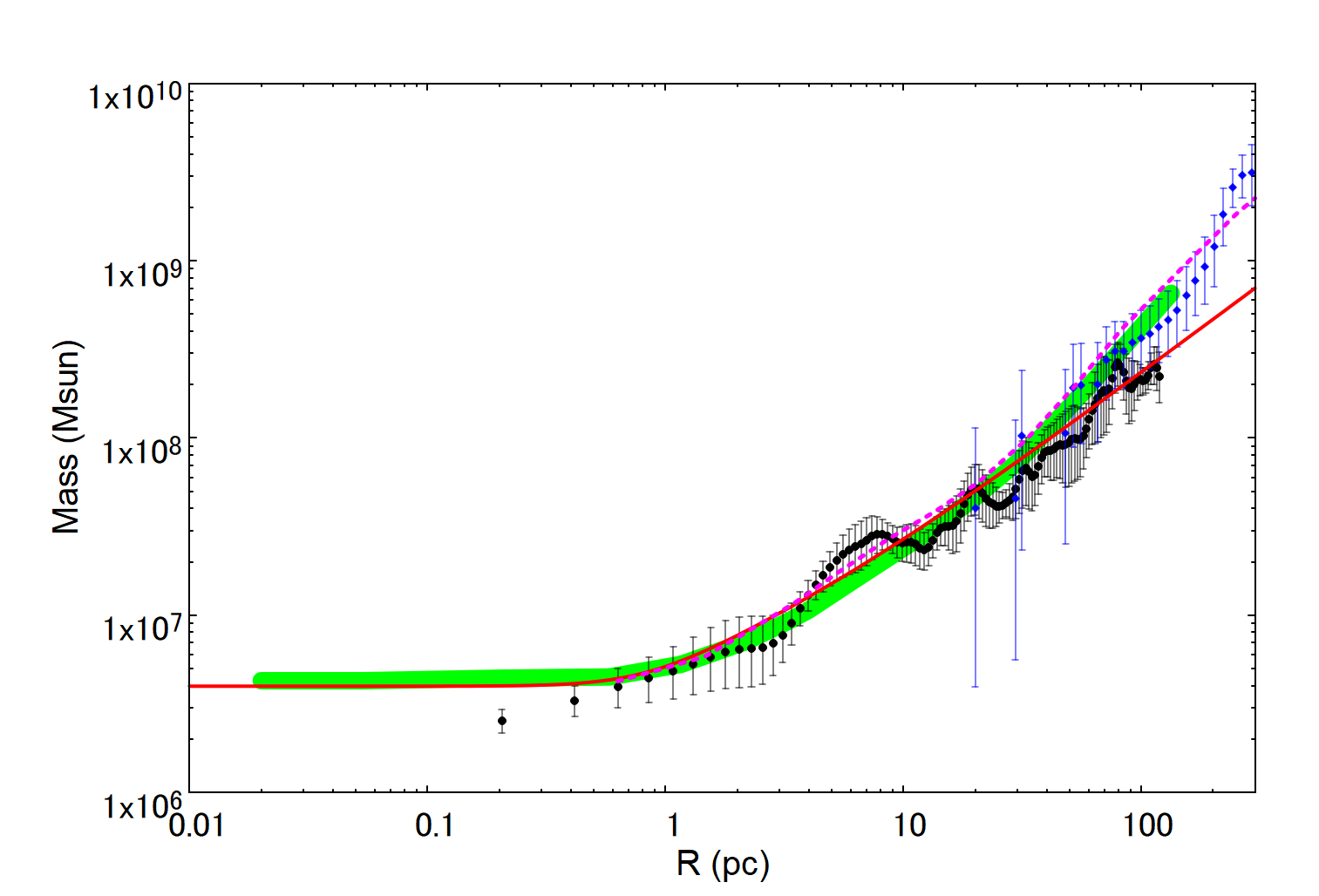}  
\includegraphics[width=.85\lw]{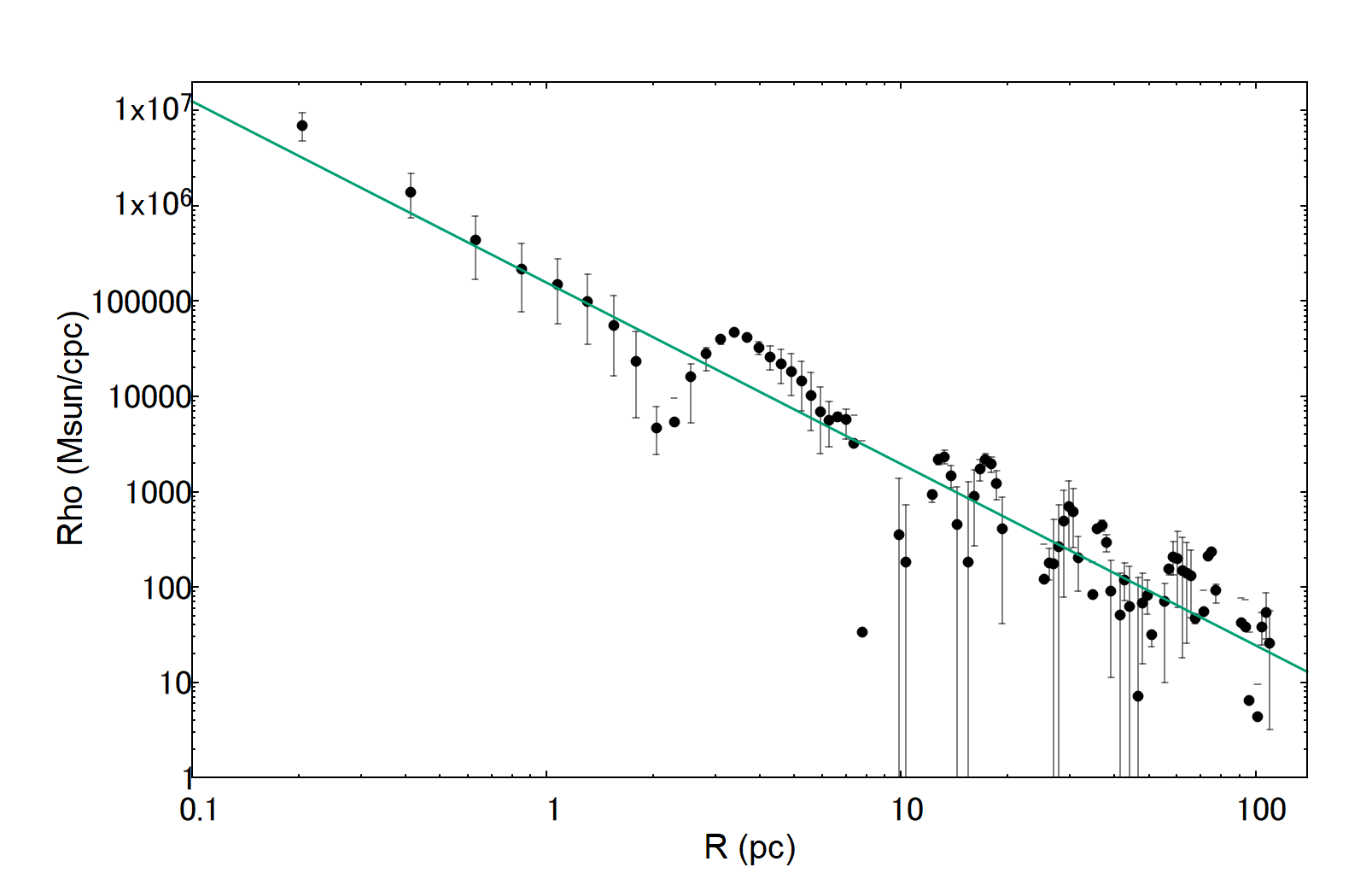}  
\includegraphics[width=.85\lw]{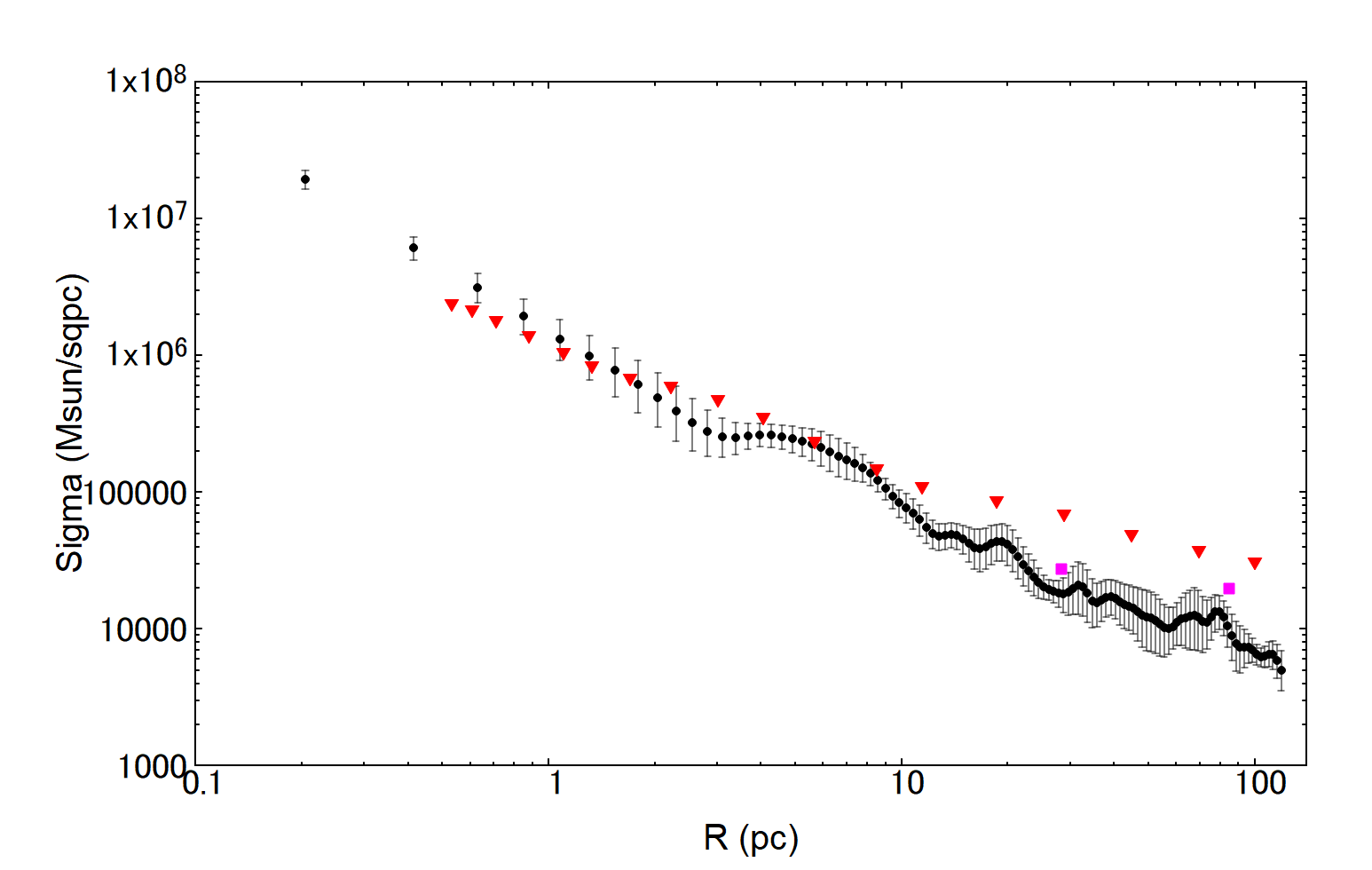}    
 \end{center} 
 \caption{[Top] Enclosed mass $M(R)$ calculated for the RC in figure \ref{rc} (black dots).
Measurements from the literature are summarized by \citet{2016ApJ...821...44F}, falling along the green thick line. 
The mass distribution for the innermost rotation curve is plotted by diamonds \citep{sofue+2025RC}.
The magenta dotted line is the cusp-type model No. 2 + nuclear cluster of the theoretical models by \citet{sorma+20mass}.
 [Middle] Volume mass density $\rho_{\rm mass}$ calculated for $M(R)$.
 The straight line is a least-squares fit to the plot with equal weighting as given by equation \ref{eqrhofit}.
 [Bottom] 
Surface mass density $\Sigma_{\rm mass}$  approximately calculated using equation \ref{eqsigpir2} comapred with the infrared photometric measurement \citep{2016ApJ...821...44F}.
 {Alt text: Radial distributions of the total mass, volume and surface densities in the CMZ.}
 }
\label{fig-mass}
 \end{figure}

Figure \ref{fig-mass} shows the calculated distributions of the total mass, volume density and surface density, using equations \ref{eqmass}, \ref{eqrho} and \ref{eqsigpir2}, respectively.
The total mass distribution is consistent with the current photometric measurements of the circum nuclear stellar cluster (CNS) and central bulge in infrared emissions shown by the thick green line as plotted for comparison from the literature
\citep{2016ApJ...821...44F,2020A&A...634A..71G}.
It also smoothly continued to the Galactic dynamical mass distribution calculated for the inner RC of the Milky Way plotted by triangles \citep{sofue+2025RC}.
The red line represents the mass distribution calculated for the model RC given by equation \ref{eqvrot}.
The middle panel shows the density distribution calculated using the integrated mass in the top panel.
The least squares fit to the log-log plot yields 
\be
\rho=10^{5.194\pm 0.122}(R/1\epc)^{-1.904\pm 0.090}.
\label{eqrhofit}
\ee
The bottom panel shows the surface density calculated approximately using equation \ref{eqsigpir2} compared with the infrared photometric measurements \citep{2016ApJ...821...44F}.

\ss{Comparison with model potentials assumed in the simulations}

There are two basic models to represent the central gravitational potential: the cusp type and the finite-density type.

An example of a cusp ($\Phi \sim \log~ R$) type assumes a potential of the form given by \ref{eqbar}, which produces a constant circular velocity (flat RC) and a cuspy profile of the central density as $\rho\propto R^{-2}$ \citep{2019MNRAS.486.3307D,kruijssen+2019}.

The finite-density type includes the Plummer potential and exponential, de Vaucouleurs, and / or Sersic-type density profiles (\citet{2017MNRAS.469.2251R} and the articles cited in Section 1). 
This type of potential yields a mild and finite density at the nucleus.
The rotation velocity is zero at the center and increases linearly with radius, showing a rigid-body RC.

\begin{figure}
\begin{center}    
\includegraphics[width=0.85\lw]{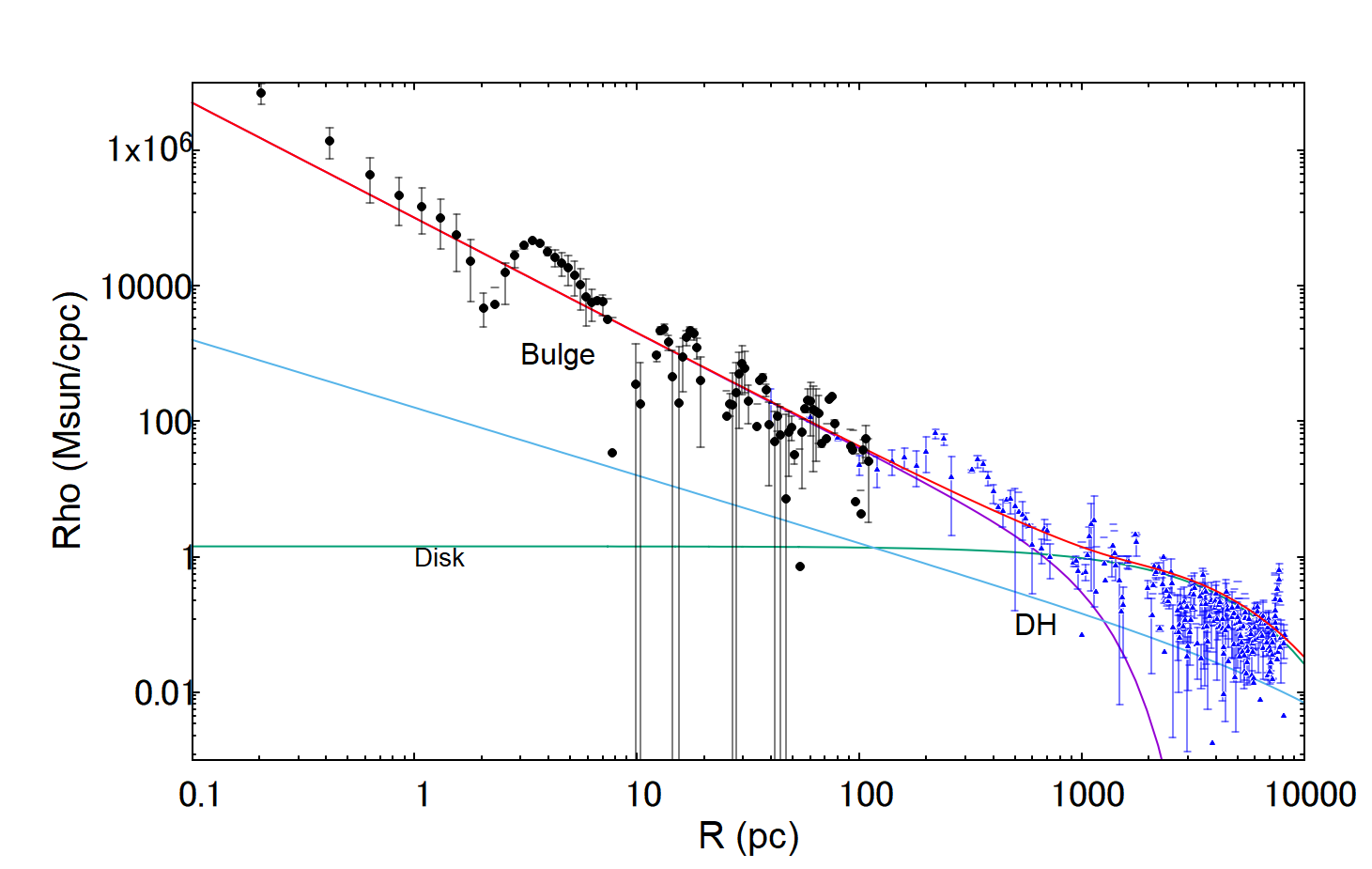}     
\end{center}
 \caption{Derived mass density distribution (dots) compared with the cusp-type potential used in the hydrodynamical simulation \citep{2017MNRAS.469.2251R}.
 Small triangles are density distribution in the Milky Way \citep{sofue+2025RC}.
 {Alt text: Derived mass distribution compared with a cusp-type potential used in a simulaiton.}
 }
\label{fig-compare}
\end{figure}

Figure \ref{fig-compare} shows the comparison of the observed distribution of the dynamical mass inside the CMZ with the theoretical models adopted in the current simulations \citep{2017MNRAS.465...76M,2017MNRAS.469.2251R}.
The dots are plots as calculated in this paper using the observed rotation curve. 
In terms of the comparison here concerned, the cusp model ($\rho \propto R^{-1}$) seems preferable to represent the potential in the central region of the CMZ.

\section{Stability of the rotating gas disk in the derived potential}
\label{sec-jeans}
 
Using the rotation curve and potential, we argue that the gas disk is dynamically stable against star formation in the central $\sim 10$ pc.

\def\LJ{\lambda_{\rm J}} \def\MJ{M_{\rm J}}

\ss{Tidal enhancement vs suppression of star formation}

A rotating gas disk in the Galactic gravitational potential suffers from various external forces that either compress the gas and enhance star formation or dilute it to suppress star formation. 

Various origins of protostellar clumps are proposed, such as gravitational instability in a uniform medium with small perturbations, external compression by galactic shock waves, compressing trigger by explosive events, or cloud-cloud collision.
However, regardless of the origin of a gas clump, the final decision for the clump to become a 'proto-stellar cloud' depends simply on the gravitational stability of the clump and on whether the self-gravitational force exceeds the thermal pressure and external disturbances.
This stability is most commonly examined by the Jeans criterion. 
It is stressed that the Jeans instability gives the necessary condition for the clump to grow to a star, but is not the sufficient condition.
On the other hand, Jeans stability is a sufficient condition for the clump not to form a star.

\ss{Tidal disruption of a cloud}

We study the tidal effect of the Galactic potential on a molecular cloud, by performing an order-of-magnitude estimation,
The Roche radius for an object of mass $m$ orbiting in the potential is given by
\be
r \sim [m/M(R)]^{1/3}R.
\ee
In case of a molecular cloud of mass $m\sim 10^4\Msun$ located at $R\sim 3$ pc, for example, the Roche radius is $r\sim 0.5$ pc and the critical density is $\sim 2\times 10^4$ \Htwocc.
In figure \ref{fig-cloudevo} we show the evolution of the tidal deformation and disruption of a cloud orbiting at $R=3$ pc, which initially was given a radius $r=0.1R=0.3$ pc and a velocity dispersion $\sigma_v = 0.1\vrot=10$ \kms representing a cloud of mass $m= r \sigma_v^2/G \sim 0.7\times 10^4 \Msun$ and $\nH2\sim 10^3$ \Htwocc. 
The simulation reveals that a cloud is easily stretched and destroyed in the deep potential.

\ss{Tidal-Jeans (TJ) instability}

\def\c{c_{\rm s}}

The Jeans length and mass of a proto-stellar molecular cloud formed in a gas cloud of density $\rho\sim 10^5$ \Htwocc are of the order of $\lambda_{\rm J}\sim 0.1$ pc and $M_{\rm J}\sim 1\Msun$.
However, the tidal force due to the central bulge suppresses its growth.
This is simply because the orbital period in the background potential is shorter than the Jeans time in the cloud. 
This is equivalent to the comparison of the Roche radius to the Jeans length, or to the Coriolis force stronger than the self-gravitational force.

We here discuss a modification of the linear gravitational instability of a molecular cloud using the dispersion relation for the Jeans criterion in a rotating system \citep{1954ApJ...119....7C,chandra1961}, which is written as
\begin{equation}
\omega^2=\c^2 k^2-4\pi G\rho+4\Omega^2,  
\label{eq_disp}
\end{equation}
where $\omega$ and $k$ are the angular frequency and wavenumber, respectively, of the linear perturbation represented by $\delta \rho\propto{\rm exp}(\omega t-kr)$, $\rho$ is the gas density, $\c$ is the sound velocity, and $\Omega=R/\vrot=2\pi/t_{\rm rot}$ is the angular velocity of the rotating system with $t_{\rm rot}$ being the period of rotation.
This equation is equivalent to the dispersion relation taking into account the tidal effect \citep{2013MNRAS.434L..56J} written as
\begin{equation}
\omega^2=\c^2 k^2-4\pi G\rho+T, 
\label{eq_tide}
\end{equation}
where $T=-\partial^2 \Phi /\partial r^2$ is the external tidal acceleration per unit distance and $\Phi$ is the gravitational potential in which the cloud orbits, and if the orbit is circular its angular speed is equal to $\Omega$.

We comment that equations \ref{eq_disp} and \ref{eq_tide} treat the tide as purely radial (spherical), which is assumed in the present circumstance.
However, if the potential is not spherical, for example a disk \citep{kruijssen+2019}, the cloud is compressed in the vertical direction by the vertical gravity of the disk and is further tidally streched vertically, the former enhancing star formation, but the latter suppressing.

Equation \ref{eq_disp} yields the modified critical wavelength corresponding to $\omega^2=0$, which we call the tidal-Jeans (TJ) wavelength,
\begin{equation}
    \lambda_{\rm TJ}=2\pi/k_{\rm TJ}=\lambda_{\rm J} \[1-Q_{\rm TJ}^2\]^{-1/2},
    \label{eq_lamtj}
\end{equation}
where 
\be
Q_{\rm TJ}=2\Omega t_{\rm J}
\ee
is a tidal-Jeans Q value defined by the ratio of the free-fall time, or Jeans time, of the cloud
\be
t_{\rm J}=1/\sqrt{4 \pi G \rho}
\ee
to the free-fall time $t_{\rm ff}\sim 1/\Omega$ in the external potential of the galaxy, and   
\be
\lambda_{\rm J}=2\pi\c t_{\rm J}
\ee
is the Jeans wavelength.
The tidal-Jeans mass is then given by
\be 
M_{\rm TJ} =M_{\rm J} \[1-Q_{\rm TJ}^2\]^{-3/2},
\label{eq_mtj}
\ee
where 
\be
M_{\rm J}=(4\pi/3) (\lambda_{\rm J}/2)^3 \rho.
\ee
is the Jeans mass.

When the wavelength $\lambda$ and the mass $M$ are smaller than these critical values, specifically if $\lambda < \lambda_{\rm J}$ or $M < M_{\rm J}$, then $\omega$ is imaginary, leading to a stable oscillation of the perturbation.
If vise versa, $\omega$ is real and the perturbation grows, the system is unstable.
It is stressed that equation \ref{eq_lamtj} has no solution, or the system is stable when
\be
Q_{\rm TJ}>1, ~ {\rm or} ~
t_{\rm J}>1/2\Omega,
\label{eq_tJomega}
\ee
or when the Jeans time is longer than the galactic rotation period, regardless of the sound velocity or regardless of wavelength and mass.
Equation \ref{eq_tJomega} can be read as 
\be
R< R_{\rm T}=2t_{\rm J}\vrot=2\vrot/\sqrt{4 \pi G\rho},
\ee
so, the gas disk is stable within a critical radius $R_{\rm T}$, depending on density and rotation velocity. For $\vrot=120$ \kms, we have 
\be
R_{\rm T}\sim 14 (\rho/10^5 {\rm \mu H_2~cm^{-3}})^{-1/2}~[{\rm pc}],
\ee
where $\mu=1.38$ is the correction for heavy elements.

\begin{figure}
\begin{center}   
\includegraphics[width=0.8\lw]{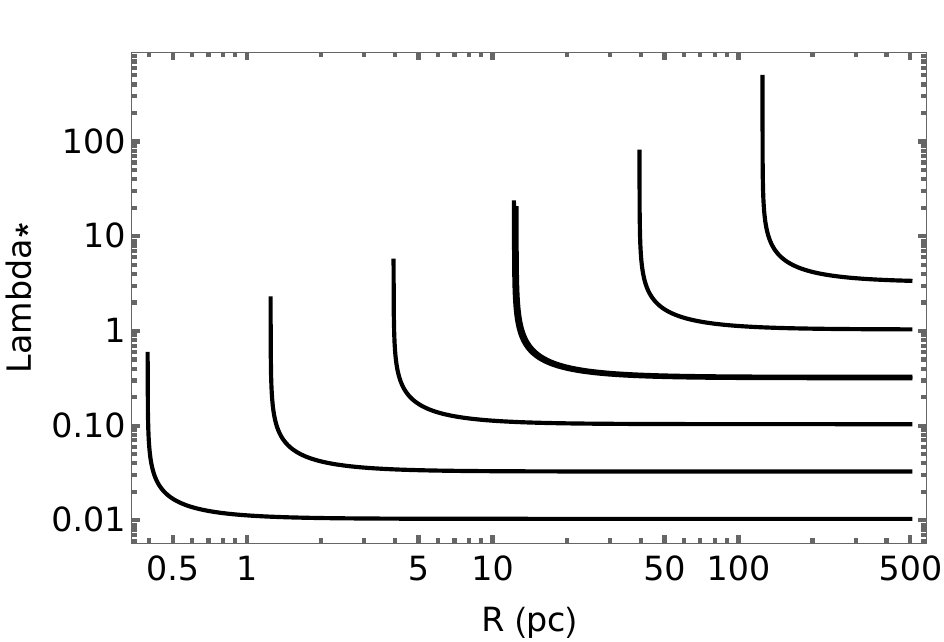}  
\includegraphics[width=0.8\lw]{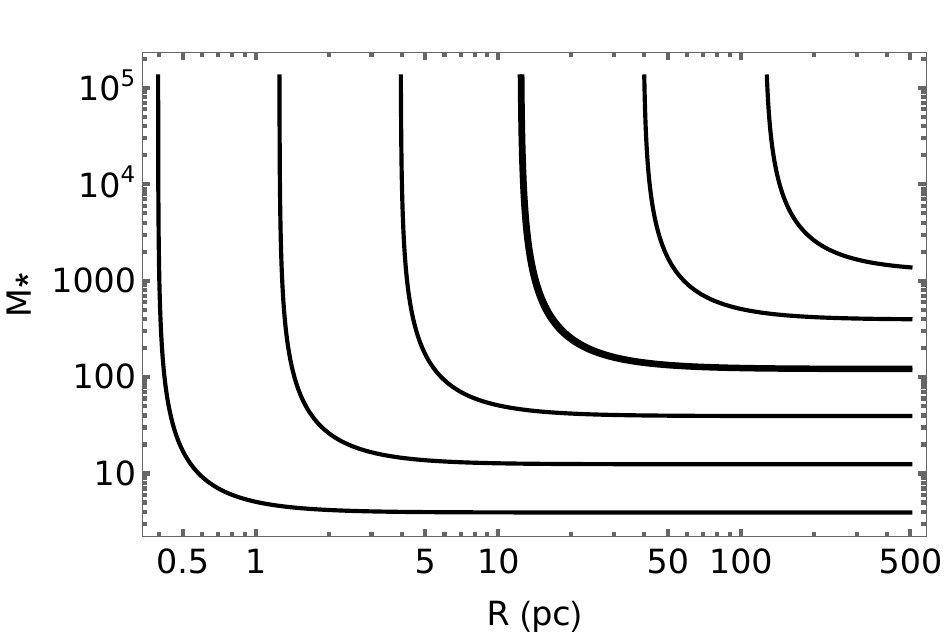} 
 \end{center} 
 \caption{[Top] 
 $L_*=\lambda_{\rm TJ}/(\c/{\rm km~s^{-1}})$ and [bottom] 
 $M_*=M_{\rm TJ}/(\c/{\rm km~s^{-1}})^3$ for gas densities from 
 $n_{\rm H_2}=10^3$ (top right curve) to $10^8 \mu$ (bottom left curve) \Htwocc, with thick curve for $10^5 \mu$ \Htwocc.
Galactic rotation velocity is $\vrot=120 \ekms$. 
Note the tilted belt covered by the strongly curved region from outer horizontal to inner vertical lines, which we call the "T-Jeans transition region".
 {Alt text: Normalized TJ length and mass against $R$ for various gas densities.}
 }
\label{fig-TJeans}
\end{figure}

For convenience in presenting the analyzed results, we introduce the parametric wavelength $\lambda^*$ and mass $M^*$ normalized by the critical tidal-Jeans values corresponding to a sound velocity of $\c=1 \ekms$, which are used in the plots in figure \ref{fig-TJeans}: 
\be
 \lambda^*(\rho)=\lambda_{\rm TJ}/ (\c/1 \ekms),
\ee
and 
\be
 M^*(\rho)=M_{\rm TJ}/(\c/1\ekms)^3.
\ee

 \ss{Tidal-Jeans criterion in Keplerian potential around the SMBH}

\begin{figure}
\begin{center}   
\includegraphics[width=0.7\lw]{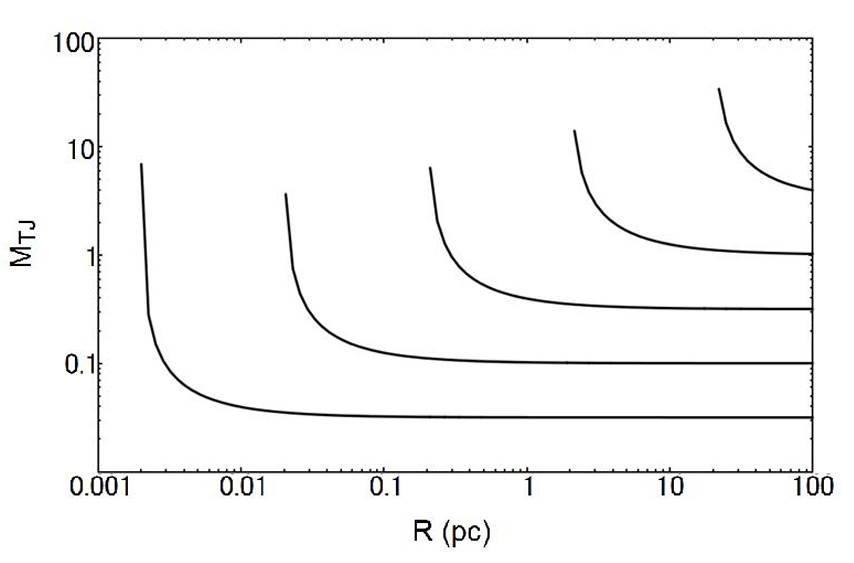}  
\includegraphics[width=0.7\lw]{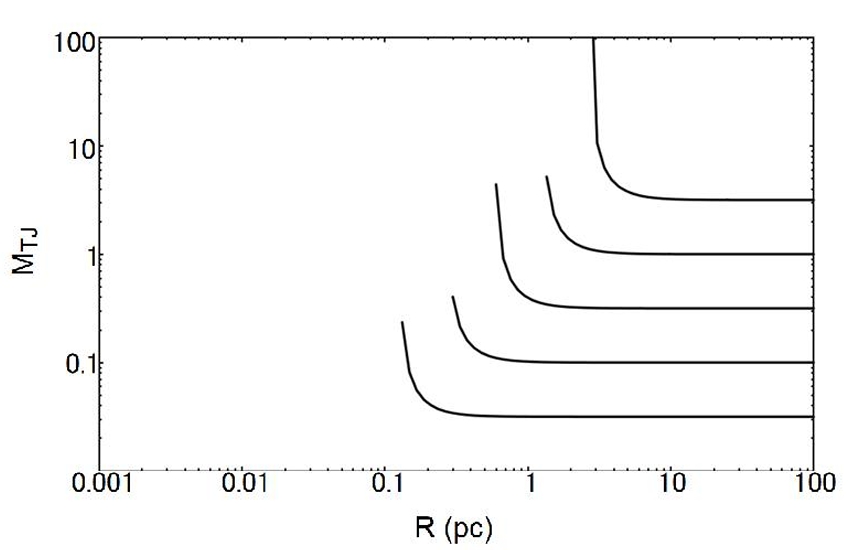}
 \includegraphics[width=0.7\lw]{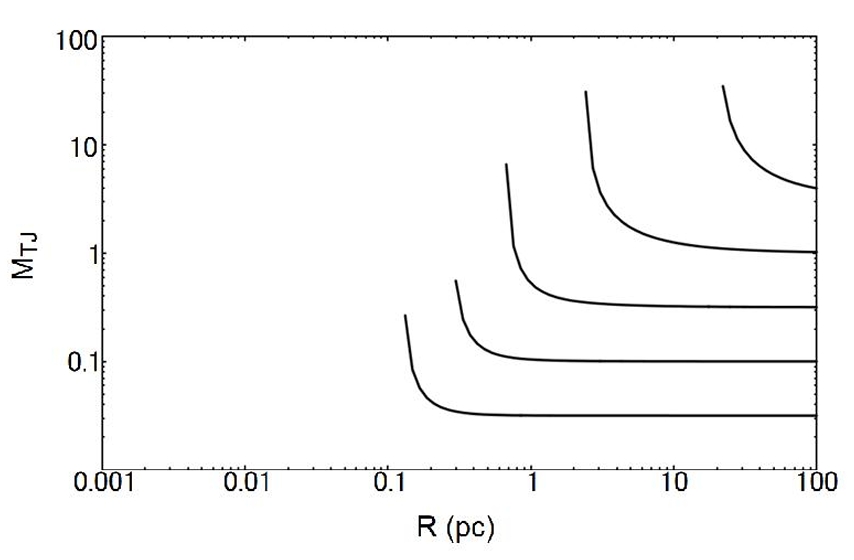}
 \end{center} 
 \caption{[Top] Normalized TJ mass of a gas cloud with density from $10^3$ (uppermost curve) to $10^7$ (lowermost curve) normalized density, orbiting in a potential with flat rotation curve.
 [Middle] Same, but around a point mass with Keplerian orbit.
[Bottom] Same, but a Keplerian by point mass + flat rotation. 
 {Alt text: Normalized TJ mass for flat rotation curve and Keplerian rotation curve.}
 }
\label{fig-MTJ-flat-kep}
 \end{figure}

 Although trivial, it is interesting to point out that the radial transition of the TJ criterion becomes more sensitive to the GC radius if we assume a steeper potential such as that by a point mass.
 This effect becomes critical near the central SMBH at $R\lesssim 2$ pc.
 In figure \ref{fig-MTJ-flat-kep} we compare the dependency of the $M_{\rm TJ}$ curve between the flat rotation as discussed in this paper and that calculated for a Keplerian rotation around the SMBH.
 The curves are shown for five different gas densities, increasing every 10 times from the top to bottom curves for 5 orders of magnitude.
 
\ss{Toomre's stability of a rotating disc}

Although essentially the same as the Jeans criterion, the stability of a self-gravitating disk in rotation has been analyzed in terms of surface density \citep{1964ApJ...139.1217T}.
Although an exact analysis does not apply here because the external potential is stronger, here we try to estimate the effect of the tide on the stability of the disk.
The dispersion relation for an axisymmetric density perturbation in a thin disk is given by
\be
\omega^2=\c^2 k^2 - 2 \pi G \Sigma_{\rm g} k + \kappa^2
\label{dispQ}
\ee
where $\kappa$ is the epicyclic frequency and $\Sigma_{\rm g}$ is the surface density of the gas disk \citep{BT2008}.
The critical wavelength is given by
\be
\lambda_{\rm Q}=2\pi/k=(2\c^2/G\Sigma_{\rm g})[1\pm\sqrt{1-Q^2}]^{-1},
\ee
where
\be
Q=\c \kappa/\pi G\Sigma_{\rm g}
\ee
is the Toomre's $Q$ value.
If $Q > 1$, equation \ref{dispQ} has no solution and the disk is stable (oscillation as $\propto \exp i |\omega| t$), while if $Q < 1$ the disk is unstable (growing as $\propto \exp |\omega| t$) and radially fragmented by self-gravity. 

\begin{figure}
\begin{center}    
\includegraphics[width=0.8\lw]{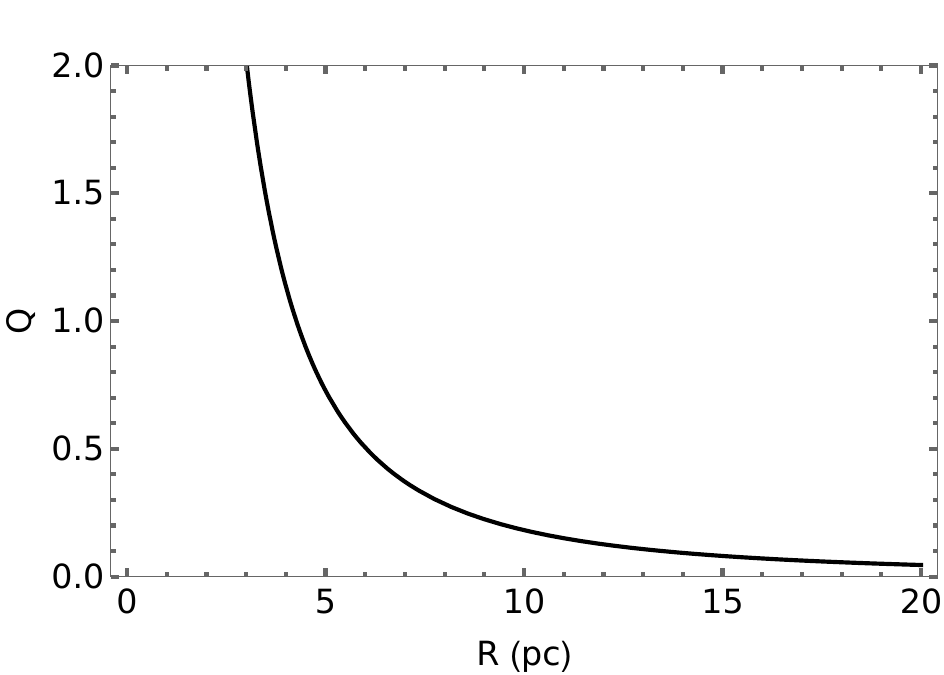}    
\includegraphics[width=0.8\lw]{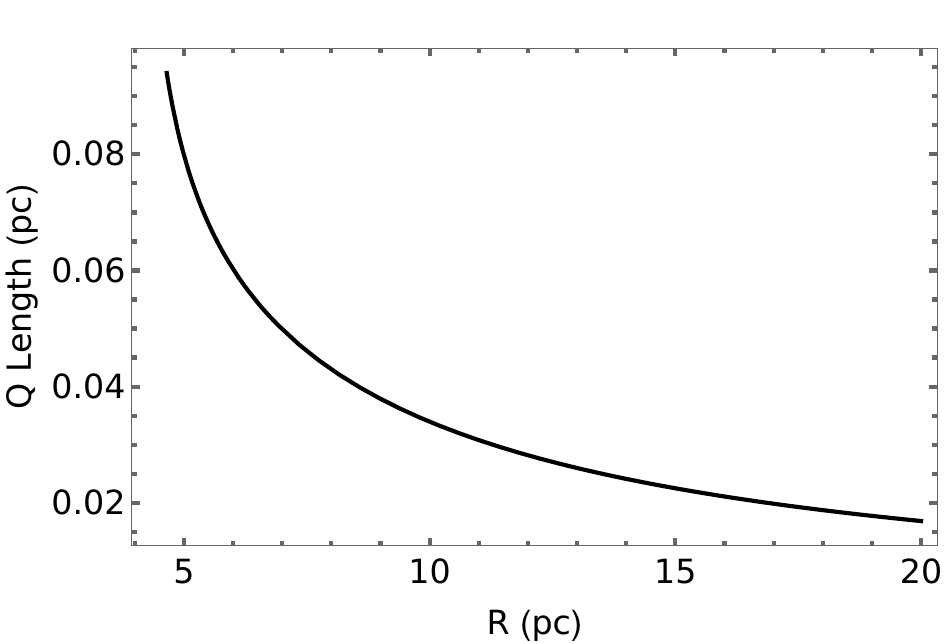}    
 \end{center} 
 \caption{[Top] Toomre's $Q$ value for a constant rotation velocity $\vrot=120$ \kms, $n_{\rm H_2}=10^5$ \Htwocc, $\c=1$ \kms, and the disk thickness is one tenth the radius, $h=0.1R$ ($\Sigma_{\rm g}=\rho h$).
 The disk is stable inside $R\sim 4$ pc where $Q\ge 1$ for any wavelength. 
 [2nd] critical wavelength $\lambda_Q$ against $R$. The disk is stable for a wavelength below the line.  
 {Alt text: Toomre's $Q$ plotted against $R$.}
 }
\label{fig-Q}
 \end{figure}

\begin{figure}
\begin{center}   
\includegraphics[width=0.8\lw]{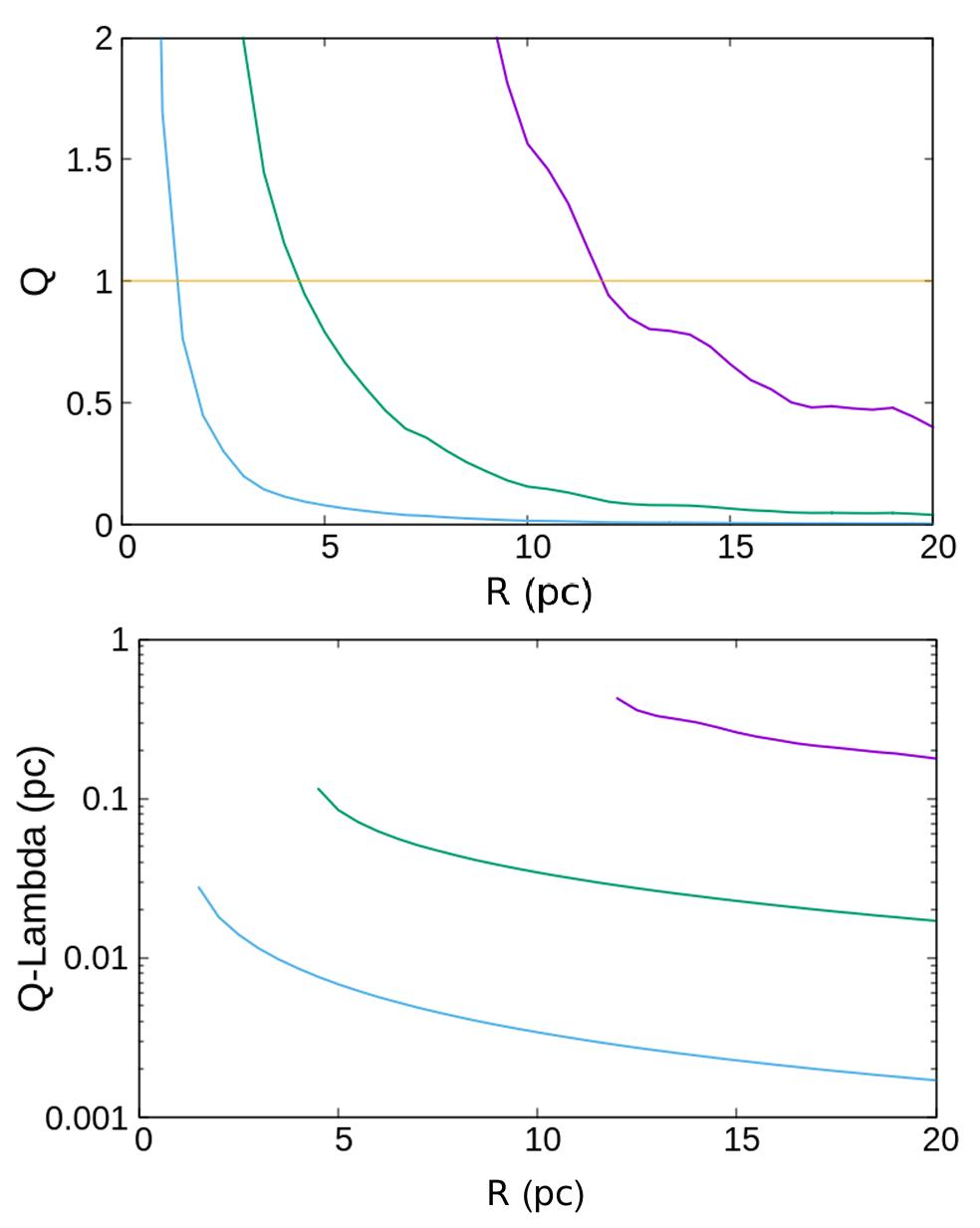}      
 \end{center} 
 \caption{ 
 [Top] Same as figure \ref{fig-Q}, but using the observed RC for $\rho=10^4, 10^5, 10^6$ \Htwocc.
 [Bottom] Same, but $\lambda_{Q}$. 
 {Alt text: Toomre's $Q$ plotted against $R$ for observed rotation curve.}
 }
\label{fig-Qrc}
 \end{figure}

In the present case of CND, where the rotation curve is flat, we may approximate the epicyclic frequency by
$\kappa\sim \sqrt{2}\Omega=\sqrt{2}\vrot/R$.
In figure \ref{fig-Q} we plot the $Q$ value as a function of $R$ for a gas density $\rho\sim 10^5$ \Htwocc, assuming that the thickness of the disk varies as $h\sim 0.1 R$, which yields $\Sigma_{\rm g}=\rho h \propto R$.
Figure \ref{fig-Qrc} shows the same, but using the observed rotation curve for three different gas densities.

The gas disk is shown to be stable inside $R\lesssim 4$ pc, where $Q\ge 1$.
Outside 4 pc, the disk is unstable and the critical wavelength of growing instability decreases with the radius as shown in the bottom panel of figure \ref{fig-Q}.

However, we note that the $Q$ value analysis is valid only for an infinitesimally thin disk with axisymmetric fragmentation into rings.
In addition, the external potential, which is treated mixed, may act to further stabilize the disk.
In fact, the calculated value of the wavelength is unreasonably small to be on the order of $\sim 10^{-3}$ pc, which is due to the assumed condition that the sound velocity is small, the disk is thin, and there are no other velocity dispersions. 
Furthermore, in order to discuss star formation, we need to further test the fragmentation of individual rings into proto-stellar clouds in the azimuthal direction.
So, we must be careful to apply it to the present circumstance, where the disk thickness is comparable to the radial extent,

\ss{Suppression of star formation}

In order to visualize the Jeans stability of a molecular cloud due to the tidal effect, we plot in figure \ref{fig-TJeans} the normalized TJ length and mass, $\lambda^*$ and $M^*$, as a function of the galactocentric distance, $R$, for various densities from $\rho=10^4$ to $10^8$ \htwocc, assuming a constant rotation velocity of $\vrot=120$ \kms.

The figures tell us that the density perturbation does not grow at any wavelength and mass, if the rotation period of the system is sufficiently shorter than the Jeans time, or $t_{\rm J} >  t_{\rm rot}$, and regardless of the sound velocity.

On the other hand, if $t_{\rm J} < t_{\rm rot}$, the critical tidal-Jeans length and mass tend to the Jeans wavelength and mass, and the values become dependent on the sound velocity.

For example, if we take $\rho=10^5 \mu$ \Htwocc, as shown by the thick line in figure \ref{fig-TJeans}, the gaseous system inside a critical radius of $R_{\rm T}\sim 14$ pc is stable at any sound velocity. 
Outside of the critical radius, the system is unstable at a wavelength of $\lambda^*>3$ pc and a mass of $M^*>100 \Msun$ for $\rho=10^5$ \Htwocc.
If the sound velocity is $\c\sim 0.1$ \kms, clouds with a mass greater than $\sim 0.1\Msun$ can grow.

We stress that the tidal suppression is effective only inside the critical radius, e.g. at $R\lesssim 14$ pc for a cloud $10^5$ \Htwocc, but outside the radius the disk stability/instability is subject to the usual Jeans criterion.
In fact, a moderate or slightly higher star formation efficiency has been obtained in the entire CMZ \citep{2025ApJ...984..157B}.

Also note that the present analysis applies to a spherical system.
If a disk potential is assumed, as simulated for the mid-CMZ region \citep{kruijssen+2019,2019MNRAS.486.3307D},
the vertical tide, or the vertical gravitational force, acts to compress the cloud/disk in the vertical direction and enhances star formation.

\ss{Effect on the circumnuclear IMF}

\begin{figure}
\begin{center}   
\includegraphics[width=0.8\lw]{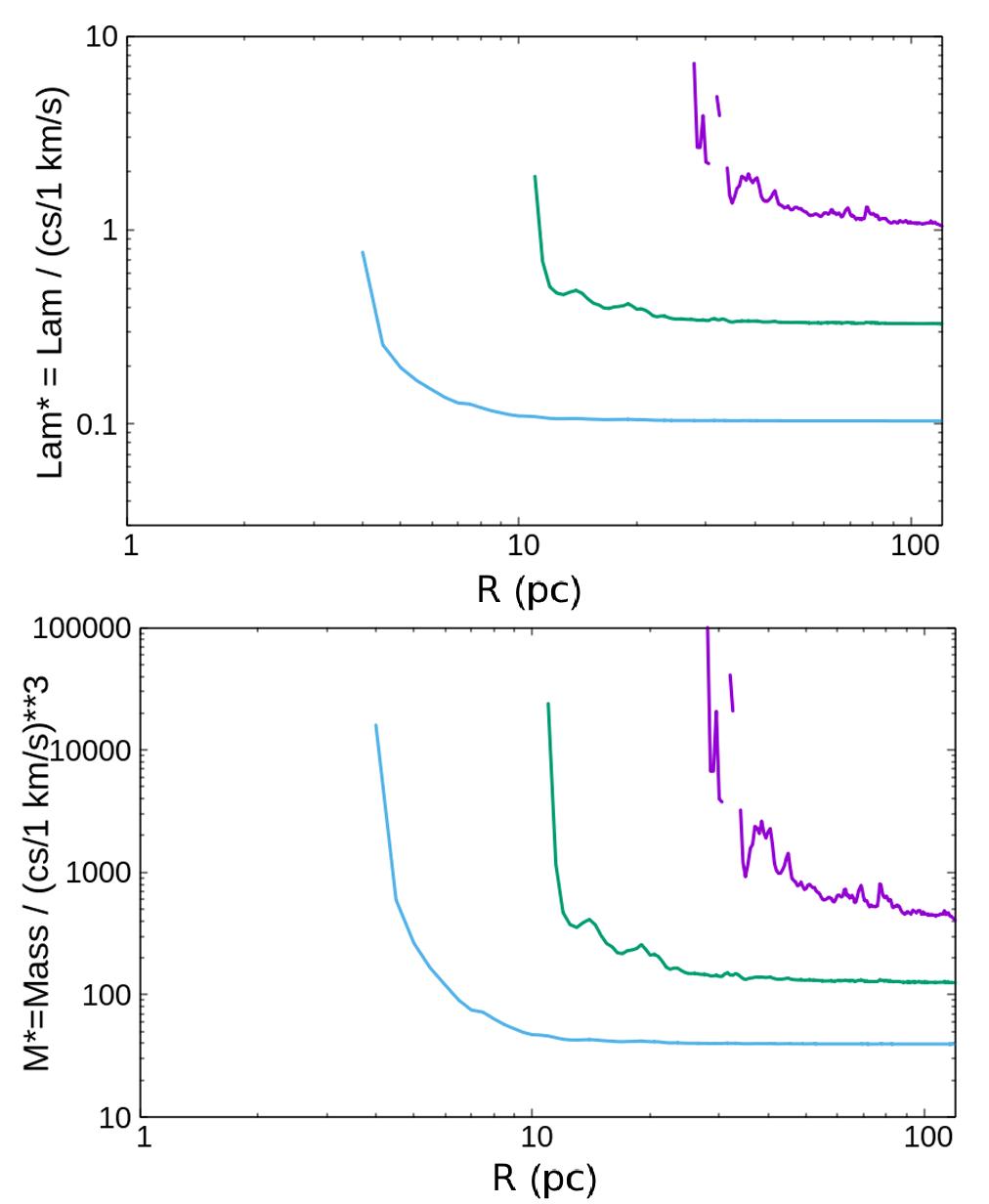}  
 \end{center} 
 \caption{[Top] 
 $L_*=\lambda_{\rm TJ}/(\c/{\rm km~s^{-1}})$ and [bottom] 
 $M_*=M_{\rm TJ}/(\c/{\rm km~s^{-1}})^3$ for gas densities from 
 $n_{\rm H_2}=10^4$ (top right curve) to $10^6 \mu$ (bottom left curve) \Htwocc using the observed rotation curve. 
 {Alt text: Normalized TJ length and mass for various gas densities using the observed rotation curve.}
 }
\label{fig-TJeans-RC}
 \end{figure}

Another interesting issue is the unpredictable behavior of the tidal effect in the ``transition region" in the plots $\lambda-R$ and $M_{\rm TJ}-R$ in figures \ref{fig-TJeans} and \ref{fig-TJeans-RC}, where the curves suddenly turn from horizontal (outside $R_{\rm T}$) to vertical (inside $R_{\rm T}$).
In this region, the molecular gas is neither stable nor unstable regardless of its high density.
It happens that $M_{\rm TJ}$ varies inside a proto-cluster cloud if its size is comparable to $R$.

Suppose an extremely high density cloud of $\rho\sim 10^6-10^8$ \Htwocc\ and $\c \sim 0.1$ \kms with a size comparable to orbital radius of $R\sim 1-10$ pc, the T-Jeans mass, which defines the cutoff mass, is variable inside the cloud from solar to high mass.
This variable cut-off point for lower-mass clumps in some parts of the same cloud leads to a shallower initial mass function (IMF).

It is interesting to point out that a ``top-heavy" IMF is indeed observed in the GC Arches Cluster G+0.12+0.02 \citep{2019ApJ...870...44H} and in the Quintuplet cluster G+0.15-0.05 \citep{2012A&A...540A..57H}.
However, it may be less likely that these cases are explained only by tides because the clusters are located at projected distances of $\sim 25-30$ pc from \sgrastar, outside the tidal transition radius.

\ss{Magnetic effect}
\def\ca{c_{\rm A}}

In the presence of a magnetic field $B$, the dispersion relation for a perturbation perpendicular to the field lines is written as \citep{chandra1961}
\be
\omega^2=(\c^2+\ca^2) k^2-4\pi G\rho=0, 
\ee
where $\ca=\cos ~\theta ~B/\sqrt{4\pi \rho}$ is the Alfv\'en velocity with $\theta$ being the angle between the magnetic field and the direction of contraction. 
This increases the critical wavelength in the direction perpendicular to the magnetic field without change of the horizontal lengths, and the critical mass increases accordingly.
The magnetic strength in the GC has been measured from radio continuum observations to be $B\sim 1$ mG \citep{2022ApJ...925..165H}, which yields $\ca\sim 4$ \kms, leading to a horizontal wavelength $\lambda_{\rm BJ}\sim 0.4$ pc and mass $M_{\rm mag-J}\sim 45\Msun$ for $\nH2\sim 10^5$ \Htwocc.
So, the formation of solar-mass stars is possible from a sheet produced by an initial vertical condensation of wavelength $\lambda_{\rm J}$, which increases the local density to cause perpendicular fragmentation of a sub-solar mass clump.

\ss{Suppression of cloud collision}

Cloud collision is often argued to trigger star formation in interstellar space \citep{2021PASJ...73S...1F}. 
The collision requires a head-on orbit between the two clouds with zero angular momentum.
However, a straight orbit is prohibited in the GC because of the strong Coriolis force.
In the rotating disk, an object orbits around its guiding center at the epicyclic frequency $\kappa\sim \sqrt{2}\Omega$ and radius $r\sim R(v/\sqrt{2}V)$ with $v$ being the local velocity about the guiding center (velocity dispersion).
Once a collision has occurred, the two clouds have had to meet repeatedly in the past every $t_{\rm col}\sim t_{\rm epi}=2\pi/(\sqrt{2}\Omega$).
For a velocity dispersion $v\sim 20$ \kms\ and $R\sim 10$ pc, for example, we have $r\sim 1.2$ pc and $t_{\rm col}\sim 0.4$ My.
Therefore, the two clouds have been binary and the motion is a straight oscillation when they are observed in the coordinate system rotating at $\kappa$.
This analysis applies to any cloud-cloud collision candidate in the Galaxy, but it is particularly serious in such a rapidly rotating disk as is considered here.
So, the triggered star formation by cloud collision is not effective in the GC.

\ss{Feedback from the environment}

In addition to tidal and magnetic suppression of density fluctuations growth, various mechanisms are considered to act to disturb and heat the molecular gas against gravitational contraction in the central few pc of the nucleus, such as stellar winds from the extremely large number of stars inside each fluctuation, supernova explosions, and AGN activities around \sgrastar.
Although these are beyond the scope of this paper, we mention that the formation of a star in the circumnuclear region requires an extraordinarily strong mechanism to compress the protostellar gas clump.

\ss{Tidal dam toward starburst}
Due to suppression of star formation, the gas that has become stagnant is trapped in the circumnuclear region, forming a super-high-density ring-shaped torus. 
Tidal forces act as embankments in a dam. 
However, when the amount of gas in the torus exceeds its limit and this dam breaks, a starburst occurs, whereas some non-gravitational compression to conquer the tide has to be activated to form a star.
This is a contradictory requirement because the collision that is anticipated for shock compression is prohibited, as discussed above.
Thermal instability may be a possible source of compression, although it is beyond the scope of this paper.

\section{Discussion: consistency with the current works} 
\label{sec-consis} 

\ss{Global vs. innermost gas dynamics} 
 
Our analysis is consistent with the current overall picture of the CMZ, but introduces new aspects regarding its more detailed structure based on the high-resolution data from ACES.
In particular, the innermost dynamics and ISM conditions in the CND within $R\sim 20$ pc have been analyzed in the most detail from the perspective of galactic dynamics. 

We do not specifically address the individual structures of the entire CMZ, but refer to related studies which are categorized into two types of models: 
The first is the "single structure model" where the CMZ is assumed to be composed of a single ellipse \citep{2011ApJ...735L..33M,lon13} or an eccentric stream \citep{k15,h16,2025ApJ...984..159L},
and the other is the "multi-arm model" \citep{1995PASJ...47..527S,2022MNRAS.516..907S,sofue+25a}.

The former takes the perspective that all the named clouds (Sgr B, C, Brick, 50 and 20 \kms clouds, etc) orbit on a single twisted, open (eccentic streams) or closed (twisted ellipse) eccentric trajectory, with galactocentric radii between $R\sim 45$ and $\sim$130 pc.
The latter assumes several arms in the framework of the density wave theory, where Arms I and II compose the 120 pc ring and Arm III to VI are inside arms. The 20 and 50 \kms\ clouds belong to III and V at $R\sim 40$ and $\sim 10$ pc, respectively, as measured using the $dv/dl$ method on the LVDs.

This study does not aim to distinguish between these models, but to explore the physical picture of the multi-arm model. However, since we use the terminal velocities (rotation curve) measured on the LVDs, the result depends little on the models.

\ss{Rotation and gravitational potential} 

The mass distribution derived using the rotation curve is consistent with current photometric measurements (figure \ref{fig-mass}).
The gravitational potential at $R\lesssim 20$ pc favors the cusp-shaped density profile as $\propto R^{-2}$ with flat rotation as used in \citet{binney+1991}.
It generally prefers a density model with the central cusp that includes a term $\propto R^{-\gamma} ~(\gamma > 0)$ to the plateau type (exponential or Plummer type) tending to a constant density at the center (see \citet{sorma+20mass}). 
Figure \ref{fig-compare} demonstrates this, comparing the cusp type density model adopted in the hydrodynamical simulation of the CMZ by \citet{2017MNRAS.469.2251R} with the present ACES result.
\ss{CND stability and low star formation efficiency}
The tidal stability analysis supports the recent detailed analysis of the CND using ALMA observations by \citet{hs21}, who report that half of the CND's total gas mass is capable of star formation, but no clear evidence of ongoing star formation has been found. They argue that this is because tidal forces dominate the gravitational contraction of the molecular clumps, strongly inhibiting star formation. 

\section{Summary}
\label{summary}

We have shown that the observed characteristics of the circumnuclear gas disk in the \cs\ line are explained only when the gravitational potential of the central bulge is nearly spherical, where the rosette orbits are coplanar, but allowing for inclination from the galactic plane.
We determined the rotation curve in the CMZ, which is nearly flat at $\vrot\sim 100$ \kms, and derived the distribution of the mass, which is approximately represented by
$M(R) \sim 3\times 10^6 (R/1~{\rm pc}) \Msun$ at $R\gtrsim 2$ pc.
This made it possible to analyze the stability of the molecular gas in the GC using the modified Jeans instability in a rotating system.
We showed that the molecular gas is stable against gravitational fragmentation for star formation in the circumnuclear region inside the threshold radius ($R\lesssim 14 (\rho_{\rm gas}/10^5 \mu {\rm H_2 cm^{-3}})^{-1/2}$ pc), where the star formation is suppressed and the SF law may be modified to have a top-heavy IMF.

\scriptsize{
\begin{ack}
This paper makes use of the following ALMA data: ADS/JAO.ALMA$\#$2021.1.00172. 
 
ALMA is a partnership of ESO (representing its member states), NSF (USA) and NINS (Japan), together with NRC (Canada), NSTC and ASIAA (Taiwan), and KASI (Republic of Korea), in cooperation with the Republic of Chile.

The Joint ALMA Observatory is operated by ESO, AUI/NRAO and NAOJ.
  
  The data analysis in this paper was performed partially at the Astronomical Data Center of the National Astronomical Observatories of Japan.
  
This study was supported by JSPS KAKENHI Grant Number JP24H00004.



A.G. acknowledges support from the NSF under AAG 2206511 and CAREER 2142300.

FNL gratefully acknowledges financial support from grant PID2024-162148NA-I00, funded by MCIN/AEI/10.13039/501100011033 and the European Regional Development Fund (ERDF) “A way of making Europe”, from the Ramón y Cajal programme (RYC2023-044924-I) funded by MCIN/AEI/10.13039/501100011033 and FSE+, and from the Severo Ochoa grant CEX2021-001131-S, funded by MCIN/AEI/10.13039/501100011033.

 
 
L.C., V.M.R. and I.J.-S. acknowledge support from the grant PID2022-136814NB-I00 by the Spanish Ministry of Science, Innovation and Universities/State Agency of Research MICIU/AEI/10.13039/501100011033 and by ERDF, UE. 
 
V.M.R. also acknowledges support from the grant RYC2020-029387-I funded by MICIU/AEI/10.13039/501100011033 and by "ESF, Investing in your future", from the Consejo Superior de Investigaciones Cient{\'i}ficas (CSIC) and the Centro de Astrobiolog{\'i}a (CAB) through the project 20225AT015 (Proyectos intramurales especiales del CSIC); and from the grant CNS2023-144464 funded by MICIU/AEI/10.13039/501100011033 and by “European Union NextGenerationEU/PRTR”.

K.F and M.C.S acknowledge financial support from the European Research Council under the ERC Starting Grant “GalFlow” (grant 101116226). M.C.S further acknowledges financial support from the Fondazione Cariplo under the grant ERC attrattivit\`{a} n. 2023-3014.


 



A.S.-M.\ acknowledges support from the RyC2021-032892-I grant funded by MCIN/AEI/10.13039/501100011033 and by the European Union `Next GenerationEU’/PRTR, as well as the program Unidad de Excelencia María de Maeztu CEX2020-001058-M, and support from the PID2023-146675NB-I00 (MCI-AEI-FEDER, UE).



\end{ack}

 
\section*{Data availability} 
The interferometer data were taken from the ACES internal release version of the 12m+7m+TP (Total Power)-mode data from the ALMA cycle 8 Large Program "ALMA Central Molecular Zone Exploration Survey" (ACES, 2021.1.00172.L).
 
\section*{Conflict of interest}
The authors declare that there is no conflict of interest.
} 


\begin{appendix}

\section{Longitude-velocity (LV) and latitude-velocity (BV) channel diagrams in the \cs\ line.}

\begin{figure*}   
\begin{center}    
\cs\ channel LVD \hskip 8cm chnnel BVD\\
\includegraphics[width=.49\lw]{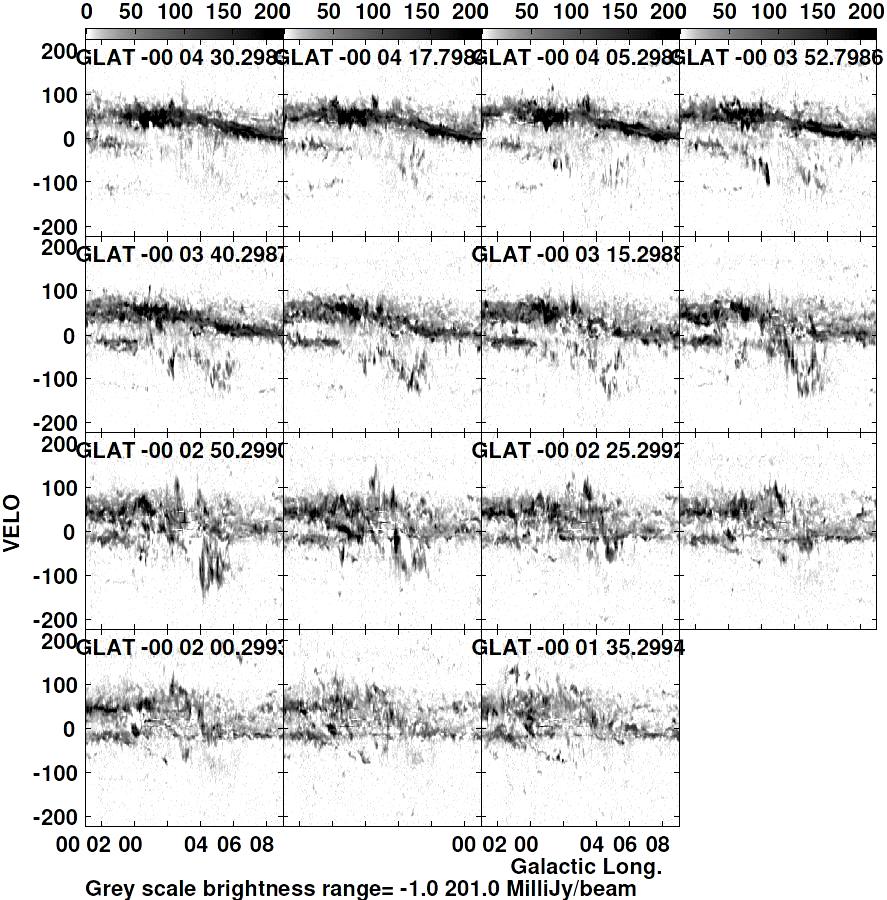}    
\includegraphics[width=.49\lw]{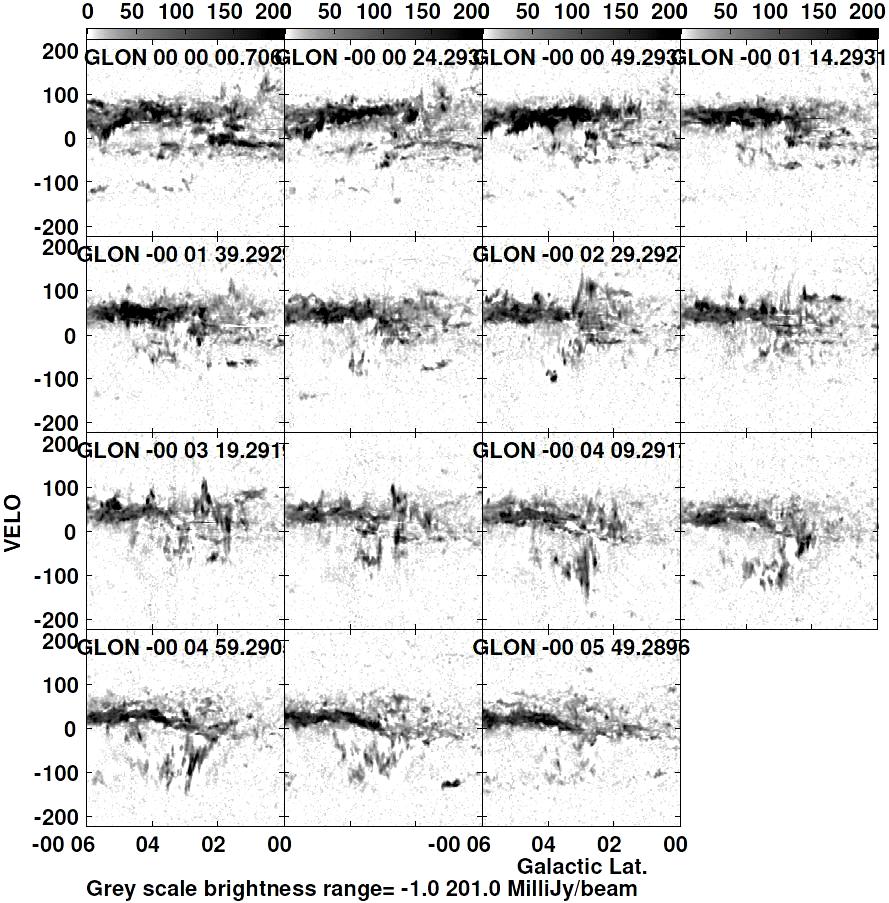}    
\h40\ channel LVD\\
\includegraphics[width=.49\lw]{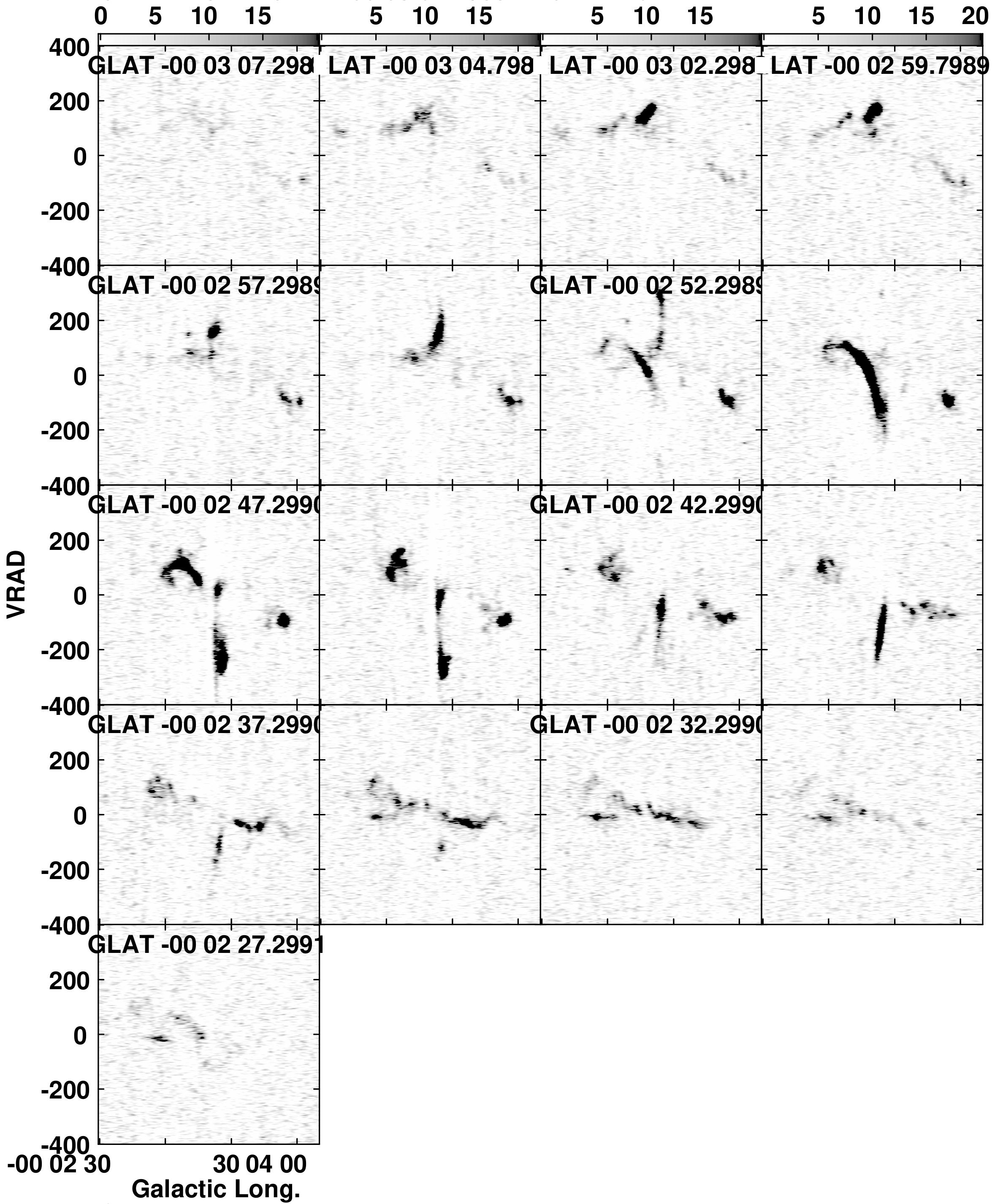}  
\end{center} 
\caption{[Top]Channel LVD (left) and BVD (right) of the circum-nuclear region in the \cs\ line from ACES.
[Bottom] Channel maps LVD of the minisrpial in the \h40\ line.
{Alt text: [Top] Channel LVD and BVD of CND in \cs\ and [Bottom] LVD of minispiral in \h40.}}
\label{fig-chan}  
\end{figure*}

Channel maps of the LVD and BVD in the \cs\ line are presented in figures \ref{fig-chan}.

\end{appendix}
\end{document}